\documentclass[aps,pra,twocolumn,amsmath,amssymb,showpacs,superscriptaddress]{revtex4-1}
\usepackage{graphicx}
\usepackage{subcaption} 
\usepackage{dcolumn}
\usepackage[mathlines]{lineno}
\usepackage{physics}
\usepackage{hyperref}
\usepackage{multirow}
\usepackage{bm}
\usepackage{epstopdf}
\usepackage{epsfig}
\usepackage{bbold}
\usepackage[normalem]{ulem}
\usepackage{caption}
\usepackage[dvipsnames]{xcolor}

\usepackage{amsmath}

\newcommand{\rr}{\mathbf{r}}

\begin{document}
\title{Revealing the invariance of vectorial structured light in perturbing media}

\author{Isaac Nape}
\email[author contributions:]{ These authors contributed equally to the work}
\affiliation{School of Physics, University of the Witwatersrand, Private Bag 3, Wits 2050, South Africa}

\author{Keshaan Singh}
\email[author contributions:]{ These authors contributed equally to the work}
\affiliation{School of Physics, University of the Witwatersrand, Private Bag 3, Wits 2050, South Africa}

\author{Asher Klug}
\affiliation{School of Physics, University of the Witwatersrand, Private Bag 3, Wits 2050, South Africa}

\author{Wagner Buono}
\affiliation{School of Physics, University of the Witwatersrand, Private Bag 3, Wits 2050, South Africa}

\author{Carmelo Rosales-Guzman}
\affiliation{Centro de Investigaciones en \'Optica, A. C., Loma del Bosque 115, Col. Lomas del Campestre, 37150, Le\'on, Gto., M\'exico}
\affiliation{Wang Da-Heng Collaborative Innovation Center for Quantum Manipulation and Control, Harbin University of Science and Technology, Harbin 150080, China}

\author{Sonja Franke-Arnold}
\affiliation{School of Physics and Astronomy, University of Glasgow, Glasgow G12 8QQ, Scotland}

\author{Angela Dudley}
\affiliation{School of Physics, University of the Witwatersrand, Private Bag 3, Wits 2050, South Africa}

\author{Andrew Forbes}
\email[email:]{ andrew.forbes@wits.ac.za}
\affiliation{School of Physics, University of the Witwatersrand, Private Bag 3, Wits 2050, South Africa}
\email[Corresponding author: ]{andrew.forbes@wits.ac.za}
\date{\today}

\begin{abstract}
\noindent \textbf{Optical aberrations have been studied for centuries, placing fundamental limits on the achievable resolution in focusing and imaging. In the context of structured light, the spatial pattern is distorted in amplitude and phase, often arising from optical imperfections, element misalignment, or even from dynamic processes due to propagation through 
perturbing media such as living tissue, free-space, underwater and optical fibre.  Here we show that the polarisation inhomogeneity that defines vectorial structured light is immune to all such perturbations, provided they are unitary.  By way of example, we study the robustness of vector vortex beams to tilted lenses and atmospheric turbulence, both highly asymmetric aberrations, demonstrating that the inhomogeneous nature of the polarisation remains unaltered from the near-field to far-field, even as the structure itself changes.  The unitary nature of the channel allows us to undo this change through a simple lossless operation, tailoring light that appears robust in all its spatial structure regardless of the medium.  Our insight highlights the overlooked role of measurement in describing classical vectorial light fields, in doing so resolving prior contradictory reports on the robustness of vector beams in complex media. This paves the way to the versatile application of vectorial structured light, even through non-ideal optical systems, crucial in applications such as imaging deep into tissue and optical communication across noisy channels.} 
\end{abstract}
\maketitle

\noindent Non-paraxial light is vectorial in 3D and has given rise to exotic states of structured light \cite{forbes2021structured} such as optical skyrmions \cite{shen2021supertoroidal, gao2020paraxial}, knotted strands of light \cite{larocque2018reconstructing,galvez2014generation}, flying donuts \cite{zdagkas2021observation,keren2019generation} and M\"obius strips \cite{bauer2015observation}.  Paraxial light too is vectorial, in 2D, characterised by an inhomogeneous polarisation structure across the transverse plane \cite{brown2011unconventional}.  Vectorial structured light in 2D and 3D has been instrumental in a range of applications (see Refs.~\cite{wang2020vectorial,otte2018polarization,vector-review,forbes2019quantum} and references therein), for example, to drive currents with a direction dictated by the vectorial nature of the optical field \cite{sederberg2020vectorized,fang2021photoelectronic}, imprinting the spatial structure into matter \cite{el2021sensitive}, enhanced metrological measurements \cite{Hawley2019,fang2021vectorial}, 
probing single molecules \cite{curcio2020birefringent}, and to encode more information for larger bandwidths \cite{Milione2015,zhang2021fiber,zhu2021compensation,zhao2015high}.  They are easy to create in the laboratory using simple glass cones \cite{radwell2016achromatic}, stressed optics \cite{beckley2010full} and GRIN lenses \cite{he2019complex}, as well as from spatial light modulators and digital micro-mirror devices \cite{rosales2020polarisation,chen2021compact}, non-linear crystals \cite{wu2019vectorial,tang2020harmonic}, geometric phase elements \cite{Marrucci2006,nassiri2018multispectral}, metasurfaces \cite{devlin2017arbitrary} and directly from lasers \cite{forbes2019structured}. 

Given the importance of these structured light fields, much attention has focused on their propagation through optical systems that are paraxial \cite{beckley2012full}, guided \cite{ma2020propagation} and tight focussing \cite{biss2004primary,youngworth2000focusing}, as well as in perturbing media such as turbid \cite{mamani2018transmission,gianani2020transmission,biton2021oam,suprano2020propagation}, turbulent \cite{cox2020structured,gu2009scintillation,cheng2009propagation,cai2008average,ji2010propagation,wang2008propagation,Cox:16,lochab2018designer} and underwater \cite{hufnagel2020investigation,bouchard2018quantum,ren2016orbital}. The conclusions are seemingly contradictory, with compelling evidence that the vectorial structure is stable, and equally compelling evidence that it is not, while the specific nature of each study prohibits making general conclusions on the robustness of vectorial light in arbitrary complex media.

Here we show that the inhomogeneity of a vectorial light field is impervious to all aberrations so long as they are unitary.  We demonstrate this with two examples featuring strong aberrations:
a tilted lens and atmospheric turbulence, treating each as a unitary channel.  Our quantum-inspired framework explains the robustness of these light fields by virtue of unitary operations on the vectorial state, manifesting as an intact inhomogeneity even if the vectorial pattern appears spatially distorted due to modal scattering in the component scalar spatial modes.  We show that the channel action can be reversed by a reciprocal unitary process applied either pre-channel or post-channel, demonstrating full scalar and vector restoration, as well as cross-talk free detection of vectorial fields through complex media.  Our study highlights the importance of measurement in the context of vectorial light fields, in doing so resolving a standing paradox on the robustness of vectorial light to perturbations, and provides a general framework for understanding the impact of arbitrary optical aberrations on vectorial structured light fields.

\section*{Results}

\noindent \textbf{Vectorial light and unitary channels.}  A vectorial structured light field can be written compactly in a quantum notation as the unnormalised state $\ket{\Psi} =  \ket{e_1}_A \ket{u_1}_B + \ket{e_2}_A\ket{u_2}_B$, highlighting the non-separable nature of the two degrees of freedom (DoFs), $A$ and $B$, denoting the polarization and the spatial degree of freedom, respectively. In this way, the vectorial field is treated as a quantum-like state (but not quantum and without non-local correlations), by virtue of its non-separable DoFs, akin to a locally entangled state \cite{Spreeuw1998,forbes2019classically,kagalwala2013bell,qian2011entanglement}.  The polarisation DoF is expressed as any pair of orthonormal states, $\{ \ket{e_1}$ and $\ket{e_2} \}$ while the spatial mode DoF is given by the orthonormal basis states $\{ \ket{u_1}$ and $\ket{u_2} \}$. In a quantum sense, this vectorial structured field would be called a pure state. The vectorial nature can be quantified through a measure of its non-separability \cite{McLaren2015}, a Vector Quality Factor (VQF) (equivalently concurrence) which for succinctness we will call its ``vectorness'', ranging from 0 (homogeneous polarisation structure of scalar light) to 1 (ideally inhomogeneous vectorial polarisation structure).  

\begin{figure}[h]
	\includegraphics[width=\linewidth]{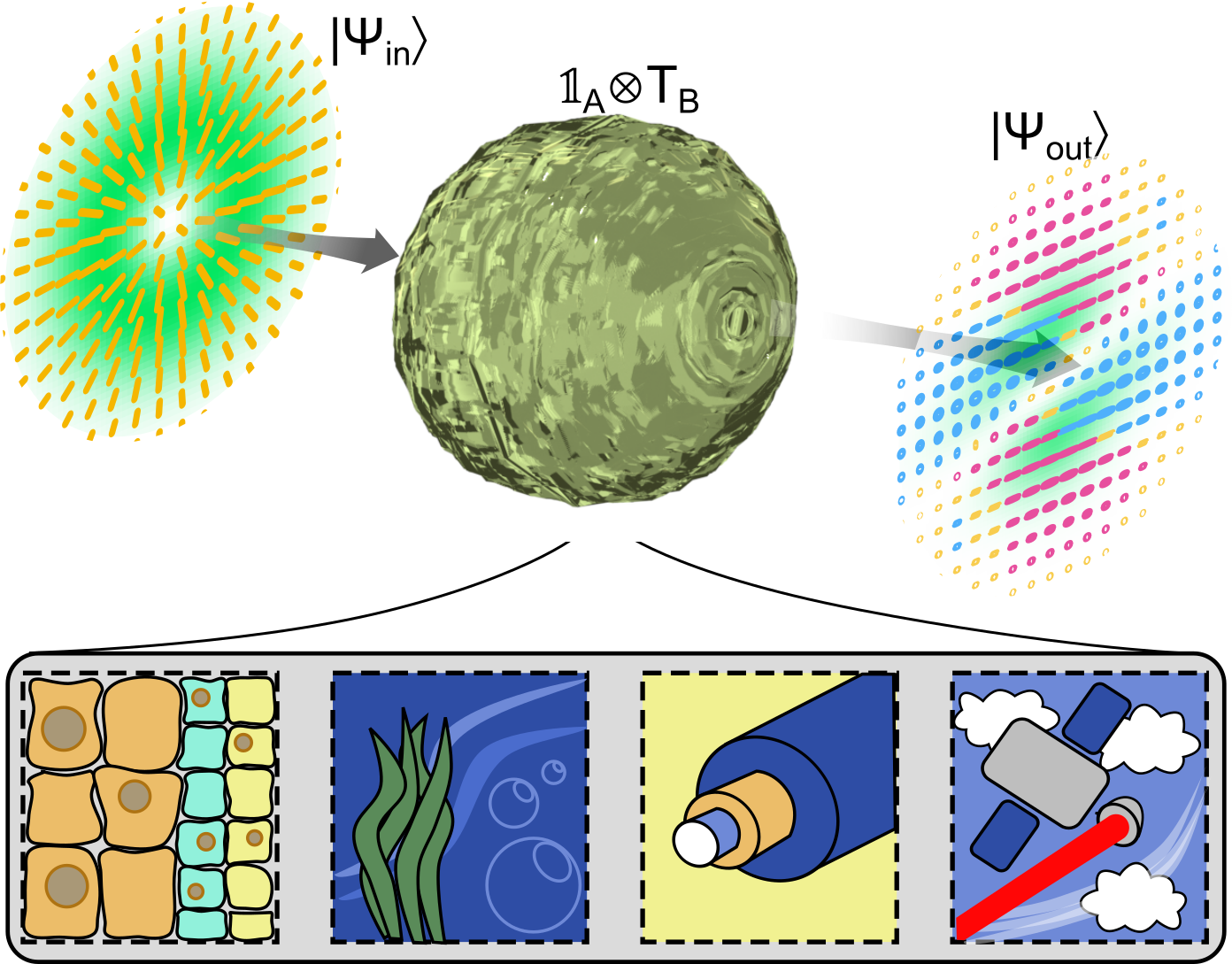}
	\caption{\textbf{Concept of vectorial fields undergoing a unitary transformation.} A vectorial light field, $\ket{\Psi_{\text{in}}}$, propagates through a perturbing medium and undergoes a unitary transformation, $\mathbb{1}_A \otimes T_B$, acting only on the spatial components. Such perturbing media can be organic tissue, underwater, optical fiber or the turbulent atmosphere (bottom panels). The resulting field, $\ket{\Psi_{\text{out}}}$, has a distorted intensity and polarisation structure.}
	\label{fig:conceptfig}
\end{figure}
\begin{figure*}[t!]
	\includegraphics[width=\linewidth]{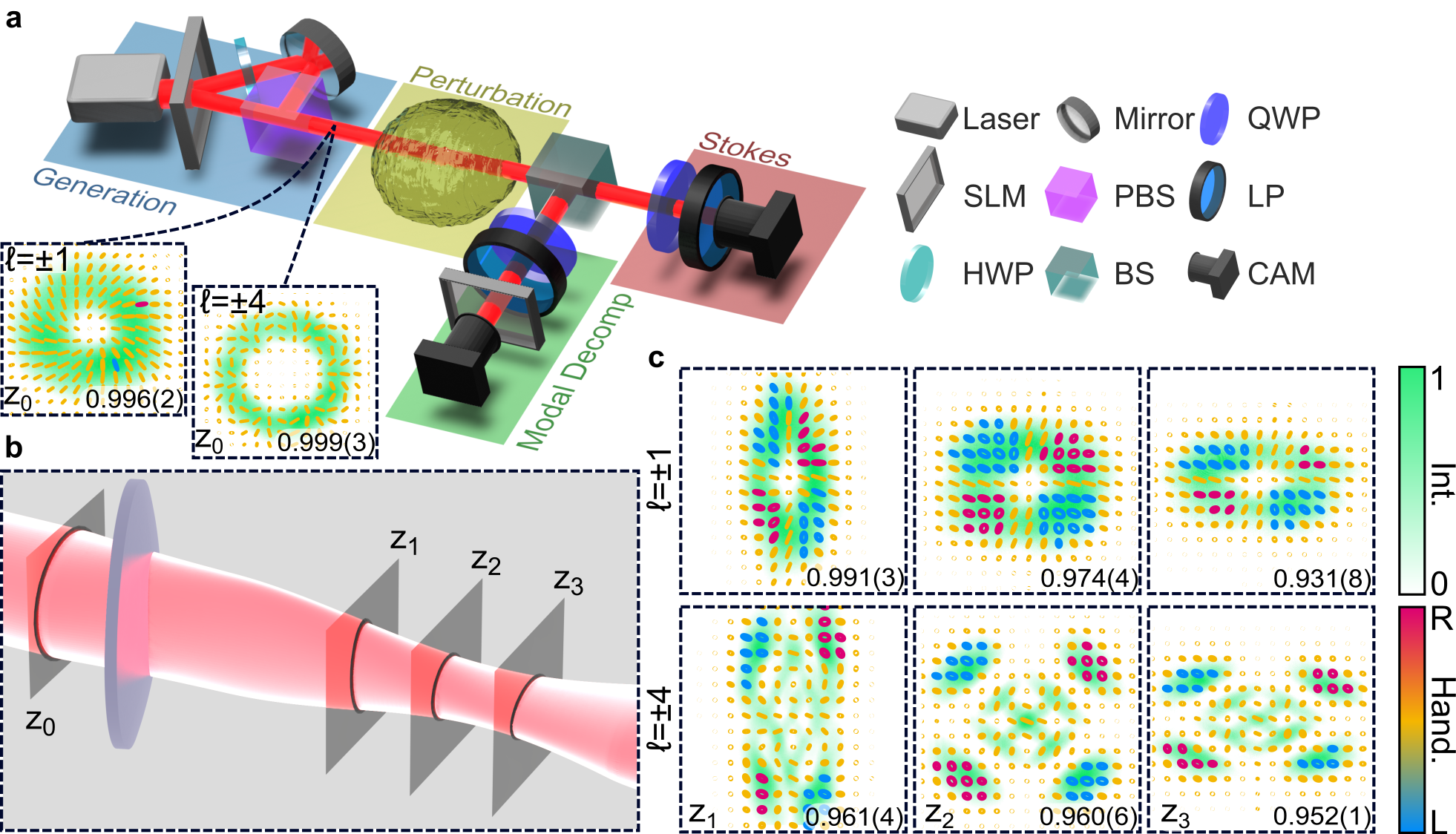} 

	\caption{\textbf{Vectorial light through a tilted lens.} (a) The experiment has four stages: a generation stage to create the vectorial fields, a perturbation stage to pass it through a perturbing medium, and two detection stages to perform Stokes projections and modal decomposition.  The insets show the initial $\ell = \pm 1$ (left panel) and $\ell = \pm 4$ (right panel) beams at plane $Z_0$.  (b) Illustration of the propagation through a tilted lens, with exemplary measurement planes indicated as $Z_1$ through to $Z_3$. (c) The output $\ell = \pm 1$ and $\ell = \pm 4$ beams at these distances, confirm that the spatial structure alters with distance but that the vectorness (the inset number to each beam frame) does not, remaining above $93\%$. The optical elements comprise a HWP: half-wave plate, PBS: polarising beam splitter, Pol: polariser, BS: beam splitter, QWP: quarter-wave plate, and CAM: camera. All beam profiles are shown as false colour intensity and polarisation maps.}
	\label{fig:lens}
\end{figure*}

Now imagine that our vectorial light field, $\ket{\Psi_\text{in}}$, passes through an arbitrary aberrated optical channel, as illustrated in Figure ~\ref{fig:conceptfig}, e.g., imperfect optics or a perturbing medium such as tissue, turbulence or underwater channels.  If there is no birefringence (often there is not, but see discussion later) then this can be considered a one-sided channel since the spatial mode (DoF $B$) is affected (distorted), while the polarisation (DoF $A$) is not. An open question is whether this implies that the entire polarisation structure of the state is likewise unaffected?  To answer this, we note that such a one-sided channel is unitary to any input vectorial (pure) state since it may be written as a positive trace-preserving map, ensuring that the output must also be a vectorial (pure) state, with full details provided in the Supplementary Information (SI).  The Choi-Jamiolkowski isomorphism \cite{jiang2013channel} establishes a correspondence between the channel operator, $T_B$, and a quantum state, so that a measurement on one returns the other, invaluable for characterising quantum channels \cite{konrad2008evolution,valencia2020unscrambling}.  Applying this to our classical beam, justified because of its non-separability \cite{Ndagano2017}, the vectorial state after the channel is then $\ket{\Psi_\text{out}} = (\mathbb{1}_A \otimes T_B)\ket{\Psi_\text{in}}$, where $\mathbb{1}_A$ is the identity operator for DoF $A$ with the subscript in $\ket{\Psi}$ indicating the input and output states.  Importantly, the spatial components remain orthogonal even after a distorting medium:  if the channel operator results in new (``distorted'') modes in DoF $B$ given by $\ket{v_1}$ and $\ket{v_2}$, then one finds that $\braket{v_1}{ v_2 } = \bra{u_1} T_B^{\dagger}  T_B \ket{u_2} = \braket{u_1}{u_2} = 0$.  For this reason, the output remains a non-separable vector beam with the same vectorness as the initial beam, although altered in its spatial structure, and now expressed in what we will call an adjusted basis, $\{ \ket{v_1}$ and $\ket{v_2} \}$.  Thus our description of the vectorial state has changed, it is now $\ket{e_1}_A \ket{v_1}_B + \ket{e_2}_A\ket{v_2}_B$, and (in general) its spatial structure too has altered in amplitude, phase and polarisation, but not its vectorial inhomogeneity (see SI for the full proof).  

\vspace{0.5cm}
\begin{figure*}
	\includegraphics[width=0.95\linewidth]{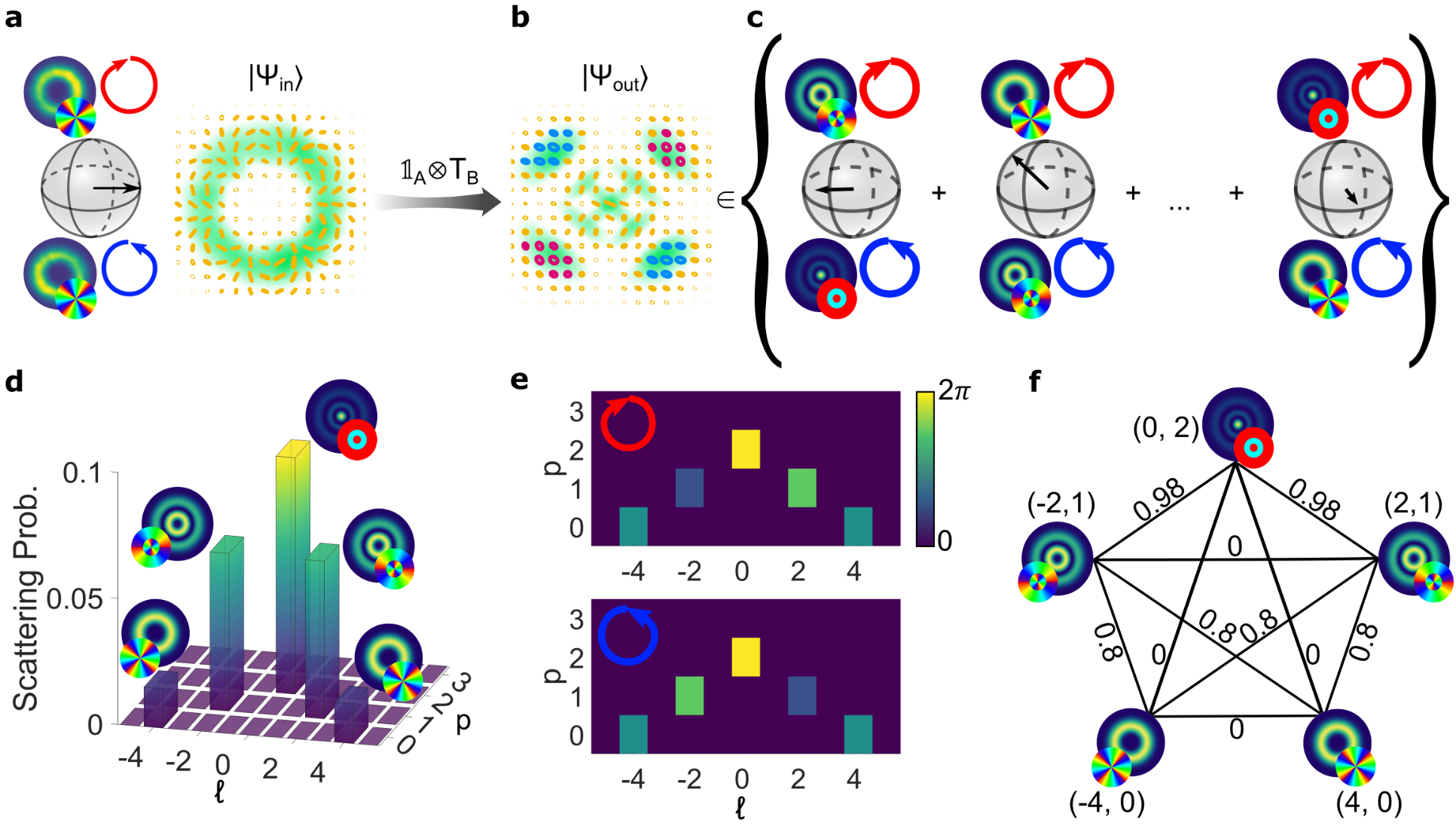}
		\caption{\textbf{Impact of scattering across multiple subspaces}. (a)  The initial vector beam is geometrically represented by an equatorial state vector of unit length on a single $\ell = \pm 4$ HOP Sphere, indicative of a maximally non-separable pure state. The experimental basis states are shown on the poles (as intensities with their respective polarisation state on the right) and the vectorial sum (amplitude and polarisation structure) is shown on the equator. The tilted lens deforms the initial mode into (b), and this new state can be mapped to the sum of multiple HOP Spheres, each spanned by a different OAM and radial mode (c). Each of these potentially describes a mixed state of some degree of non-separability (non-unit and/or non-equatorial state vectors).  The (d) scattering probabilities and (e) phases over the OAM ($\ell$) and radial  ($p$) modes for the right (left) ($R(L)$)  circular polarisation components after the $\ell = \pm 4$ vectorial field traverses the tilted lens channel. The right and left components have equivalent scattering probabilities. There are $N = 5$ states, $(\ell, p)$, with non-zero probabilities and $N(N-1)/2 = 10$ pairs of xcmultiple subspaces that can be formed. (f) The pairs of states constituting the subspace and corresponding VQF for the vector field in that particular subspace can be represented with a graph having a set of vertices (as the contributing states) connected by weighted edges, respectively. The weights of the edges correspond to the VQF.}
	\label{fig:basisscattering}
\end{figure*}

Because the channel (with or without propagation) is unitary, the channel operator is unitary too, so that its Hermitian adjoint represents the inverse process
\begin{equation}
\ket{\Psi_\text{in}} = (\mathbb{1}_A \otimes T_B)^\dagger \ket{\Psi_\text{out}},
\label{eq:channel}
\end{equation}

\noindent pointing to a recipe that will unravel the vectorial light back to its original form without adaptive optics (see SI).  We will demonstrate this correction both pre- and post-channel.

\vspace{0.5cm}

\noindent \textbf{Experimental demonstration: the tilted lens.}  To validate this perspective, we built the set-up shown in Figure~\ref{fig:lens} (a), first creating our test vectorial fields before passing them through some perturbing medium.  Without any loss of generality, we chose the left- and right-circular basis, $\{ \ket{e_1}\equiv \ket{R}$ and $\ket{e_2} \equiv \ket{L} \}$, for the polarisation degree of freedom and spatial modes imbued with orbital angular momentum (OAM) following $\{ \ket{u_1} \equiv \ket{\ell}$ and $\ket{u_2} \equiv \ket{-\ell} \}$, with $\ell$ the topological charge, forming the topical cylindrical vector vortex beams \cite{zhan2009cylindrical}. The resulting vectorial field was then analysed by both Stokes measurements and  modal decomposition  (see Methods).  The superimposed intensity and polarisation (ellipse) profile of the initial beams (no perturbation) are shown in the inset of Figure~\ref{fig:lens} (a) for $\ell = \pm 1$ and $\ell = \pm 4$, both with a radial mode of $p=0$, with corresponding mode numbers $N=2p+|\ell|+1$ of 2 and 5 respectively.

Next, we pass these beams through a highly aberrated system, a tilted lens, illustrated in Figure~\ref{fig:lens} (b).  This is known to severely distort OAM modes and is routinely used as an OAM detector by breaking the beam into countable fringes \cite{Vaity2013}.  The results at illustrative distances ($Z_1$ to $Z_3$) after the lens are shown in Figure~\ref{fig:lens} (c).  The superimposed intensity and polarisation profiles reveal that while the vectorial structure distorts as one moves towards the far-field, the inhomogeneity as measured by the vectorness does not, corroborated by the vectorness of each beam (reported in the insets), all remaining above $ 93\%$, i.e., remaining fully vectorial as predicted by the unitary nature of the channel. 


 In contrast we see that the intensity profiles change morphology, and concomitantly polarisation structure. This change can be explained by the coupling of modes outside the original subspace by virtue of the channel operator: the channel scatters the original OAM modes 
 into new mode sets that maintain the same mode number as the input modes.
 We can visualise this using the Higher Order Poincar{\'e} (HOP) Sphere, a geometric representation of vectorial structured light \cite{milione2011higher,holleczek2011classical}. The case for $\ell = \pm 4$ is shown in Figure~\ref{fig:basisscattering} as an illustrative example.  The  initial cylindrical vector vortex beam is visualised as an equatorial vector of unit length (a pure state), shown in Figure~\ref{fig:basisscattering} (a).  The channel (our tilted lens) maps this initial state to a new field, shown in Figure~\ref{fig:basisscattering} (b), expressed across multiple HOP Spheres, illustrated in Figure~\ref{fig:basisscattering} (c). 
 
 




\begin{figure*}[t]
	\includegraphics[width=\linewidth]{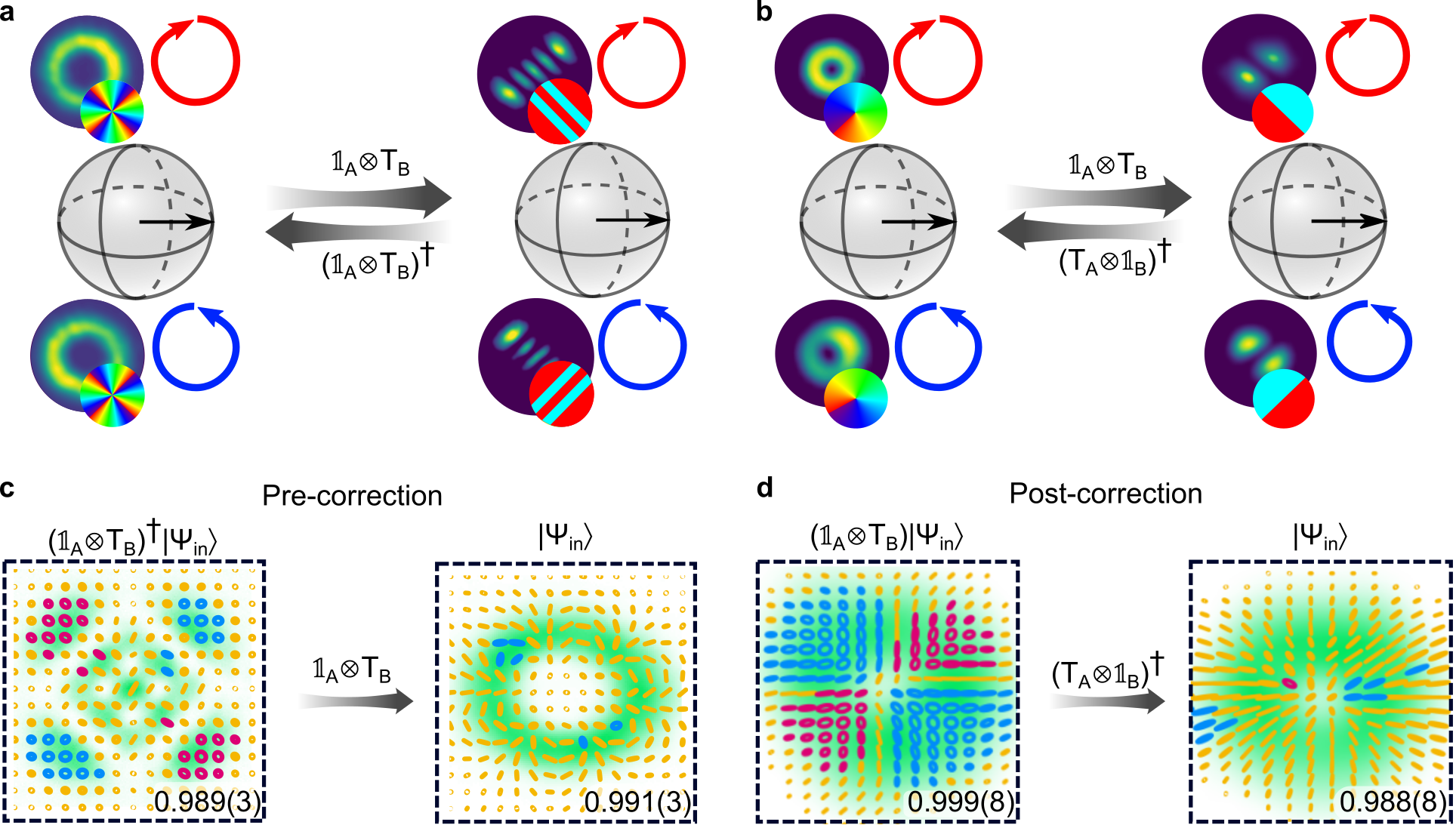}
	\caption{\textbf{The unitary channel mapping and its inversion.} (a) The unitary channel (tilted lens) $\mathbb{1}_A \otimes T_B$ maps the initial $\ell\pm4$ vector field from the equator of its s corresponding subspace onto a (b) new HOP Sphere spanned by an adjusted basis (shown as experimental images on the poles) where the vectorial structure is also a maximally non-separable pure state. Because the unitary channel is a change of basis, the inverse $(\mathbb{1}_A \otimes T_B)^\dagger$ can be applied to map the field back to the original sphere. The inversion can be done (c) pre-channel by inserting the state $(\mathbb{1}_A \otimes T_B)^\dagger \ket{\Psi_\text{in}}$, the conjugated field in the new basis, resulting in the desired corrected field $\ket{\Psi_\text{in}}$ after transmission. For (d) post-channel correction, the unitary nature of the channel is inverted after the channel. We show this for the $\ell=1$ vector field, where the operation is simply a quarter waveplate, since $\mathbb{1}_{A} \otimes T_{B} = T_{A} \otimes  \mathbb{1}_{B}$ for this subspace.}
	\label{fig:adjustedbasis}
\end{figure*}

 The new HOP Spheres are made up of all modal pairings that have non-zero modal powers (scattering probabilities). 
 We quantify this by reporting the scattering probabilities and phases for every subspace, shown in Figures \ref{fig:basisscattering} (d) and (e), respectively. The initial modes $(\ell_\text{in} = \pm4, p_\text{in} = 0)$ now only contain $\approx$12\% of the total modal power, with new modal subspaces emerging to carry the rest.  The HOP Spheres are made of \textit{pairings} of these modes, one on each pole, but not all contribute to the vectorness.  To determine the contributing pairs, we measure the vectorness for every possible pairing (there are $N(N-1)/2 = 10$ possibilities), with the results shown as a graph in Figure \ref{fig:adjustedbasis} (f). In the graph, each vertex corresponds to a non-zero state from (d) and the edges represent possible pairings to form a HOP Sphere. The weight of each edge corresponds to the vectorness of that pairing. The initial subspace $(\pm4, 0)$ is no longer a non-separable state, with a vectorness of 0, while some of the new subspaces (new HOP Spheres) can be as high as 98\%, i.e., pure vectorial states.  The graph can be re-arranged with the zero weighted edges removed to reveal a $K_{3,2}$ bipartite graph structure with two independent vertex sets, $\mathcal{U} =\{ (\pm4, 0), (2, 0)\}$ and $\mathcal{V} = \{ (\pm2, 1) \}$, connected by 6 edges (see SI). Thus of the 10 unique subspaces created from the five scattered modes, only six are non-separable states, occurring for mapping combinations of states between $\mathcal{U}$ and $\mathcal{V}$. One can appreciate that this becomes ever more complicated as the complexity of the medium (and modal scattering) increases.

From this conventional perspective the situation is complicated, but only because the original (OAM) basis modes ($\{ \ket{u_1} \equiv \ket{\ell}$ and $\ket{u_2} \equiv \ket{-\ell} \}$) are used to form the HOP Spheres. We offer a simplification by recognising the unitary nature of the transformation:  one can visualise the action of the channel as a mapping (after the channel operation $(\mathbb{1}_{A} \otimes T_B$) to a \textit{single} HOP Sphere spanned by the new vectors forming an \textit{adjusted basis}, $\ket{v_1}$ and $\ket{v_2}$, shown in Figure~\ref{fig:adjustedbasis} (a). This mapping is a result of channel state duality, where the new spatial basis states that the non-separable vectorial field maps onto are isomorphic to the basis states of the unitary channel operation \cite{jiang2013channel}.  In this new ``adjusted'' HOP Sphere the state vector is again maximally non-separable and pure. The new adjusted basis states are complex structured light fields (one may say ``distorted'' modes due to the noisy channel), shown as experimental images on the poles. In some special cases the mode may be recognisable: for the $\ell = \pm 1$ state shown in Figure~\ref{fig:adjustedbasis} (b) the mapping returns the Hermite-Gaussian modes as the adjusted basis.  Pertinently, the initial state vector in the original HOP Sphere, and the new state vector in the adjusted HOP Sphere are both equatorial and of unit length, representing a maximally non-separable pure state.  This explains why the polarisation structure appears to change (a change in HOP Sphere) yet the inhomogeneity does not (the same state vector in each HOP Sphere).

Since the mapping is unitary it is possible to invert the action of the channel either as a pre-channel action or post-channel correction, both lossless; conveniently, the experimental steps to determine the channel unitary are straightforward: modal and Stokes projections (see SI and Methods). Using a pre-channel corrective step, we insert the adjusted basis vectorial mode into the tilted lens and allow the aberrations to unravel it back to the original initial state, shown in Figure~\ref{fig:adjustedbasis} (c).  For the $\ell = \pm 1$ example the required inversion operator is just a quarter wave plate, which can be derived analytically from Eq.~\ref{eq:channel} (see SI).  The counter-intuitive notion that a polarisation element can be used to correct a channel that acts only on the spatial mode DoF is explained by the fact that the channel correction, $ (\mathbb{1}_{A} \otimes T_B)^\dagger (\mathbb{1}_{A} \otimes T_B) $ can be rewritten as $T_{A}^{\dagger} T_{A}  \otimes T_{B}^{\dagger} T_{B}  = ( T_{A}^\dagger \otimes  \mathbb{1}_{B}) ( T_{A} \otimes \mathbb{1}_{B})$, so that the post-channel unitary, $T_B^{\dagger} = T_A^{\dagger}$, can be applied to the polarisation degree of freedom for $\ell = \pm 1$.  The impact of this unitary is shown in Figure~\ref{fig:adjustedbasis} (d) as a post-channel correction, restoring the full vectorial initial state.  
\begin{figure}[h!]
	\includegraphics[width=\linewidth]{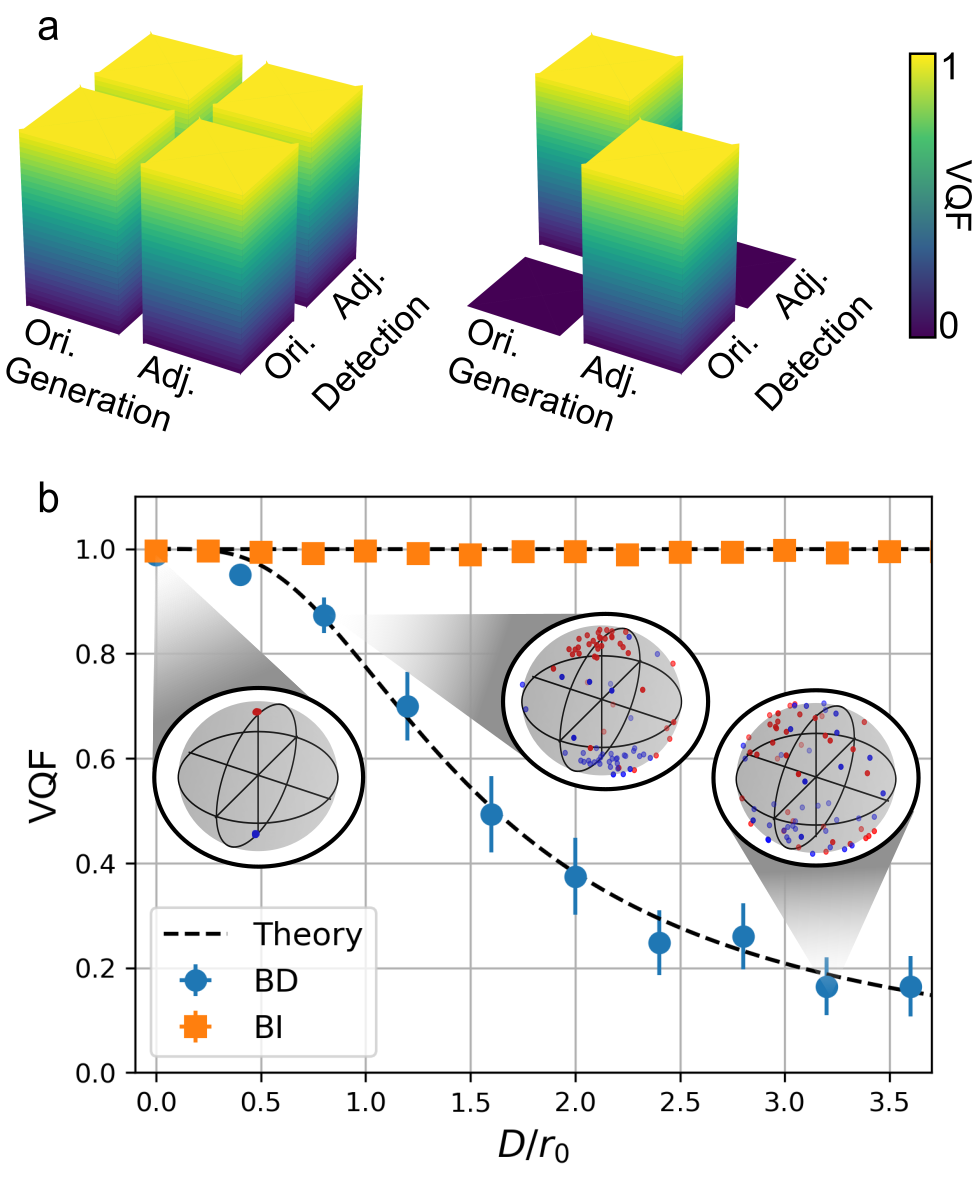}
	\caption{\textbf{The choice of measurement basis.} (a) Measurement of the vectorness in the original (Ori.) and adjusted (Adj.) basis $\ell = \pm 1$ (left) and $\ell = \pm 4$ (right) in the far-field of the tilted lens.  (b) Measurement of the vectorness of an initial $\ell = \pm 1$ mode through experimentally simulated atmospheric turbulence as a function of the turbulence strength ($D/r_0$).  A basis-dependent (BD) measurement in the original basis ($\ell = \pm 1$) shows a decay in vectorness, while the basis-independent (BI) approach reveals an invariance. Experimental data (points) show excellent agreement with theory (dashed curves).  The insets show the left (red) and right (blue) polarisation projections on the $\ell = \pm 1$ modal sphere for three example turbulence strengths (low, medium and strong). Each point on the spheres represents one instance from the experimental turbulence ensemble. Error bars are plotted as standard deviations from 50 instances of simulated turbulence.}
	\label{fig:measure}
\end{figure}

\vspace{0.5cm}
\noindent \textbf{The role of measurement.}  Given that the state vector after the channel lives on many HOP Spheres in the original basis, $\{ \ket{u_1}$ and $\ket{u_2} \}$, but only one HOP Sphere in the adjusted basis, $\{ \ket{v_1}$ and $\ket{v_2} \}$, it is pertinent to ask in which basis (or HOP Sphere) 
should one make the vectorial measurement?  In the quoted vectorness values thus far, we have circumvented this problem by using a Stokes measurement approach to extract the degree of non-separability \cite{selyem2019basis}, with the benefit of sampling in a basis-independent fashion (see Methods).  In contrast, many measurements of structured light are basis dependent, e.g., in classical and quantum communication, where the basis elements form the communication alphabet.  In  Figure~\ref{fig:measure} (a) we show the outcome of a basis-dependent vectorness measurement (see Methods) in the original basis and in the adjusted basis, using the tilted lens as the channel.  We see that for some symmetries the choice of basis has no impact on the outcome, as in the case of $\ell = \pm 1$, while for other vectorial fields the impact is significant ($\ell = \pm 4$).  It should be noted that the $\ell = \pm 1$ beam through a tilted lens is a special case, because in modal space the state vector has simply been rotated.  In general, it is the adjusted basis that always reveals the invariance of the vectorness.  The failure of the measurement in the original basis is easily explained by the concatenation of the measurement subspace to just one of the many HOP Spheres in which the state resides, and the consequent reduction of the state vector to a mixed and separable state because information is lost to other OAM subspaces. 

To make clear that the tilted lens is not a special case, we alter the channel to atmospheric turbulence, and this time measure the vectorness as a function of the turbulence strength in the original basis and in a basis independent fashion, with the result shown in Figure~\ref{fig:measure} (b). We find that a measurement in the original basis shows a decay in the vectorial nature of the light as the turbulence strength increases, whereas the basis-independent approach reveals the unitary nature of turbulence: while the spatial structure is complex and altered, its vectorness remains intact and invariant to the turbulence strength. Here, all the measurements were performed in the far-field, so that the phase-only perturbations manifest as phase and amplitude effects.  We clearly see the paradox: the vectorial structure can appear robust or not depending on how the measurement was done.  This result highlights the important role of measurement in determining the salient properties of vectorial light fields. 

The inset spheres in Figure~\ref{fig:measure} (b) show the left (red) and right (blue) state vector projections on the $\ell = \pm 1$ modal sphere \cite{padgett1999poincare} for low, medium and strong turbulence.  Each instance of a turbulence strength is a red (blue) point, the scatter of which and deviation from the poles is indicative of modal cross-talk.  This is a visual representation of why the vectorness decays when one considers only one sphere in the original basis: the original states are orthogonal (points on opposite poles with little scatter) but as turbulence increases they disperse across the sphere, resulting on superpositions of OAM, which become maximally mixed.  For example, looking only at the $\ell = \pm 1$ subspace, the state may evolve following (ignoring normalisation): $\ket{R}\ket{1} \rightarrow \ket{R}\ket{1} + i \ket{R}\ket{-1}$ and likewise $\ket{L}\ket{-1} \rightarrow \ket{L} \ket{-1} -i\ket{L} \ket{1}$.  The original vectorial state becomes  $\ket{R}\ket{1} + \ket{L}\ket{-1} \rightarrow \ket{L}\ket{-1} + i \ket{L}\ket{1} + \ket{R}\ket{1} + \ket{R}\ket{-1} = (\ket{R} +i \ket{L})(\ket{1} + \ket{-1})$, which is a scalar, diagonally polarised, Hermite-Gaussian beam with a vectorness of 0. One can deduce from this simple example that if only some modal spaces are considered in the beam analysis, then vectorial modes can ``decay'' to become scalar, but scalar modes cannot ``decay'' to become vectorial. 

\begin{figure*}[t]
	\includegraphics[width=\linewidth]{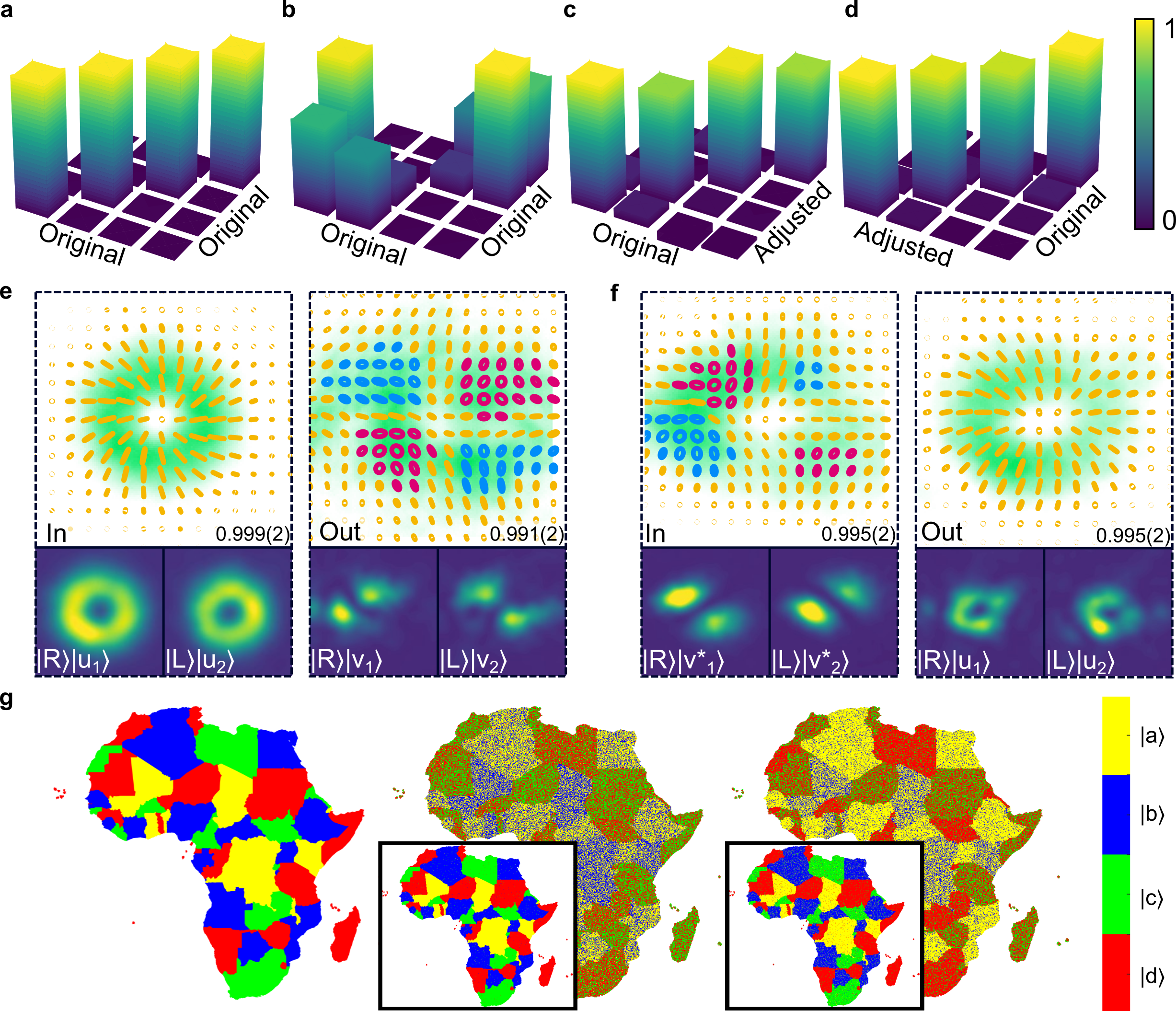}
	\caption{\textbf{Unravelling turbulence.} Cross-talk matrices for preparation and measurement in the original basis for (a) no turbulence and (b) medium turbulence ($D/r_0=1.6$).  (c) Measuring or (d) preparing in the adjusted removes the cross-talk. (e) The unitary action of the channel distorts the polarization structure of the initial beam but not its vectorial homogeneity, with the vectorness given in the insets. Below the polarisation structure are the spatial modes composing the original basis ($u_{1,2}$) and perturbed basis ($v_{1,2}$) along with their polarization states $\ket{R}$ and $\ket{L}$. (f) The unravelling of the turbulence by a lossless unitary applied pre-channel, restoring the initial beam after the turbulence. The adjusted basis modes 
	($v^*_{1,2}$) and the corrected basis modes 
	($u_{1,2}$) are shown below the polarisation structures with their respective polarisation states. In (g), an image (left) encoded in the original basis is distorted after transmission through moderately weak (middle) and medium (right) turbulence. The same image is transmitted through the channel but decoded in the adjusted basis, for distortion-free communication, shown in the insets. }
	\label{fig:turb}
\end{figure*}

\vspace{0.5cm}
\noindent \textbf{Reversing turbulence distortions.}  In our final example we keep atmospheric turbulence as our unitary channel because of its complex aberration profile.  In Figure~\ref{fig:turb} (a) and (b) we show typical cross-talk matrices with and without turbulence, respectively, where the input and output modes are both expressed in the original basis, $\{\ket{R}\ket{u_1}, \ket{R}\ket{u_2}, \ket{L}\ket{u_1},\ket{L}\ket{u_2} \}$.  The cross-talk in (b) is deleterious for both classical and quantum communication.  However, the unitary nature of the channel means that there is an adjusted basis, $\{\ket{R}\ket{v_1}, \ket{R}\ket{v_2}, \ket{L}\ket{v_1},\ket{L}\ket{v_2} \}$, where the state vector is pure.  Consequently, a post- or pre-channel unitary can undo the action of the channel, removing the cross-talk, as shown in Figure~\ref{fig:turb} (c) and (d).  The post-channel unitary is simply a measurement in the new adjusted basis, requiring nothing more than a change to the detection optics (holograms in our example) based on the channel under study. In Figure~\ref{fig:turb} (d), the preparation optics are programmed to prepare the state in the adjusted basis, but measure it in the original basis, once again returning a cross-talk free result.  While the action of the channel is to distort the initial beam, as shown in Figure~\ref{fig:turb} (e), the channel action can be reversed as shown in Figure~\ref{fig:turb} (f), restoring the initial beam.  This is a visualisation of the low cross-talk matrix in part (d): sending in the adjusted basis but measuring in the original basis.  The scalar version of the basis modes are shown in Figures~\ref{fig:turb} (e) and (f) as insets below each polarisation profile, starting with the original modes, which become perturbed due to turbulence.  The adjusted basis, with scalar versions shown in the insets of Figure~\ref{fig:turb} (f), maintains the orthogonality of the modes, and shows the key to the restoration of the initial field.  When the adjusted basis is the input, the output is the corrected mode in the original basis. These results suggest that cross-talk free communication is possible with a judiciously selected basis set for the preparation or measurement, exploiting the fact that the vectorial state is intact in the adjusted basis. We use this fact in (g) to encode graphical information using our modal set, send it across the channel, and decode it on the other side. Turbulence causes cross-talk, distorting the image, but this can easily be overcome by simply decoding (measuring) in the adjusted basis, with results shown for medium and strong turbulence. A measurement in the adjusted basis reveals no modal decay, no cross-talk and high-fidelity information transfer across this noisy channel.

\vspace{0.5cm}
\section*{Discussion and Conclusion}

Careful inspection of the studies that report vectorial robustness in noisy channels reveal that the distances propagated were short and the strength of perturbation low, mimicking a phase-only near-field effect where indeed little change is expected, and hence these are not true tests for robustness or invariance.  Studies that claim enhanced resilience of vector modes over distances comparable to the Rayleigh length \cite{lochab2018designer,gu2009scintillation} have used the variance in the field's intensity as a measure, a quantity that one would expect to be robust due to the fact that each polarisation component behaves independently and so will have a low covariance. 

Our results show that vectorial structured light in complex media will evolve from the near-field to the far-field, generally appearing spatially distorted in amplitude, phase and polarisation structure, although unaltered in vectorial inhomogeneity.  This is explained by the unitary nature of such channels, mapping the state from a HOP Sphere spanned by the original basis to a new HOP Sphere spanned by an adjusted basis, as if only our perspective has altered.  Any measurement in the original basis will show an apparent "decay" of the vectorial structure in strongly perturbing media even though the degree of polarisation remains intact - a hidden invariance that can be observed through a judicious measurement. The role of measurement in quantum studies is well appreciated, and here too the vectorness of a classical beam can be found to be high \textit{and} low, seemingly contradictory outcomes, yet both equally valid based on the choice of measurement.  This is not only of fundamental importance but also of practical relevance: we have shown how to make a basis choice for preparation and/or measurement to negate modal cross-talk, with obvious benefits in classical and quantum communication across noisy channels, as well as in imaging through complex media. 

The argument for robustness of vectorial light due to the fact that the polarisation is not affected in a non-birefringent channel is egregious: our quantum notation makes clear that the entire state is altered since its two DoFs are non-separable, in the same way that a true bi-photon quantum state is altered if just one of the two photons is perturbed - both examples of one-sided channels.  Ironically, the incorrect statement is true if the initial mode is \textit{not} vectorial, since separable states (scalar homogenously polarised light) cannot become less scalar in a non-birefringent medium.  Our statement is corroborated here by theoretical examples and experimental proof, particularly illustrated by the example of operating on the ``unaffected'' polarisation DoF to correct the entire vectorial state.

Our analysis has considered non-birefringent channels, but can be generalised by allowing the polarisation DoF to be perturbed while selecting the other DoF to be invariant, or by adjusting the theory to treat the medium as a two-sided channel to allow both DoFs to alter.  We remark too that not all channels are unitary.  In non-linear systems the vectorness can improve or degrade depending on the medium \cite{hu2021collapse,sroor2018purity,meyer2020steady}, and our results are not in contradiction to these findings.  Finally, the notion of robustness implies that there must also be states that are not robust, and here we wish to clarify the role of a unitary operator versus a filter.  It is always possible to convert noise into loss by a filtering process \cite{farias2015resilience,ndagano2019entanglement}, but this does not make the initial state robust to the medium, while the loss induced correction is non-unitary.   

In conclusion, we have provided a general framework for understanding the impact of aberrations on vectorial light fields, in doing so revealing the unitary nature of many complex media, where the perturbation is understood as a simple unitary mapping.  Because the mapping is unitary, it can be pre- or post-corrected in a lossless manner, armed only with simple Stokes and modal decomposition projections.  Our work resolves a standing debate on the robustness of vectorial light in complex media, and will be invaluable to the exploding community working with vectorial structured light and its applications.

\vspace{0.5cm}
\newpage
\section*{Materials and correspondence}
Correspondence and requests for materials should be addressed to AF.

\section*{Acknowledgments}
AF thanks the NRF-CSIR Rental Pool Programme.

\section*{Authors' contribution}
IN, WB, AK and KS performed the experiments and AF supervised the project.  All authors contributed to data analysis and writing of the manuscript.

\section*{Competing financial interests}
The authors declare no financial interests.

\newpage

\section*{Methods}

\textbf{Vector beam generation.} To generate vector beams, a horizontally polarized Gaussian beam from a HeNe laser (wavelength $\lambda$ = 633 nm) was expanded and collimated using a 10$\times$ objective and a 250 mm focal length lens. The expanded beam was then passed through a half-wave plate (HWP) before being separated into its horizontally (H) and vertically (V) polarized components using a Wollaston prism (WP). The plane at the WP was then imaged onto the screen of a digital micro-mirror device (DMD - DLP6500) using a 4f system. The separation angle of the polarization components from the WP of $\approx1\,^\circ$ resulted in the diffracted 0$^{th}$ order components overlapping. To create a desired scalar component mode of the form $U(\rr) = |U(\rr)|\exp(i\Phi)$, where $\Phi$ is the phase of the field and $U$ has maximum unit amplitude, the DMD was programmed with a hologram given by 

\begin{align}
    H &= \frac{1}{2}  - \frac{1}{2}\text{sign}\left( \cos\left(\pi\phi+2\pi G \right)-\cos(\pi A)\right)
\end{align}

\noindent with $G(\rr) = \mathbf{g}\cdot\rr$ a phase ramp function with grating frequencies $\mathbf{g} = (g_x,g_y)$, and $A(\rr) = \arcsin(|U(\rr)|)/\pi$ and $\phi(\rr) = \Phi/\pi$ are the respective, appropriately enveloped, amplitude and phase of the desired complex fields at pixel positions $\rr = (x,y)$. Holograms for each polarization component (denoted as $H_A$ and $H_B$) were multiplexed on the DMD, where the grating frequencies $(g_x,g_y)$ could be chosen to cause the H polarized 1$^{st}$ diffraction order from $H_A$ and the V polarized 1$^{st}$ diffraction order from $H_B$ to spatially overlap. This combined 1$^{st}$ order contained our vector field which was subsequently spatially filtered at the focal plane of a 4f imaging system and imaged onto a second DMD. The second DMD was addressed with a single hologram of the same form as $H_{A/B}$ onto which the turbulence phase screens, along with any correlation filters were encoded (a quater-wave plate, QWP, was used to convert H and V polarization to right (R) and left (L) circular respectively). Polarization projections were made using a linear polarizer (LP) and a QWP before the second DMD. The intensities at the Fourier plane were captured using a CCD (FLIR Grasshopper 3) placed at the focal plane of a 2f imaging system. 

\vspace{0.5cm}
\textbf{Non-separability measurements.} We measured the non-separability of the vector beams in a basis dependent and independent approach using Stokes parameters and modal decomposition, respectively (see SI for further details ).  Firstly, to measure the Stokes parameters, we used the reduced set of four Stokes intensities, $I_H,I_D,I_R$ and $I_L$ corresponding to the linearly polarized horizontal (H), diagonal (D) and the circular right (R) and left (L) polarizations. From these measurements we extracted the Stokes parameters, 
\begin{align}
    S'_0 &= I_R + I_L \\
    S'_1 &= 2I_H - S_0 \\
    S'_2 &= 2I_D - S_0\\
    S'_3 &= I_R - I_L.
\end{align}

The four intensity projections were acquired through the use of a linear polarizer (for $I_H$ and $I_D$) rotated by $0$ and $\pi/4$ radians together with a quarter-wave plate (for $I_R$ and $I_L$), oriented at $\pm \pi/4$ radians relative to the fast axis. Subsequently, the basis independent VQF (equivalently non-separability) was determined from $V = \sqrt{1-(S_1^2+S_2^2+S_3^2)/S_0^2}$, with $S_i=\int S'_i(\mathbf{r}) d \mathbf{r}$ being the global Stokes parameters integrated over the transverse plane.

For the basis dependent approach, the overlap between orthogonally polarized projections of the field in question was used as a measure of non-separability, with unity overlap signalling that the field is completely scalar, while a zero overlap indicated a maximally non-separable vector field. This overlap can be characterized by the magnitude of the Bloch vector, $s$, lying on the surface of a sphere spanned by superpositions of a chosen pair of basis states $\ket{\psi_{1,2}}$. The magnitude can then be seen as a sum-in-quadrature of the Pauli matrix expectation values $\langle \sigma_i \rangle$, as their operation on the basis states gives the unit vectors of the sphere.  We can determine these expectation values using projections $\langle P|$ into superpositions of the spatial basis components described by

\begin{equation}
    \langle P_j| = \alpha_j\langle\psi_1| + \beta_j\langle\psi_2|\,.
\end{equation}

With $(\alpha,\beta)=\{(1,0),(0,1),\frac{1}{\sqrt{2}}(1,1),\frac{1}{\sqrt{2}}(1,-1),\\\frac{1}{\sqrt{2}}(1,i),\frac{1}{\sqrt{2}}(1,-i)\}$ for both $|R\rangle$ and $|L\rangle$. These 12 on axis intensity projections are used to calculate the length of the Bloch vector according to

\begin{align}
    \langle\sigma_1\rangle &= (I_{13}+I_{23}) - (I_{14}+I_{24}) \\
    \langle\sigma_2\rangle &= (I_{15}+I_{25}) - (I_{16}+I_{26})  \\
    \langle\sigma_3\rangle &= (I_{11}+I_{21}) - (I_{11}+I_{22}) 
\end{align}

where the $i$ index of $I_{ij}$ corresponds to the $\langle R,L|$ polarization projections and the $j$ index represents the spatial mode projections defined above. The nonseparability is then given by $V = \sqrt{1-s^2}$. In this work the projections into the right- and left-circular polarization components was achieved using a linear polarizer and QWP. The subsequent spatial mode projections were performed using a correlation filter encoded into a digital micro-mirror device (DMD), and a Fourier lens to produce on-axis intensities $I_{ij}$.

\vspace{0.5cm}
\textbf{Adjusted basis measurement.} To determine the adjusted basis, the complex amplitude of an aberrated probe field $\ket{\Psi}_\text{probe}$ needs to be measured (see SI). This field was approximated using a maximum likelihood estimation procedure, where far field intensities of the right- and left-polarized components of the ideal vector beam through turbulence were captured:

\begin{equation}
    I_\text{probe}^{R,L} = \bra{R,L}\ket{\psi(\mathbf{k} = 0)}_\text{probe}
\end{equation}

Where $\ket{\psi}_\text{probe}$ denotes the Fourier transform of $\ket{\Psi}_\text{probe}$ and $\mathbf{k}=(k_x,k_y)$. Simulated Fourier intensities, $\ket{\psi}_\text{Zern}$, of ideal beams modulated by a phase (modelled by possible weighted combinations of Zernike polynomials $Z_j$),
\begin{equation}
    \ket{\Psi_{\pm1}}_\text{Zern} = \text{LG}_0^{\pm1}\exp(i\sum\limits_j\,c_j Z_j)\,,
\end{equation}
were generated (spatial dependence has been omitted). The set of coefficients $c_j$ which lead to the lowest square difference in intensity between the experimental and simulated cases,

\begin{equation}
    \chi^2 = (I_\text{probe}^{R} - I_\text{Zern}^{R})^2 + (I_\text{probe}^{L} - I_\text{Zern}^{L})^2
\end{equation}

was used to determine the required basis for recovery of the initial beam.

\clearpage
\newpage

\bibliography{mybib_2}

\begin{thebibliography}{73}%
\makeatletter
\providecommand \@ifxundefined [1]{%
 \@ifx{#1\undefined}
}%
\providecommand \@ifnum [1]{%
 \ifnum #1\expandafter \@firstoftwo
 \else \expandafter \@secondoftwo
 \fi
}%
\providecommand \@ifx [1]{%
 \ifx #1\expandafter \@firstoftwo
 \else \expandafter \@secondoftwo
 \fi
}%
\providecommand \natexlab [1]{#1}%
\providecommand \enquote  [1]{``#1''}%
\providecommand \bibnamefont  [1]{#1}%
\providecommand \bibfnamefont [1]{#1}%
\providecommand \citenamefont [1]{#1}%
\providecommand \href@noop [0]{\@secondoftwo}%
\providecommand \href [0]{\begingroup \@sanitize@url \@href}%
\providecommand \@href[1]{\@@startlink{#1}\@@href}%
\providecommand \@@href[1]{\endgroup#1\@@endlink}%
\providecommand \@sanitize@url [0]{\catcode `\\12\catcode `\$12\catcode
  `\&12\catcode `\#12\catcode `\^12\catcode `\_12\catcode `\%12\relax}%
\providecommand \@@startlink[1]{}%
\providecommand \@@endlink[0]{}%
\providecommand \url  [0]{\begingroup\@sanitize@url \@url }%
\providecommand \@url [1]{\endgroup\@href {#1}{\urlprefix }}%
\providecommand \urlprefix  [0]{URL }%
\providecommand \Eprint [0]{\href }%
\providecommand \doibase [0]{http://dx.doi.org/}%
\providecommand \selectlanguage [0]{\@gobble}%
\providecommand \bibinfo  [0]{\@secondoftwo}%
\providecommand \bibfield  [0]{\@secondoftwo}%
\providecommand \translation [1]{[#1]}%
\providecommand \BibitemOpen [0]{}%
\providecommand \bibitemStop [0]{}%
\providecommand \bibitemNoStop [0]{.\EOS\space}%
\providecommand \EOS [0]{\spacefactor3000\relax}%
\providecommand \BibitemShut  [1]{\csname bibitem#1\endcsname}%
\let\auto@bib@innerbib\@empty
\bibitem [{\citenamefont {Forbes}\ \emph {et~al.}(2021)\citenamefont {Forbes},
  \citenamefont {de~Oliveira},\ and\ \citenamefont
  {Dennis}}]{forbes2021structured}%
  \BibitemOpen
  \bibfield  {author} {\bibinfo {author} {\bibfnamefont {A.}~\bibnamefont
  {Forbes}}, \bibinfo {author} {\bibfnamefont {M.}~\bibnamefont {de~Oliveira}},
  \ and\ \bibinfo {author} {\bibfnamefont {M.~R.}\ \bibnamefont {Dennis}},\
  }\href@noop {} {\bibfield  {journal} {\bibinfo  {journal} {Nature Photonics}\
  }\textbf {\bibinfo {volume} {15}},\ \bibinfo {pages} {253} (\bibinfo {year}
  {2021})}\BibitemShut {NoStop}%
\bibitem [{\citenamefont {Shen}\ \emph {et~al.}(2021)\citenamefont {Shen},
  \citenamefont {Hou}, \citenamefont {Papasimakis},\ and\ \citenamefont
  {Zheludev}}]{shen2021supertoroidal}%
  \BibitemOpen
  \bibfield  {author} {\bibinfo {author} {\bibfnamefont {Y.}~\bibnamefont
  {Shen}}, \bibinfo {author} {\bibfnamefont {Y.}~\bibnamefont {Hou}}, \bibinfo
  {author} {\bibfnamefont {N.}~\bibnamefont {Papasimakis}}, \ and\ \bibinfo
  {author} {\bibfnamefont {N.~I.}\ \bibnamefont {Zheludev}},\ }\href@noop {}
  {\bibfield  {journal} {\bibinfo  {journal} {arXiv preprint arXiv:2103.08431}\
  } (\bibinfo {year} {2021})}\BibitemShut {NoStop}%
\bibitem [{\citenamefont {Gao}\ \emph {et~al.}(2020)\citenamefont {Gao},
  \citenamefont {Speirits}, \citenamefont {Castellucci}, \citenamefont
  {Franke-Arnold}, \citenamefont {Barnett},\ and\ \citenamefont
  {G{\"o}tte}}]{gao2020paraxial}%
  \BibitemOpen
  \bibfield  {author} {\bibinfo {author} {\bibfnamefont {S.}~\bibnamefont
  {Gao}}, \bibinfo {author} {\bibfnamefont {F.~C.}\ \bibnamefont {Speirits}},
  \bibinfo {author} {\bibfnamefont {F.}~\bibnamefont {Castellucci}}, \bibinfo
  {author} {\bibfnamefont {S.}~\bibnamefont {Franke-Arnold}}, \bibinfo {author}
  {\bibfnamefont {S.~M.}\ \bibnamefont {Barnett}}, \ and\ \bibinfo {author}
  {\bibfnamefont {J.~B.}\ \bibnamefont {G{\"o}tte}},\ }\href@noop {} {\bibfield
   {journal} {\bibinfo  {journal} {Physical Review A}\ }\textbf {\bibinfo
  {volume} {102}},\ \bibinfo {pages} {053513} (\bibinfo {year}
  {2020})}\BibitemShut {NoStop}%
\bibitem [{\citenamefont {Larocque}\ \emph {et~al.}(2018)\citenamefont
  {Larocque}, \citenamefont {Sugic}, \citenamefont {Mortimer}, \citenamefont
  {Taylor}, \citenamefont {Fickler}, \citenamefont {Boyd}, \citenamefont
  {Dennis},\ and\ \citenamefont {Karimi}}]{larocque2018reconstructing}%
  \BibitemOpen
  \bibfield  {author} {\bibinfo {author} {\bibfnamefont {H.}~\bibnamefont
  {Larocque}}, \bibinfo {author} {\bibfnamefont {D.}~\bibnamefont {Sugic}},
  \bibinfo {author} {\bibfnamefont {D.}~\bibnamefont {Mortimer}}, \bibinfo
  {author} {\bibfnamefont {A.~J.}\ \bibnamefont {Taylor}}, \bibinfo {author}
  {\bibfnamefont {R.}~\bibnamefont {Fickler}}, \bibinfo {author} {\bibfnamefont
  {R.~W.}\ \bibnamefont {Boyd}}, \bibinfo {author} {\bibfnamefont {M.~R.}\
  \bibnamefont {Dennis}}, \ and\ \bibinfo {author} {\bibfnamefont
  {E.}~\bibnamefont {Karimi}},\ }\href@noop {} {\bibfield  {journal} {\bibinfo
  {journal} {Nature Physics}\ }\textbf {\bibinfo {volume} {14}},\ \bibinfo
  {pages} {1079} (\bibinfo {year} {2018})}\BibitemShut {NoStop}%
\bibitem [{\citenamefont {Galvez}\ \emph {et~al.}(2014)\citenamefont {Galvez},
  \citenamefont {Rojec}, \citenamefont {Kumar},\ and\ \citenamefont
  {Viswanathan}}]{galvez2014generation}%
  \BibitemOpen
  \bibfield  {author} {\bibinfo {author} {\bibfnamefont {E.~J.}\ \bibnamefont
  {Galvez}}, \bibinfo {author} {\bibfnamefont {B.~L.}\ \bibnamefont {Rojec}},
  \bibinfo {author} {\bibfnamefont {V.}~\bibnamefont {Kumar}}, \ and\ \bibinfo
  {author} {\bibfnamefont {N.~K.}\ \bibnamefont {Viswanathan}},\ }\href@noop {}
  {\bibfield  {journal} {\bibinfo  {journal} {Physical Review A}\ }\textbf
  {\bibinfo {volume} {89}},\ \bibinfo {pages} {031801} (\bibinfo {year}
  {2014})}\BibitemShut {NoStop}%
\bibitem [{\citenamefont {Zdagkas}\ \emph {et~al.}(2021)\citenamefont
  {Zdagkas}, \citenamefont {Shen}, \citenamefont {McDonnell}, \citenamefont
  {Deng}, \citenamefont {Li}, \citenamefont {Ellenbogen}, \citenamefont
  {Papasimakis},\ and\ \citenamefont {Zheludev}}]{zdagkas2021observation}%
  \BibitemOpen
  \bibfield  {author} {\bibinfo {author} {\bibfnamefont {A.}~\bibnamefont
  {Zdagkas}}, \bibinfo {author} {\bibfnamefont {Y.}~\bibnamefont {Shen}},
  \bibinfo {author} {\bibfnamefont {C.}~\bibnamefont {McDonnell}}, \bibinfo
  {author} {\bibfnamefont {J.}~\bibnamefont {Deng}}, \bibinfo {author}
  {\bibfnamefont {G.}~\bibnamefont {Li}}, \bibinfo {author} {\bibfnamefont
  {T.}~\bibnamefont {Ellenbogen}}, \bibinfo {author} {\bibfnamefont
  {N.}~\bibnamefont {Papasimakis}}, \ and\ \bibinfo {author} {\bibfnamefont
  {N.}~\bibnamefont {Zheludev}},\ }\href@noop {} {\bibfield  {journal}
  {\bibinfo  {journal} {arXiv preprint arXiv:2102.03636}\ } (\bibinfo {year}
  {2021})}\BibitemShut {NoStop}%
\bibitem [{\citenamefont {Keren-Zur}\ \emph {et~al.}(2019)\citenamefont
  {Keren-Zur}, \citenamefont {Tal}, \citenamefont {Fleischer}, \citenamefont
  {Mittleman},\ and\ \citenamefont {Ellenbogen}}]{keren2019generation}%
  \BibitemOpen
  \bibfield  {author} {\bibinfo {author} {\bibfnamefont {S.}~\bibnamefont
  {Keren-Zur}}, \bibinfo {author} {\bibfnamefont {M.}~\bibnamefont {Tal}},
  \bibinfo {author} {\bibfnamefont {S.}~\bibnamefont {Fleischer}}, \bibinfo
  {author} {\bibfnamefont {D.~M.}\ \bibnamefont {Mittleman}}, \ and\ \bibinfo
  {author} {\bibfnamefont {T.}~\bibnamefont {Ellenbogen}},\ }\href@noop {}
  {\bibfield  {journal} {\bibinfo  {journal} {Nature communications}\ }\textbf
  {\bibinfo {volume} {10}},\ \bibinfo {pages} {1} (\bibinfo {year}
  {2019})}\BibitemShut {NoStop}%
\bibitem [{\citenamefont {Bauer}\ \emph {et~al.}(2015)\citenamefont {Bauer},
  \citenamefont {Banzer}, \citenamefont {Karimi}, \citenamefont {Orlov},
  \citenamefont {Rubano}, \citenamefont {Marrucci}, \citenamefont {Santamato},
  \citenamefont {Boyd},\ and\ \citenamefont {Leuchs}}]{bauer2015observation}%
  \BibitemOpen
  \bibfield  {author} {\bibinfo {author} {\bibfnamefont {T.}~\bibnamefont
  {Bauer}}, \bibinfo {author} {\bibfnamefont {P.}~\bibnamefont {Banzer}},
  \bibinfo {author} {\bibfnamefont {E.}~\bibnamefont {Karimi}}, \bibinfo
  {author} {\bibfnamefont {S.}~\bibnamefont {Orlov}}, \bibinfo {author}
  {\bibfnamefont {A.}~\bibnamefont {Rubano}}, \bibinfo {author} {\bibfnamefont
  {L.}~\bibnamefont {Marrucci}}, \bibinfo {author} {\bibfnamefont
  {E.}~\bibnamefont {Santamato}}, \bibinfo {author} {\bibfnamefont {R.~W.}\
  \bibnamefont {Boyd}}, \ and\ \bibinfo {author} {\bibfnamefont
  {G.}~\bibnamefont {Leuchs}},\ }\href@noop {} {\bibfield  {journal} {\bibinfo
  {journal} {Science}\ }\textbf {\bibinfo {volume} {347}},\ \bibinfo {pages}
  {964} (\bibinfo {year} {2015})}\BibitemShut {NoStop}%
\bibitem [{\citenamefont {Brown}(2011)}]{brown2011unconventional}%
  \BibitemOpen
  \bibfield  {author} {\bibinfo {author} {\bibfnamefont {T.~G.}\ \bibnamefont
  {Brown}},\ }\href@noop {} {\bibfield  {journal} {\bibinfo  {journal}
  {Progress in Optics}\ }\textbf {\bibinfo {volume} {56}},\ \bibinfo {pages}
  {81} (\bibinfo {year} {2011})}\BibitemShut {NoStop}%
\bibitem [{\citenamefont {Wang}\ \emph {et~al.}(2020)\citenamefont {Wang},
  \citenamefont {Castellucci},\ and\ \citenamefont
  {Franke-Arnold}}]{wang2020vectorial}%
  \BibitemOpen
  \bibfield  {author} {\bibinfo {author} {\bibfnamefont {J.}~\bibnamefont
  {Wang}}, \bibinfo {author} {\bibfnamefont {F.}~\bibnamefont {Castellucci}}, \
  and\ \bibinfo {author} {\bibfnamefont {S.}~\bibnamefont {Franke-Arnold}},\
  }\href@noop {} {\bibfield  {journal} {\bibinfo  {journal} {AVS Quantum
  Science}\ }\textbf {\bibinfo {volume} {2}},\ \bibinfo {pages} {031702}
  (\bibinfo {year} {2020})}\BibitemShut {NoStop}%
\bibitem [{\citenamefont {Otte}\ \emph {et~al.}(2018)\citenamefont {Otte},
  \citenamefont {Alpmann},\ and\ \citenamefont {Denz}}]{otte2018polarization}%
  \BibitemOpen
  \bibfield  {author} {\bibinfo {author} {\bibfnamefont {E.}~\bibnamefont
  {Otte}}, \bibinfo {author} {\bibfnamefont {C.}~\bibnamefont {Alpmann}}, \
  and\ \bibinfo {author} {\bibfnamefont {C.}~\bibnamefont {Denz}},\ }\href@noop
  {} {\bibfield  {journal} {\bibinfo  {journal} {Laser \& Photonics Reviews}\
  }\textbf {\bibinfo {volume} {12}},\ \bibinfo {pages} {1700200} (\bibinfo
  {year} {2018})}\BibitemShut {NoStop}%
\bibitem [{\citenamefont {Rosales-Guzm{\'a}n}\ \emph
  {et~al.}(2018)\citenamefont {Rosales-Guzm{\'a}n}, \citenamefont {Ndagano},\
  and\ \citenamefont {Forbes}}]{vector-review}%
  \BibitemOpen
  \bibfield  {author} {\bibinfo {author} {\bibfnamefont {C.}~\bibnamefont
  {Rosales-Guzm{\'a}n}}, \bibinfo {author} {\bibfnamefont {B.}~\bibnamefont
  {Ndagano}}, \ and\ \bibinfo {author} {\bibfnamefont {A.}~\bibnamefont
  {Forbes}},\ }\href {http://stacks.iop.org/2040-8986/20/i=12/a=123001}
  {\bibfield  {journal} {\bibinfo  {journal} {Journal of Optics}\ }\textbf
  {\bibinfo {volume} {20}},\ \bibinfo {pages} {123001} (\bibinfo {year}
  {2018})}\BibitemShut {NoStop}%
\bibitem [{\citenamefont {Forbes}\ and\ \citenamefont
  {Nape}(2019)}]{forbes2019quantum}%
  \BibitemOpen
  \bibfield  {author} {\bibinfo {author} {\bibfnamefont {A.}~\bibnamefont
  {Forbes}}\ and\ \bibinfo {author} {\bibfnamefont {I.}~\bibnamefont {Nape}},\
  }\href@noop {} {\bibfield  {journal} {\bibinfo  {journal} {AVS Quantum
  Science}\ }\textbf {\bibinfo {volume} {1}},\ \bibinfo {pages} {011701}
  (\bibinfo {year} {2019})}\BibitemShut {NoStop}%
\bibitem [{\citenamefont {Sederberg}\ \emph {et~al.}(2020)\citenamefont
  {Sederberg}, \citenamefont {Kong}, \citenamefont {Hufnagel}, \citenamefont
  {Zhang}, \citenamefont {Karimi},\ and\ \citenamefont
  {Corkum}}]{sederberg2020vectorized}%
  \BibitemOpen
  \bibfield  {author} {\bibinfo {author} {\bibfnamefont {S.}~\bibnamefont
  {Sederberg}}, \bibinfo {author} {\bibfnamefont {F.}~\bibnamefont {Kong}},
  \bibinfo {author} {\bibfnamefont {F.}~\bibnamefont {Hufnagel}}, \bibinfo
  {author} {\bibfnamefont {C.}~\bibnamefont {Zhang}}, \bibinfo {author}
  {\bibfnamefont {E.}~\bibnamefont {Karimi}}, \ and\ \bibinfo {author}
  {\bibfnamefont {P.~B.}\ \bibnamefont {Corkum}},\ }\href@noop {} {\bibfield
  {journal} {\bibinfo  {journal} {Nature Photonics}\ }\textbf {\bibinfo
  {volume} {14}},\ \bibinfo {pages} {680} (\bibinfo {year} {2020})}\BibitemShut
  {NoStop}%
\bibitem [{\citenamefont {Fang}\ \emph
  {et~al.}(2021{\natexlab{a}})\citenamefont {Fang}, \citenamefont {Han},
  \citenamefont {Ge}, \citenamefont {Guo}, \citenamefont {Yu}, \citenamefont
  {Deng}, \citenamefont {Wu}, \citenamefont {Gong},\ and\ \citenamefont
  {Liu}}]{fang2021photoelectronic}%
  \BibitemOpen
  \bibfield  {author} {\bibinfo {author} {\bibfnamefont {Y.}~\bibnamefont
  {Fang}}, \bibinfo {author} {\bibfnamefont {M.}~\bibnamefont {Han}}, \bibinfo
  {author} {\bibfnamefont {P.}~\bibnamefont {Ge}}, \bibinfo {author}
  {\bibfnamefont {Z.}~\bibnamefont {Guo}}, \bibinfo {author} {\bibfnamefont
  {X.}~\bibnamefont {Yu}}, \bibinfo {author} {\bibfnamefont {Y.}~\bibnamefont
  {Deng}}, \bibinfo {author} {\bibfnamefont {C.}~\bibnamefont {Wu}}, \bibinfo
  {author} {\bibfnamefont {Q.}~\bibnamefont {Gong}}, \ and\ \bibinfo {author}
  {\bibfnamefont {Y.}~\bibnamefont {Liu}},\ }\href@noop {} {\bibfield
  {journal} {\bibinfo  {journal} {Nature Photonics}\ }\textbf {\bibinfo
  {volume} {15}},\ \bibinfo {pages} {115} (\bibinfo {year}
  {2021}{\natexlab{a}})}\BibitemShut {NoStop}%
\bibitem [{\citenamefont {El~Ketara}\ \emph {et~al.}(2021)\citenamefont
  {El~Ketara}, \citenamefont {Kobayashi},\ and\ \citenamefont
  {Brasselet}}]{el2021sensitive}%
  \BibitemOpen
  \bibfield  {author} {\bibinfo {author} {\bibfnamefont {M.}~\bibnamefont
  {El~Ketara}}, \bibinfo {author} {\bibfnamefont {H.}~\bibnamefont
  {Kobayashi}}, \ and\ \bibinfo {author} {\bibfnamefont {E.}~\bibnamefont
  {Brasselet}},\ }\href@noop {} {\bibfield  {journal} {\bibinfo  {journal}
  {Nature Photonics}\ }\textbf {\bibinfo {volume} {15}},\ \bibinfo {pages}
  {121} (\bibinfo {year} {2021})}\BibitemShut {NoStop}%
\bibitem [{\citenamefont {Hawley}\ \emph {et~al.}(2019)\citenamefont {Hawley},
  \citenamefont {Cork}, \citenamefont {Radwell},\ and\ \citenamefont
  {Franke-Arnold}}]{Hawley2019}%
  \BibitemOpen
  \bibfield  {author} {\bibinfo {author} {\bibfnamefont {R.~D.}\ \bibnamefont
  {Hawley}}, \bibinfo {author} {\bibfnamefont {J.}~\bibnamefont {Cork}},
  \bibinfo {author} {\bibfnamefont {N.}~\bibnamefont {Radwell}}, \ and\
  \bibinfo {author} {\bibfnamefont {S.}~\bibnamefont {Franke-Arnold}},\ }\href
  {\doibase 10.1038/s41598-019-39118-0} {\bibfield  {journal} {\bibinfo
  {journal} {Scientific Reports}\ }\textbf {\bibinfo {volume} {9}} (\bibinfo
  {year} {2019}),\ 10.1038/s41598-019-39118-0}\BibitemShut {NoStop}%
\bibitem [{\citenamefont {Fang}\ \emph
  {et~al.}(2021{\natexlab{b}})\citenamefont {Fang}, \citenamefont {Wan},
  \citenamefont {Forbes},\ and\ \citenamefont {Wang}}]{fang2021vectorial}%
  \BibitemOpen
  \bibfield  {author} {\bibinfo {author} {\bibfnamefont {L.}~\bibnamefont
  {Fang}}, \bibinfo {author} {\bibfnamefont {Z.}~\bibnamefont {Wan}}, \bibinfo
  {author} {\bibfnamefont {A.}~\bibnamefont {Forbes}}, \ and\ \bibinfo {author}
  {\bibfnamefont {J.}~\bibnamefont {Wang}},\ }\href@noop {} {\bibfield
  {journal} {\bibinfo  {journal} {Nature Communications}\ }\textbf {\bibinfo
  {volume} {12}},\ \bibinfo {pages} {1} (\bibinfo {year}
  {2021}{\natexlab{b}})}\BibitemShut {NoStop}%
\bibitem [{\citenamefont {Curcio}\ \emph {et~al.}(2020)\citenamefont {Curcio},
  \citenamefont {Alem{\'a}n-Casta{\~n}eda}, \citenamefont {Brown},
  \citenamefont {Brasselet},\ and\ \citenamefont
  {Alonso}}]{curcio2020birefringent}%
  \BibitemOpen
  \bibfield  {author} {\bibinfo {author} {\bibfnamefont {V.}~\bibnamefont
  {Curcio}}, \bibinfo {author} {\bibfnamefont {L.~A.}\ \bibnamefont
  {Alem{\'a}n-Casta{\~n}eda}}, \bibinfo {author} {\bibfnamefont {T.~G.}\
  \bibnamefont {Brown}}, \bibinfo {author} {\bibfnamefont {S.}~\bibnamefont
  {Brasselet}}, \ and\ \bibinfo {author} {\bibfnamefont {M.~A.}\ \bibnamefont
  {Alonso}},\ }\href@noop {} {\bibfield  {journal} {\bibinfo  {journal} {Nature
  communications}\ }\textbf {\bibinfo {volume} {11}},\ \bibinfo {pages} {1}
  (\bibinfo {year} {2020})}\BibitemShut {NoStop}%
\bibitem [{\citenamefont {Milione}\ \emph {et~al.}(2015)\citenamefont
  {Milione}, \citenamefont {Lavery}, \citenamefont {Huang}, \citenamefont
  {Ren}, \citenamefont {Xie}, \citenamefont {Nguyen}, \citenamefont {Karimi},
  \citenamefont {Marrucci}, \citenamefont {Nolan}, \citenamefont {Alfano} \emph
  {et~al.}}]{Milione2015}%
  \BibitemOpen
  \bibfield  {author} {\bibinfo {author} {\bibfnamefont {G.}~\bibnamefont
  {Milione}}, \bibinfo {author} {\bibfnamefont {M.~P.}\ \bibnamefont {Lavery}},
  \bibinfo {author} {\bibfnamefont {H.}~\bibnamefont {Huang}}, \bibinfo
  {author} {\bibfnamefont {Y.}~\bibnamefont {Ren}}, \bibinfo {author}
  {\bibfnamefont {G.}~\bibnamefont {Xie}}, \bibinfo {author} {\bibfnamefont
  {T.~A.}\ \bibnamefont {Nguyen}}, \bibinfo {author} {\bibfnamefont
  {E.}~\bibnamefont {Karimi}}, \bibinfo {author} {\bibfnamefont
  {L.}~\bibnamefont {Marrucci}}, \bibinfo {author} {\bibfnamefont {D.~A.}\
  \bibnamefont {Nolan}}, \bibinfo {author} {\bibfnamefont {R.~R.}\ \bibnamefont
  {Alfano}},  \emph {et~al.},\ }\href@noop {} {\bibfield  {journal} {\bibinfo
  {journal} {Optics Letters}\ }\textbf {\bibinfo {volume} {40}},\ \bibinfo
  {pages} {1980} (\bibinfo {year} {2015})}\BibitemShut {NoStop}%
\bibitem [{\citenamefont {Zhang}\ \emph {et~al.}(2021)\citenamefont {Zhang},
  \citenamefont {Wu}, \citenamefont {Li}, \citenamefont {Lu}, \citenamefont
  {Tu}, \citenamefont {Li},\ and\ \citenamefont {Lu}}]{zhang2021fiber}%
  \BibitemOpen
  \bibfield  {author} {\bibinfo {author} {\bibfnamefont {J.}~\bibnamefont
  {Zhang}}, \bibinfo {author} {\bibfnamefont {X.}~\bibnamefont {Wu}}, \bibinfo
  {author} {\bibfnamefont {J.}~\bibnamefont {Li}}, \bibinfo {author}
  {\bibfnamefont {L.}~\bibnamefont {Lu}}, \bibinfo {author} {\bibfnamefont
  {J.}~\bibnamefont {Tu}}, \bibinfo {author} {\bibfnamefont {Z.}~\bibnamefont
  {Li}}, \ and\ \bibinfo {author} {\bibfnamefont {C.}~\bibnamefont {Lu}},\
  }\href@noop {} {\bibfield  {journal} {\bibinfo  {journal} {Journal of
  Lightwave Technology}\ } (\bibinfo {year} {2021})}\BibitemShut {NoStop}%
\bibitem [{\citenamefont {Zhu}\ \emph {et~al.}(2021)\citenamefont {Zhu},
  \citenamefont {Janasik}, \citenamefont {Fyffe}, \citenamefont {Hay},
  \citenamefont {Zhou}, \citenamefont {Kantor}, \citenamefont {Winder},
  \citenamefont {Boyd}, \citenamefont {Leuchs},\ and\ \citenamefont
  {Shi}}]{zhu2021compensation}%
  \BibitemOpen
  \bibfield  {author} {\bibinfo {author} {\bibfnamefont {Z.}~\bibnamefont
  {Zhu}}, \bibinfo {author} {\bibfnamefont {M.}~\bibnamefont {Janasik}},
  \bibinfo {author} {\bibfnamefont {A.}~\bibnamefont {Fyffe}}, \bibinfo
  {author} {\bibfnamefont {D.}~\bibnamefont {Hay}}, \bibinfo {author}
  {\bibfnamefont {Y.}~\bibnamefont {Zhou}}, \bibinfo {author} {\bibfnamefont
  {B.}~\bibnamefont {Kantor}}, \bibinfo {author} {\bibfnamefont
  {T.}~\bibnamefont {Winder}}, \bibinfo {author} {\bibfnamefont {R.~W.}\
  \bibnamefont {Boyd}}, \bibinfo {author} {\bibfnamefont {G.}~\bibnamefont
  {Leuchs}}, \ and\ \bibinfo {author} {\bibfnamefont {Z.}~\bibnamefont {Shi}},\
  }\href@noop {} {\bibfield  {journal} {\bibinfo  {journal} {Nature
  communications}\ }\textbf {\bibinfo {volume} {12}},\ \bibinfo {pages} {1}
  (\bibinfo {year} {2021})}\BibitemShut {NoStop}%
\bibitem [{\citenamefont {Zhao}\ and\ \citenamefont
  {Wang}(2015)}]{zhao2015high}%
  \BibitemOpen
  \bibfield  {author} {\bibinfo {author} {\bibfnamefont {Y.}~\bibnamefont
  {Zhao}}\ and\ \bibinfo {author} {\bibfnamefont {J.}~\bibnamefont {Wang}},\
  }\href@noop {} {\bibfield  {journal} {\bibinfo  {journal} {Optics Letters}\
  }\textbf {\bibinfo {volume} {40}},\ \bibinfo {pages} {4843} (\bibinfo {year}
  {2015})}\BibitemShut {NoStop}%
\bibitem [{\citenamefont {Radwell}\ \emph {et~al.}(2016)\citenamefont
  {Radwell}, \citenamefont {Hawley}, \citenamefont {G{\"o}tte},\ and\
  \citenamefont {Franke-Arnold}}]{radwell2016achromatic}%
  \BibitemOpen
  \bibfield  {author} {\bibinfo {author} {\bibfnamefont {N.}~\bibnamefont
  {Radwell}}, \bibinfo {author} {\bibfnamefont {R.}~\bibnamefont {Hawley}},
  \bibinfo {author} {\bibfnamefont {J.}~\bibnamefont {G{\"o}tte}}, \ and\
  \bibinfo {author} {\bibfnamefont {S.}~\bibnamefont {Franke-Arnold}},\
  }\href@noop {} {\bibfield  {journal} {\bibinfo  {journal} {Nature
  communications}\ }\textbf {\bibinfo {volume} {7}},\ \bibinfo {pages} {1}
  (\bibinfo {year} {2016})}\BibitemShut {NoStop}%
\bibitem [{\citenamefont {Beckley}\ \emph {et~al.}(2010)\citenamefont
  {Beckley}, \citenamefont {Brown},\ and\ \citenamefont
  {Alonso}}]{beckley2010full}%
  \BibitemOpen
  \bibfield  {author} {\bibinfo {author} {\bibfnamefont {A.~M.}\ \bibnamefont
  {Beckley}}, \bibinfo {author} {\bibfnamefont {T.~G.}\ \bibnamefont {Brown}},
  \ and\ \bibinfo {author} {\bibfnamefont {M.~A.}\ \bibnamefont {Alonso}},\
  }\href@noop {} {\bibfield  {journal} {\bibinfo  {journal} {Optics express}\
  }\textbf {\bibinfo {volume} {18}},\ \bibinfo {pages} {10777} (\bibinfo {year}
  {2010})}\BibitemShut {NoStop}%
\bibitem [{\citenamefont {He}\ \emph {et~al.}(2019)\citenamefont {He},
  \citenamefont {Chang}, \citenamefont {Hu}, \citenamefont {Wang},
  \citenamefont {Antonello}, \citenamefont {He}, \citenamefont {Liu},
  \citenamefont {Lin}, \citenamefont {Dai}, \citenamefont {Elson} \emph
  {et~al.}}]{he2019complex}%
  \BibitemOpen
  \bibfield  {author} {\bibinfo {author} {\bibfnamefont {C.}~\bibnamefont
  {He}}, \bibinfo {author} {\bibfnamefont {J.}~\bibnamefont {Chang}}, \bibinfo
  {author} {\bibfnamefont {Q.}~\bibnamefont {Hu}}, \bibinfo {author}
  {\bibfnamefont {J.}~\bibnamefont {Wang}}, \bibinfo {author} {\bibfnamefont
  {J.}~\bibnamefont {Antonello}}, \bibinfo {author} {\bibfnamefont
  {H.}~\bibnamefont {He}}, \bibinfo {author} {\bibfnamefont {S.}~\bibnamefont
  {Liu}}, \bibinfo {author} {\bibfnamefont {J.}~\bibnamefont {Lin}}, \bibinfo
  {author} {\bibfnamefont {B.}~\bibnamefont {Dai}}, \bibinfo {author}
  {\bibfnamefont {D.~S.}\ \bibnamefont {Elson}},  \emph {et~al.},\ }\href@noop
  {} {\bibfield  {journal} {\bibinfo  {journal} {Nature communications}\
  }\textbf {\bibinfo {volume} {10}},\ \bibinfo {pages} {1} (\bibinfo {year}
  {2019})}\BibitemShut {NoStop}%
\bibitem [{\citenamefont {Rosales-Guzm{\'a}n}\ \emph
  {et~al.}(2020)\citenamefont {Rosales-Guzm{\'a}n}, \citenamefont {Hu},
  \citenamefont {Selyem}, \citenamefont {Moreno-Acosta}, \citenamefont
  {Franke-Arnold}, \citenamefont {Ramos-Garcia},\ and\ \citenamefont
  {Forbes}}]{rosales2020polarisation}%
  \BibitemOpen
  \bibfield  {author} {\bibinfo {author} {\bibfnamefont {C.}~\bibnamefont
  {Rosales-Guzm{\'a}n}}, \bibinfo {author} {\bibfnamefont {X.-B.}\ \bibnamefont
  {Hu}}, \bibinfo {author} {\bibfnamefont {A.}~\bibnamefont {Selyem}}, \bibinfo
  {author} {\bibfnamefont {P.}~\bibnamefont {Moreno-Acosta}}, \bibinfo {author}
  {\bibfnamefont {S.}~\bibnamefont {Franke-Arnold}}, \bibinfo {author}
  {\bibfnamefont {R.}~\bibnamefont {Ramos-Garcia}}, \ and\ \bibinfo {author}
  {\bibfnamefont {A.}~\bibnamefont {Forbes}},\ }\href@noop {} {\bibfield
  {journal} {\bibinfo  {journal} {Scientific Reports}\ }\textbf {\bibinfo
  {volume} {10}},\ \bibinfo {pages} {1} (\bibinfo {year} {2020})}\BibitemShut
  {NoStop}%
\bibitem [{\citenamefont {Chen}\ \emph {et~al.}(2021)\citenamefont {Chen},
  \citenamefont {Wang}, \citenamefont {Wan}, \citenamefont {Lu}, \citenamefont
  {Liu},\ and\ \citenamefont {Zhan}}]{chen2021compact}%
  \BibitemOpen
  \bibfield  {author} {\bibinfo {author} {\bibfnamefont {J.}~\bibnamefont
  {Chen}}, \bibinfo {author} {\bibfnamefont {Y.}~\bibnamefont {Wang}}, \bibinfo
  {author} {\bibfnamefont {C.}~\bibnamefont {Wan}}, \bibinfo {author}
  {\bibfnamefont {K.}~\bibnamefont {Lu}}, \bibinfo {author} {\bibfnamefont
  {Y.}~\bibnamefont {Liu}}, \ and\ \bibinfo {author} {\bibfnamefont
  {Q.}~\bibnamefont {Zhan}},\ }\href@noop {} {\bibfield  {journal} {\bibinfo
  {journal} {Optics Communications}\ ,\ \bibinfo {pages} {127112}} (\bibinfo
  {year} {2021})}\BibitemShut {NoStop}%
\bibitem [{\citenamefont {Wu}\ \emph {et~al.}(2019)\citenamefont {Wu},
  \citenamefont {Yang}, \citenamefont {Rosales-Guzm{\'a}n}, \citenamefont
  {Gao}, \citenamefont {Shi}, \citenamefont {Zhu} \emph
  {et~al.}}]{wu2019vectorial}%
  \BibitemOpen
  \bibfield  {author} {\bibinfo {author} {\bibfnamefont {H.-J.}\ \bibnamefont
  {Wu}}, \bibinfo {author} {\bibfnamefont {H.-R.}\ \bibnamefont {Yang}},
  \bibinfo {author} {\bibfnamefont {C.}~\bibnamefont {Rosales-Guzm{\'a}n}},
  \bibinfo {author} {\bibfnamefont {W.}~\bibnamefont {Gao}}, \bibinfo {author}
  {\bibfnamefont {B.-S.}\ \bibnamefont {Shi}}, \bibinfo {author} {\bibfnamefont
  {Z.-H.}\ \bibnamefont {Zhu}},  \emph {et~al.},\ }\href@noop {} {\bibfield
  {journal} {\bibinfo  {journal} {Physical Review A}\ }\textbf {\bibinfo
  {volume} {100}},\ \bibinfo {pages} {053840} (\bibinfo {year}
  {2019})}\BibitemShut {NoStop}%
\bibitem [{\citenamefont {Tang}\ \emph {et~al.}(2020)\citenamefont {Tang},
  \citenamefont {Li}, \citenamefont {Zhang}, \citenamefont {Deng},
  \citenamefont {Li},\ and\ \citenamefont {Brasselet}}]{tang2020harmonic}%
  \BibitemOpen
  \bibfield  {author} {\bibinfo {author} {\bibfnamefont {Y.}~\bibnamefont
  {Tang}}, \bibinfo {author} {\bibfnamefont {K.}~\bibnamefont {Li}}, \bibinfo
  {author} {\bibfnamefont {X.}~\bibnamefont {Zhang}}, \bibinfo {author}
  {\bibfnamefont {J.}~\bibnamefont {Deng}}, \bibinfo {author} {\bibfnamefont
  {G.}~\bibnamefont {Li}}, \ and\ \bibinfo {author} {\bibfnamefont
  {E.}~\bibnamefont {Brasselet}},\ }\href@noop {} {\bibfield  {journal}
  {\bibinfo  {journal} {Nature Photonics}\ }\textbf {\bibinfo {volume} {14}},\
  \bibinfo {pages} {658} (\bibinfo {year} {2020})}\BibitemShut {NoStop}%
\bibitem [{\citenamefont {Marrucci}\ \emph {et~al.}(2006)\citenamefont
  {Marrucci}, \citenamefont {Manzo},\ and\ \citenamefont
  {Paparo}}]{Marrucci2006}%
  \BibitemOpen
  \bibfield  {author} {\bibinfo {author} {\bibfnamefont {L.}~\bibnamefont
  {Marrucci}}, \bibinfo {author} {\bibfnamefont {C.}~\bibnamefont {Manzo}}, \
  and\ \bibinfo {author} {\bibfnamefont {D.}~\bibnamefont {Paparo}},\ }\href
  {\doibase 10.1103/PhysRevLett.96.163905} {\bibfield  {journal} {\bibinfo
  {journal} {Physical Review Letters}\ }\textbf {\bibinfo {volume} {96}},\
  \bibinfo {pages} {163905} (\bibinfo {year} {2006})}\BibitemShut {NoStop}%
\bibitem [{\citenamefont {Nassiri}\ and\ \citenamefont
  {Brasselet}(2018)}]{nassiri2018multispectral}%
  \BibitemOpen
  \bibfield  {author} {\bibinfo {author} {\bibfnamefont {M.~G.}\ \bibnamefont
  {Nassiri}}\ and\ \bibinfo {author} {\bibfnamefont {E.}~\bibnamefont
  {Brasselet}},\ }\href@noop {} {\bibfield  {journal} {\bibinfo  {journal}
  {Physical Review Letters}\ }\textbf {\bibinfo {volume} {121}},\ \bibinfo
  {pages} {213901} (\bibinfo {year} {2018})}\BibitemShut {NoStop}%
\bibitem [{\citenamefont {Devlin}\ \emph {et~al.}(2017)\citenamefont {Devlin},
  \citenamefont {Ambrosio}, \citenamefont {Rubin}, \citenamefont {Mueller},\
  and\ \citenamefont {Capasso}}]{devlin2017arbitrary}%
  \BibitemOpen
  \bibfield  {author} {\bibinfo {author} {\bibfnamefont {R.~C.}\ \bibnamefont
  {Devlin}}, \bibinfo {author} {\bibfnamefont {A.}~\bibnamefont {Ambrosio}},
  \bibinfo {author} {\bibfnamefont {N.~A.}\ \bibnamefont {Rubin}}, \bibinfo
  {author} {\bibfnamefont {J.~B.}\ \bibnamefont {Mueller}}, \ and\ \bibinfo
  {author} {\bibfnamefont {F.}~\bibnamefont {Capasso}},\ }\href@noop {}
  {\bibfield  {journal} {\bibinfo  {journal} {Science}\ }\textbf {\bibinfo
  {volume} {358}},\ \bibinfo {pages} {896} (\bibinfo {year}
  {2017})}\BibitemShut {NoStop}%
\bibitem [{\citenamefont {Forbes}(2019)}]{forbes2019structured}%
  \BibitemOpen
  \bibfield  {author} {\bibinfo {author} {\bibfnamefont {A.}~\bibnamefont
  {Forbes}},\ }\href@noop {} {\bibfield  {journal} {\bibinfo  {journal} {Laser
  \& Photonics Reviews}\ }\textbf {\bibinfo {volume} {13}},\ \bibinfo {pages}
  {1900140} (\bibinfo {year} {2019})}\BibitemShut {NoStop}%
\bibitem [{\citenamefont {Beckley}\ \emph {et~al.}(2012)\citenamefont
  {Beckley}, \citenamefont {Brown},\ and\ \citenamefont
  {Alonso}}]{beckley2012full}%
  \BibitemOpen
  \bibfield  {author} {\bibinfo {author} {\bibfnamefont {A.~M.}\ \bibnamefont
  {Beckley}}, \bibinfo {author} {\bibfnamefont {T.~G.}\ \bibnamefont {Brown}},
  \ and\ \bibinfo {author} {\bibfnamefont {M.~A.}\ \bibnamefont {Alonso}},\
  }\href@noop {} {\bibfield  {journal} {\bibinfo  {journal} {Optics express}\
  }\textbf {\bibinfo {volume} {20}},\ \bibinfo {pages} {9357} (\bibinfo {year}
  {2012})}\BibitemShut {NoStop}%
\bibitem [{\citenamefont {Ma}\ and\ \citenamefont
  {Ramachandran}(2020)}]{ma2020propagation}%
  \BibitemOpen
  \bibfield  {author} {\bibinfo {author} {\bibfnamefont {Z.}~\bibnamefont
  {Ma}}\ and\ \bibinfo {author} {\bibfnamefont {S.}~\bibnamefont
  {Ramachandran}},\ }\href@noop {} {\bibfield  {journal} {\bibinfo  {journal}
  {Nanophotonics}\ }\textbf {\bibinfo {volume} {1}} (\bibinfo {year}
  {2020})}\BibitemShut {NoStop}%
\bibitem [{\citenamefont {Biss}\ and\ \citenamefont
  {Brown}(2004)}]{biss2004primary}%
  \BibitemOpen
  \bibfield  {author} {\bibinfo {author} {\bibfnamefont {D.~P.}\ \bibnamefont
  {Biss}}\ and\ \bibinfo {author} {\bibfnamefont {T.}~\bibnamefont {Brown}},\
  }\href@noop {} {\bibfield  {journal} {\bibinfo  {journal} {Optics express}\
  }\textbf {\bibinfo {volume} {12}},\ \bibinfo {pages} {384} (\bibinfo {year}
  {2004})}\BibitemShut {NoStop}%
\bibitem [{\citenamefont {Youngworth}\ and\ \citenamefont
  {Brown}(2000)}]{youngworth2000focusing}%
  \BibitemOpen
  \bibfield  {author} {\bibinfo {author} {\bibfnamefont {K.~S.}\ \bibnamefont
  {Youngworth}}\ and\ \bibinfo {author} {\bibfnamefont {T.~G.}\ \bibnamefont
  {Brown}},\ }\href@noop {} {\bibfield  {journal} {\bibinfo  {journal} {Optics
  Express}\ }\textbf {\bibinfo {volume} {7}},\ \bibinfo {pages} {77} (\bibinfo
  {year} {2000})}\BibitemShut {NoStop}%
\bibitem [{\citenamefont {Mamani}\ \emph {et~al.}(2018)\citenamefont {Mamani},
  \citenamefont {Shi}, \citenamefont {Ahmed}, \citenamefont {Karnik},
  \citenamefont {Rodr{\'\i}guez-Contreras}, \citenamefont {Nolan},\ and\
  \citenamefont {Alfano}}]{mamani2018transmission}%
  \BibitemOpen
  \bibfield  {author} {\bibinfo {author} {\bibfnamefont {S.}~\bibnamefont
  {Mamani}}, \bibinfo {author} {\bibfnamefont {L.}~\bibnamefont {Shi}},
  \bibinfo {author} {\bibfnamefont {T.}~\bibnamefont {Ahmed}}, \bibinfo
  {author} {\bibfnamefont {R.}~\bibnamefont {Karnik}}, \bibinfo {author}
  {\bibfnamefont {A.}~\bibnamefont {Rodr{\'\i}guez-Contreras}}, \bibinfo
  {author} {\bibfnamefont {D.}~\bibnamefont {Nolan}}, \ and\ \bibinfo {author}
  {\bibfnamefont {R.}~\bibnamefont {Alfano}},\ }\href@noop {} {\bibfield
  {journal} {\bibinfo  {journal} {Journal of biophotonics}\ ,\ \bibinfo {pages}
  {e201800096}} (\bibinfo {year} {2018})}\BibitemShut {NoStop}%
\bibitem [{\citenamefont {Gianani}\ \emph {et~al.}(2020)\citenamefont
  {Gianani}, \citenamefont {Suprano}, \citenamefont {Giordani}, \citenamefont
  {Spagnolo}, \citenamefont {Sciarrino}, \citenamefont {Gorpas}, \citenamefont
  {Ntziachristos}, \citenamefont {Pinker}, \citenamefont {Biton}, \citenamefont
  {Kupferman} \emph {et~al.}}]{gianani2020transmission}%
  \BibitemOpen
  \bibfield  {author} {\bibinfo {author} {\bibfnamefont {I.}~\bibnamefont
  {Gianani}}, \bibinfo {author} {\bibfnamefont {A.}~\bibnamefont {Suprano}},
  \bibinfo {author} {\bibfnamefont {T.}~\bibnamefont {Giordani}}, \bibinfo
  {author} {\bibfnamefont {N.}~\bibnamefont {Spagnolo}}, \bibinfo {author}
  {\bibfnamefont {F.}~\bibnamefont {Sciarrino}}, \bibinfo {author}
  {\bibfnamefont {D.}~\bibnamefont {Gorpas}}, \bibinfo {author} {\bibfnamefont
  {V.}~\bibnamefont {Ntziachristos}}, \bibinfo {author} {\bibfnamefont
  {K.}~\bibnamefont {Pinker}}, \bibinfo {author} {\bibfnamefont
  {N.}~\bibnamefont {Biton}}, \bibinfo {author} {\bibfnamefont
  {J.}~\bibnamefont {Kupferman}},  \emph {et~al.},\ }\href@noop {} {\bibfield
  {journal} {\bibinfo  {journal} {Advanced Photonics}\ }\textbf {\bibinfo
  {volume} {2}},\ \bibinfo {pages} {036003} (\bibinfo {year}
  {2020})}\BibitemShut {NoStop}%
\bibitem [{\citenamefont {Biton}\ \emph {et~al.}(2021)\citenamefont {Biton},
  \citenamefont {Kupferman},\ and\ \citenamefont {Arnon}}]{biton2021oam}%
  \BibitemOpen
  \bibfield  {author} {\bibinfo {author} {\bibfnamefont {N.}~\bibnamefont
  {Biton}}, \bibinfo {author} {\bibfnamefont {J.}~\bibnamefont {Kupferman}}, \
  and\ \bibinfo {author} {\bibfnamefont {S.}~\bibnamefont {Arnon}},\
  }\href@noop {} {\bibfield  {journal} {\bibinfo  {journal} {Scientific
  Reports}\ }\textbf {\bibinfo {volume} {11}},\ \bibinfo {pages} {1} (\bibinfo
  {year} {2021})}\BibitemShut {NoStop}%
\bibitem [{\citenamefont {Suprano}\ \emph {et~al.}(2020)\citenamefont
  {Suprano}, \citenamefont {Giordani}, \citenamefont {Gianani}, \citenamefont
  {Spagnolo}, \citenamefont {Pinker}, \citenamefont {Kupferman}, \citenamefont
  {Arnon}, \citenamefont {Klemm}, \citenamefont {Gorpas}, \citenamefont
  {Ntziachristos} \emph {et~al.}}]{suprano2020propagation}%
  \BibitemOpen
  \bibfield  {author} {\bibinfo {author} {\bibfnamefont {A.}~\bibnamefont
  {Suprano}}, \bibinfo {author} {\bibfnamefont {T.}~\bibnamefont {Giordani}},
  \bibinfo {author} {\bibfnamefont {I.}~\bibnamefont {Gianani}}, \bibinfo
  {author} {\bibfnamefont {N.}~\bibnamefont {Spagnolo}}, \bibinfo {author}
  {\bibfnamefont {K.}~\bibnamefont {Pinker}}, \bibinfo {author} {\bibfnamefont
  {J.}~\bibnamefont {Kupferman}}, \bibinfo {author} {\bibfnamefont
  {S.}~\bibnamefont {Arnon}}, \bibinfo {author} {\bibfnamefont
  {U.}~\bibnamefont {Klemm}}, \bibinfo {author} {\bibfnamefont
  {D.}~\bibnamefont {Gorpas}}, \bibinfo {author} {\bibfnamefont
  {V.}~\bibnamefont {Ntziachristos}},  \emph {et~al.},\ }\href@noop {}
  {\bibfield  {journal} {\bibinfo  {journal} {Optics Express}\ }\textbf
  {\bibinfo {volume} {28}},\ \bibinfo {pages} {35427} (\bibinfo {year}
  {2020})}\BibitemShut {NoStop}%
\bibitem [{\citenamefont {Cox}\ \emph {et~al.}(2020)\citenamefont {Cox},
  \citenamefont {Mphuthi}, \citenamefont {Nape}, \citenamefont {Mashaba},
  \citenamefont {Cheng},\ and\ \citenamefont {Forbes}}]{cox2020structured}%
  \BibitemOpen
  \bibfield  {author} {\bibinfo {author} {\bibfnamefont {M.~A.}\ \bibnamefont
  {Cox}}, \bibinfo {author} {\bibfnamefont {N.}~\bibnamefont {Mphuthi}},
  \bibinfo {author} {\bibfnamefont {I.}~\bibnamefont {Nape}}, \bibinfo {author}
  {\bibfnamefont {N.}~\bibnamefont {Mashaba}}, \bibinfo {author} {\bibfnamefont
  {L.}~\bibnamefont {Cheng}}, \ and\ \bibinfo {author} {\bibfnamefont
  {A.}~\bibnamefont {Forbes}},\ }\href@noop {} {\bibfield  {journal} {\bibinfo
  {journal} {IEEE Journal of Selected Topics in Quantum Electronics}\ }\textbf
  {\bibinfo {volume} {27}},\ \bibinfo {pages} {1} (\bibinfo {year}
  {2020})}\BibitemShut {NoStop}%
\bibitem [{\citenamefont {Gu}\ \emph {et~al.}(2009)\citenamefont {Gu},
  \citenamefont {Korotkova},\ and\ \citenamefont {Gbur}}]{gu2009scintillation}%
  \BibitemOpen
  \bibfield  {author} {\bibinfo {author} {\bibfnamefont {Y.}~\bibnamefont
  {Gu}}, \bibinfo {author} {\bibfnamefont {O.}~\bibnamefont {Korotkova}}, \
  and\ \bibinfo {author} {\bibfnamefont {G.}~\bibnamefont {Gbur}},\ }\href@noop
  {} {\bibfield  {journal} {\bibinfo  {journal} {Optics letters}\ }\textbf
  {\bibinfo {volume} {34}},\ \bibinfo {pages} {2261} (\bibinfo {year}
  {2009})}\BibitemShut {NoStop}%
\bibitem [{\citenamefont {Cheng}\ \emph {et~al.}(2009)\citenamefont {Cheng},
  \citenamefont {Haus},\ and\ \citenamefont {Zhan}}]{cheng2009propagation}%
  \BibitemOpen
  \bibfield  {author} {\bibinfo {author} {\bibfnamefont {W.}~\bibnamefont
  {Cheng}}, \bibinfo {author} {\bibfnamefont {J.~W.}\ \bibnamefont {Haus}}, \
  and\ \bibinfo {author} {\bibfnamefont {Q.}~\bibnamefont {Zhan}},\ }\href@noop
  {} {\bibfield  {journal} {\bibinfo  {journal} {Optics express}\ }\textbf
  {\bibinfo {volume} {17}},\ \bibinfo {pages} {17829} (\bibinfo {year}
  {2009})}\BibitemShut {NoStop}%
\bibitem [{\citenamefont {Cai}\ \emph {et~al.}(2008)\citenamefont {Cai},
  \citenamefont {Lin}, \citenamefont {Eyyubo{\u{g}}lu},\ and\ \citenamefont
  {Baykal}}]{cai2008average}%
  \BibitemOpen
  \bibfield  {author} {\bibinfo {author} {\bibfnamefont {Y.}~\bibnamefont
  {Cai}}, \bibinfo {author} {\bibfnamefont {Q.}~\bibnamefont {Lin}}, \bibinfo
  {author} {\bibfnamefont {H.~T.}\ \bibnamefont {Eyyubo{\u{g}}lu}}, \ and\
  \bibinfo {author} {\bibfnamefont {Y.}~\bibnamefont {Baykal}},\ }\href@noop {}
  {\bibfield  {journal} {\bibinfo  {journal} {Optics express}\ }\textbf
  {\bibinfo {volume} {16}},\ \bibinfo {pages} {7665} (\bibinfo {year}
  {2008})}\BibitemShut {NoStop}%
\bibitem [{\citenamefont {Ji-Xiong}\ \emph {et~al.}(2010)\citenamefont
  {Ji-Xiong}, \citenamefont {Tao}, \citenamefont {Hui-Chuan},\ and\
  \citenamefont {Cheng-Liang}}]{ji2010propagation}%
  \BibitemOpen
  \bibfield  {author} {\bibinfo {author} {\bibfnamefont {P.}~\bibnamefont
  {Ji-Xiong}}, \bibinfo {author} {\bibfnamefont {W.}~\bibnamefont {Tao}},
  \bibinfo {author} {\bibfnamefont {L.}~\bibnamefont {Hui-Chuan}}, \ and\
  \bibinfo {author} {\bibfnamefont {L.}~\bibnamefont {Cheng-Liang}},\
  }\href@noop {} {\bibfield  {journal} {\bibinfo  {journal} {Chinese Physics
  B}\ }\textbf {\bibinfo {volume} {19}},\ \bibinfo {pages} {089201} (\bibinfo
  {year} {2010})}\BibitemShut {NoStop}%
\bibitem [{\citenamefont {Wang}\ and\ \citenamefont
  {Pu}(2008)}]{wang2008propagation}%
  \BibitemOpen
  \bibfield  {author} {\bibinfo {author} {\bibfnamefont {T.}~\bibnamefont
  {Wang}}\ and\ \bibinfo {author} {\bibfnamefont {J.}~\bibnamefont {Pu}},\
  }\href@noop {} {\bibfield  {journal} {\bibinfo  {journal} {Optics
  communications}\ }\textbf {\bibinfo {volume} {281}},\ \bibinfo {pages} {3617}
  (\bibinfo {year} {2008})}\BibitemShut {NoStop}%
\bibitem [{\citenamefont {Cox}\ \emph {et~al.}(2016)\citenamefont {Cox},
  \citenamefont {Rosales-Guzm\'{a}n}, \citenamefont {Lavery}, \citenamefont
  {Versfeld},\ and\ \citenamefont {Forbes}}]{Cox:16}%
  \BibitemOpen
  \bibfield  {author} {\bibinfo {author} {\bibfnamefont {M.~A.}\ \bibnamefont
  {Cox}}, \bibinfo {author} {\bibfnamefont {C.}~\bibnamefont
  {Rosales-Guzm\'{a}n}}, \bibinfo {author} {\bibfnamefont {M.~P.~J.}\
  \bibnamefont {Lavery}}, \bibinfo {author} {\bibfnamefont {D.~J.}\
  \bibnamefont {Versfeld}}, \ and\ \bibinfo {author} {\bibfnamefont
  {A.}~\bibnamefont {Forbes}},\ }\href@noop {} {\bibfield  {journal} {\bibinfo
  {journal} {Optics Express}\ }\textbf {\bibinfo {volume} {24}},\ \bibinfo
  {pages} {18105} (\bibinfo {year} {2016})}\BibitemShut {NoStop}%
\bibitem [{\citenamefont {Lochab}\ \emph {et~al.}(2018)\citenamefont {Lochab},
  \citenamefont {Senthilkumaran},\ and\ \citenamefont
  {Khare}}]{lochab2018designer}%
  \BibitemOpen
  \bibfield  {author} {\bibinfo {author} {\bibfnamefont {P.}~\bibnamefont
  {Lochab}}, \bibinfo {author} {\bibfnamefont {P.}~\bibnamefont
  {Senthilkumaran}}, \ and\ \bibinfo {author} {\bibfnamefont {K.}~\bibnamefont
  {Khare}},\ }\href@noop {} {\bibfield  {journal} {\bibinfo  {journal}
  {Physical Review A}\ }\textbf {\bibinfo {volume} {98}},\ \bibinfo {pages}
  {023831} (\bibinfo {year} {2018})}\BibitemShut {NoStop}%
\bibitem [{\citenamefont {Hufnagel}\ \emph {et~al.}(2020)\citenamefont
  {Hufnagel}, \citenamefont {Sit}, \citenamefont {Bouchard}, \citenamefont
  {Zhang}, \citenamefont {England}, \citenamefont {Heshami}, \citenamefont
  {Sussman},\ and\ \citenamefont {Karimi}}]{hufnagel2020investigation}%
  \BibitemOpen
  \bibfield  {author} {\bibinfo {author} {\bibfnamefont {F.}~\bibnamefont
  {Hufnagel}}, \bibinfo {author} {\bibfnamefont {A.}~\bibnamefont {Sit}},
  \bibinfo {author} {\bibfnamefont {F.}~\bibnamefont {Bouchard}}, \bibinfo
  {author} {\bibfnamefont {Y.}~\bibnamefont {Zhang}}, \bibinfo {author}
  {\bibfnamefont {D.}~\bibnamefont {England}}, \bibinfo {author} {\bibfnamefont
  {K.}~\bibnamefont {Heshami}}, \bibinfo {author} {\bibfnamefont {B.~J.}\
  \bibnamefont {Sussman}}, \ and\ \bibinfo {author} {\bibfnamefont
  {E.}~\bibnamefont {Karimi}},\ }\href@noop {} {\bibfield  {journal} {\bibinfo
  {journal} {New Journal of Physics}\ }\textbf {\bibinfo {volume} {22}},\
  \bibinfo {pages} {093074} (\bibinfo {year} {2020})}\BibitemShut {NoStop}%
\bibitem [{\citenamefont {Bouchard}\ \emph {et~al.}(2018)\citenamefont
  {Bouchard}, \citenamefont {Sit}, \citenamefont {Hufnagel}, \citenamefont
  {Abbas}, \citenamefont {Zhang}, \citenamefont {Heshami}, \citenamefont
  {Fickler}, \citenamefont {Marquardt}, \citenamefont {Leuchs}, \citenamefont
  {Karimi} \emph {et~al.}}]{bouchard2018quantum}%
  \BibitemOpen
  \bibfield  {author} {\bibinfo {author} {\bibfnamefont {F.}~\bibnamefont
  {Bouchard}}, \bibinfo {author} {\bibfnamefont {A.}~\bibnamefont {Sit}},
  \bibinfo {author} {\bibfnamefont {F.}~\bibnamefont {Hufnagel}}, \bibinfo
  {author} {\bibfnamefont {A.}~\bibnamefont {Abbas}}, \bibinfo {author}
  {\bibfnamefont {Y.}~\bibnamefont {Zhang}}, \bibinfo {author} {\bibfnamefont
  {K.}~\bibnamefont {Heshami}}, \bibinfo {author} {\bibfnamefont
  {R.}~\bibnamefont {Fickler}}, \bibinfo {author} {\bibfnamefont
  {C.}~\bibnamefont {Marquardt}}, \bibinfo {author} {\bibfnamefont
  {G.}~\bibnamefont {Leuchs}}, \bibinfo {author} {\bibfnamefont
  {E.}~\bibnamefont {Karimi}},  \emph {et~al.},\ }\href@noop {} {\bibfield
  {journal} {\bibinfo  {journal} {Optics express}\ }\textbf {\bibinfo {volume}
  {26}},\ \bibinfo {pages} {22563} (\bibinfo {year} {2018})}\BibitemShut
  {NoStop}%
\bibitem [{\citenamefont {Ren}\ \emph {et~al.}(2016)\citenamefont {Ren},
  \citenamefont {Li}, \citenamefont {Wang}, \citenamefont {Kamali},
  \citenamefont {Arbabi}, \citenamefont {Arbabi}, \citenamefont {Zhao},
  \citenamefont {Xie}, \citenamefont {Cao}, \citenamefont {Ahmed} \emph
  {et~al.}}]{ren2016orbital}%
  \BibitemOpen
  \bibfield  {author} {\bibinfo {author} {\bibfnamefont {Y.}~\bibnamefont
  {Ren}}, \bibinfo {author} {\bibfnamefont {L.}~\bibnamefont {Li}}, \bibinfo
  {author} {\bibfnamefont {Z.}~\bibnamefont {Wang}}, \bibinfo {author}
  {\bibfnamefont {S.~M.}\ \bibnamefont {Kamali}}, \bibinfo {author}
  {\bibfnamefont {E.}~\bibnamefont {Arbabi}}, \bibinfo {author} {\bibfnamefont
  {A.}~\bibnamefont {Arbabi}}, \bibinfo {author} {\bibfnamefont
  {Z.}~\bibnamefont {Zhao}}, \bibinfo {author} {\bibfnamefont {G.}~\bibnamefont
  {Xie}}, \bibinfo {author} {\bibfnamefont {Y.}~\bibnamefont {Cao}}, \bibinfo
  {author} {\bibfnamefont {N.}~\bibnamefont {Ahmed}},  \emph {et~al.},\
  }\href@noop {} {\bibfield  {journal} {\bibinfo  {journal} {Scientific
  reports}\ }\textbf {\bibinfo {volume} {6}},\ \bibinfo {pages} {1} (\bibinfo
  {year} {2016})}\BibitemShut {NoStop}%
\bibitem [{\citenamefont {Spreeuw}(1998)}]{Spreeuw1998}%
  \BibitemOpen
  \bibfield  {author} {\bibinfo {author} {\bibfnamefont {R.~J.~C.}\
  \bibnamefont {Spreeuw}},\ }\href {\doibase 10.1023/A:1018703709245}
  {\bibfield  {journal} {\bibinfo  {journal} {Foundations of Physics}\ }\textbf
  {\bibinfo {volume} {28}},\ \bibinfo {pages} {361} (\bibinfo {year}
  {1998})}\BibitemShut {NoStop}%
\bibitem [{\citenamefont {Forbes}\ \emph {et~al.}(2019)\citenamefont {Forbes},
  \citenamefont {Aiello},\ and\ \citenamefont
  {Ndagano}}]{forbes2019classically}%
  \BibitemOpen
  \bibfield  {author} {\bibinfo {author} {\bibfnamefont {A.}~\bibnamefont
  {Forbes}}, \bibinfo {author} {\bibfnamefont {A.}~\bibnamefont {Aiello}}, \
  and\ \bibinfo {author} {\bibfnamefont {B.}~\bibnamefont {Ndagano}},\
  }\href@noop {} {\bibfield  {journal} {\bibinfo  {journal} {Progress in
  Optics}\ }\textbf {\bibinfo {volume} {64}},\ \bibinfo {pages} {99} (\bibinfo
  {year} {2019})}\BibitemShut {NoStop}%
\bibitem [{\citenamefont {Kagalwala}\ \emph {et~al.}(2013)\citenamefont
  {Kagalwala}, \citenamefont {Di~Giuseppe}, \citenamefont {Abouraddy},\ and\
  \citenamefont {Saleh}}]{kagalwala2013bell}%
  \BibitemOpen
  \bibfield  {author} {\bibinfo {author} {\bibfnamefont {K.~H.}\ \bibnamefont
  {Kagalwala}}, \bibinfo {author} {\bibfnamefont {G.}~\bibnamefont
  {Di~Giuseppe}}, \bibinfo {author} {\bibfnamefont {A.~F.}\ \bibnamefont
  {Abouraddy}}, \ and\ \bibinfo {author} {\bibfnamefont {B.~E.}\ \bibnamefont
  {Saleh}},\ }\href@noop {} {\bibfield  {journal} {\bibinfo  {journal} {Nature
  Photonics}\ }\textbf {\bibinfo {volume} {7}},\ \bibinfo {pages} {72}
  (\bibinfo {year} {2013})}\BibitemShut {NoStop}%
\bibitem [{\citenamefont {Qian}\ and\ \citenamefont
  {Eberly}(2011)}]{qian2011entanglement}%
  \BibitemOpen
  \bibfield  {author} {\bibinfo {author} {\bibfnamefont {X.-F.}\ \bibnamefont
  {Qian}}\ and\ \bibinfo {author} {\bibfnamefont {J.}~\bibnamefont {Eberly}},\
  }\href@noop {} {\bibfield  {journal} {\bibinfo  {journal} {Optics Letters}\
  }\textbf {\bibinfo {volume} {36}},\ \bibinfo {pages} {4110} (\bibinfo {year}
  {2011})}\BibitemShut {NoStop}%
\bibitem [{\citenamefont {McLaren}\ \emph {et~al.}(2015)\citenamefont
  {McLaren}, \citenamefont {Konrad},\ and\ \citenamefont
  {Forbes}}]{McLaren2015}%
  \BibitemOpen
  \bibfield  {author} {\bibinfo {author} {\bibfnamefont {M.}~\bibnamefont
  {McLaren}}, \bibinfo {author} {\bibfnamefont {T.}~\bibnamefont {Konrad}}, \
  and\ \bibinfo {author} {\bibfnamefont {A.}~\bibnamefont {Forbes}},\ }\href
  {\doibase 10.1103/PhysRevA.92.023833} {\bibfield  {journal} {\bibinfo
  {journal} {Physical Review A}\ }\textbf {\bibinfo {volume} {92}},\ \bibinfo
  {pages} {023833} (\bibinfo {year} {2015})}\BibitemShut {NoStop}%
\bibitem [{\citenamefont {Jiang}\ \emph {et~al.}(2013)\citenamefont {Jiang},
  \citenamefont {Luo},\ and\ \citenamefont {Fu}}]{jiang2013channel}%
  \BibitemOpen
  \bibfield  {author} {\bibinfo {author} {\bibfnamefont {M.}~\bibnamefont
  {Jiang}}, \bibinfo {author} {\bibfnamefont {S.}~\bibnamefont {Luo}}, \ and\
  \bibinfo {author} {\bibfnamefont {S.}~\bibnamefont {Fu}},\ }\href@noop {}
  {\bibfield  {journal} {\bibinfo  {journal} {Physical Review A}\ }\textbf
  {\bibinfo {volume} {87}},\ \bibinfo {pages} {022310} (\bibinfo {year}
  {2013})}\BibitemShut {NoStop}%
\bibitem [{\citenamefont {Konrad}\ \emph {et~al.}(2008)\citenamefont {Konrad},
  \citenamefont {De~Melo}, \citenamefont {Tiersch}, \citenamefont {Kasztelan},
  \citenamefont {Arag{\~a}o},\ and\ \citenamefont
  {Buchleitner}}]{konrad2008evolution}%
  \BibitemOpen
  \bibfield  {author} {\bibinfo {author} {\bibfnamefont {T.}~\bibnamefont
  {Konrad}}, \bibinfo {author} {\bibfnamefont {F.}~\bibnamefont {De~Melo}},
  \bibinfo {author} {\bibfnamefont {M.}~\bibnamefont {Tiersch}}, \bibinfo
  {author} {\bibfnamefont {C.}~\bibnamefont {Kasztelan}}, \bibinfo {author}
  {\bibfnamefont {A.}~\bibnamefont {Arag{\~a}o}}, \ and\ \bibinfo {author}
  {\bibfnamefont {A.}~\bibnamefont {Buchleitner}},\ }\href@noop {} {\bibfield
  {journal} {\bibinfo  {journal} {Nature physics}\ }\textbf {\bibinfo {volume}
  {4}},\ \bibinfo {pages} {99} (\bibinfo {year} {2008})}\BibitemShut {NoStop}%
\bibitem [{\citenamefont {Valencia}\ \emph {et~al.}(2020)\citenamefont
  {Valencia}, \citenamefont {Goel}, \citenamefont {McCutcheon}, \citenamefont
  {Defienne},\ and\ \citenamefont {Malik}}]{valencia2020unscrambling}%
  \BibitemOpen
  \bibfield  {author} {\bibinfo {author} {\bibfnamefont {N.~H.}\ \bibnamefont
  {Valencia}}, \bibinfo {author} {\bibfnamefont {S.}~\bibnamefont {Goel}},
  \bibinfo {author} {\bibfnamefont {W.}~\bibnamefont {McCutcheon}}, \bibinfo
  {author} {\bibfnamefont {H.}~\bibnamefont {Defienne}}, \ and\ \bibinfo
  {author} {\bibfnamefont {M.}~\bibnamefont {Malik}},\ }\href@noop {}
  {\bibfield  {journal} {\bibinfo  {journal} {Nature Physics}\ }\textbf
  {\bibinfo {volume} {16}},\ \bibinfo {pages} {1112} (\bibinfo {year}
  {2020})}\BibitemShut {NoStop}%
\bibitem [{\citenamefont {Ndagano}\ \emph {et~al.}(2017)\citenamefont
  {Ndagano}, \citenamefont {Perez-Garcia}, \citenamefont {Roux}, \citenamefont
  {McLaren}, \citenamefont {Rosales-Guzman}, \citenamefont {Zhang},
  \citenamefont {Mouane}, \citenamefont {Hernandez-Aranda}, \citenamefont
  {Konrad},\ and\ \citenamefont {Forbes}}]{Ndagano2017}%
  \BibitemOpen
  \bibfield  {author} {\bibinfo {author} {\bibfnamefont {B.}~\bibnamefont
  {Ndagano}}, \bibinfo {author} {\bibfnamefont {B.}~\bibnamefont
  {Perez-Garcia}}, \bibinfo {author} {\bibfnamefont {F.~S.}\ \bibnamefont
  {Roux}}, \bibinfo {author} {\bibfnamefont {M.}~\bibnamefont {McLaren}},
  \bibinfo {author} {\bibfnamefont {C.}~\bibnamefont {Rosales-Guzman}},
  \bibinfo {author} {\bibfnamefont {Y.}~\bibnamefont {Zhang}}, \bibinfo
  {author} {\bibfnamefont {O.}~\bibnamefont {Mouane}}, \bibinfo {author}
  {\bibfnamefont {R.~I.}\ \bibnamefont {Hernandez-Aranda}}, \bibinfo {author}
  {\bibfnamefont {T.}~\bibnamefont {Konrad}}, \ and\ \bibinfo {author}
  {\bibfnamefont {A.}~\bibnamefont {Forbes}},\ }\href {\doibase
  10.1038/nphys4003} {\bibfield  {journal} {\bibinfo  {journal} {Nature
  Physics}\ }\textbf {\bibinfo {volume} {13}},\ \bibinfo {pages} {397}
  (\bibinfo {year} {2017})}\BibitemShut {NoStop}%
\bibitem [{\citenamefont {Zhan}(2009)}]{zhan2009cylindrical}%
  \BibitemOpen
  \bibfield  {author} {\bibinfo {author} {\bibfnamefont {Q.}~\bibnamefont
  {Zhan}},\ }\href@noop {} {\bibfield  {journal} {\bibinfo  {journal} {Advances
  in Optics and Photonics}\ }\textbf {\bibinfo {volume} {1}},\ \bibinfo {pages}
  {1} (\bibinfo {year} {2009})}\BibitemShut {NoStop}%
\bibitem [{\citenamefont {Vaity}\ \emph {et~al.}(2013)\citenamefont {Vaity},
  \citenamefont {Banerji},\ and\ \citenamefont {Singh}}]{Vaity2013}%
  \BibitemOpen
  \bibfield  {author} {\bibinfo {author} {\bibfnamefont {P.}~\bibnamefont
  {Vaity}}, \bibinfo {author} {\bibfnamefont {J.}~\bibnamefont {Banerji}}, \
  and\ \bibinfo {author} {\bibfnamefont {R.}~\bibnamefont {Singh}},\ }\href
  {\doibase 10.1016/j.physleta.2013.02.030} {\bibfield  {journal} {\bibinfo
  {journal} {Physics Letters A}\ }\textbf {\bibinfo {volume} {377}},\ \bibinfo
  {pages} {1154} (\bibinfo {year} {2013})}\BibitemShut {NoStop}%
\bibitem [{\citenamefont {Milione}\ \emph {et~al.}(2011)\citenamefont
  {Milione}, \citenamefont {Sztul}, \citenamefont {Nolan},\ and\ \citenamefont
  {Alfano}}]{milione2011higher}%
  \BibitemOpen
  \bibfield  {author} {\bibinfo {author} {\bibfnamefont {G.}~\bibnamefont
  {Milione}}, \bibinfo {author} {\bibfnamefont {H.}~\bibnamefont {Sztul}},
  \bibinfo {author} {\bibfnamefont {D.}~\bibnamefont {Nolan}}, \ and\ \bibinfo
  {author} {\bibfnamefont {R.}~\bibnamefont {Alfano}},\ }\href@noop {}
  {\bibfield  {journal} {\bibinfo  {journal} {Physical review letters}\
  }\textbf {\bibinfo {volume} {107}},\ \bibinfo {pages} {053601} (\bibinfo
  {year} {2011})}\BibitemShut {NoStop}%
\bibitem [{\citenamefont {Holleczek}\ \emph {et~al.}(2011)\citenamefont
  {Holleczek}, \citenamefont {Aiello}, \citenamefont {Gabriel}, \citenamefont
  {Marquardt},\ and\ \citenamefont {Leuchs}}]{holleczek2011classical}%
  \BibitemOpen
  \bibfield  {author} {\bibinfo {author} {\bibfnamefont {A.}~\bibnamefont
  {Holleczek}}, \bibinfo {author} {\bibfnamefont {A.}~\bibnamefont {Aiello}},
  \bibinfo {author} {\bibfnamefont {C.}~\bibnamefont {Gabriel}}, \bibinfo
  {author} {\bibfnamefont {C.}~\bibnamefont {Marquardt}}, \ and\ \bibinfo
  {author} {\bibfnamefont {G.}~\bibnamefont {Leuchs}},\ }\href@noop {}
  {\bibfield  {journal} {\bibinfo  {journal} {Optics express}\ }\textbf
  {\bibinfo {volume} {19}},\ \bibinfo {pages} {9714} (\bibinfo {year}
  {2011})}\BibitemShut {NoStop}%
\bibitem [{\citenamefont {Selyem}\ \emph {et~al.}(2019)\citenamefont {Selyem},
  \citenamefont {Rosales-Guzm{\'a}n}, \citenamefont {Croke}, \citenamefont
  {Forbes},\ and\ \citenamefont {Franke-Arnold}}]{selyem2019basis}%
  \BibitemOpen
  \bibfield  {author} {\bibinfo {author} {\bibfnamefont {A.}~\bibnamefont
  {Selyem}}, \bibinfo {author} {\bibfnamefont {C.}~\bibnamefont
  {Rosales-Guzm{\'a}n}}, \bibinfo {author} {\bibfnamefont {S.}~\bibnamefont
  {Croke}}, \bibinfo {author} {\bibfnamefont {A.}~\bibnamefont {Forbes}}, \
  and\ \bibinfo {author} {\bibfnamefont {S.}~\bibnamefont {Franke-Arnold}},\
  }\href@noop {} {\bibfield  {journal} {\bibinfo  {journal} {Physical Review
  A}\ }\textbf {\bibinfo {volume} {100}},\ \bibinfo {pages} {063842} (\bibinfo
  {year} {2019})}\BibitemShut {NoStop}%
\bibitem [{\citenamefont {Padgett}\ and\ \citenamefont
  {Courtial}(1999)}]{padgett1999poincare}%
  \BibitemOpen
  \bibfield  {author} {\bibinfo {author} {\bibfnamefont {M.~J.}\ \bibnamefont
  {Padgett}}\ and\ \bibinfo {author} {\bibfnamefont {J.}~\bibnamefont
  {Courtial}},\ }\href@noop {} {\bibfield  {journal} {\bibinfo  {journal}
  {Optics letters}\ }\textbf {\bibinfo {volume} {24}},\ \bibinfo {pages} {430}
  (\bibinfo {year} {1999})}\BibitemShut {NoStop}%
\bibitem [{\citenamefont {Hu}\ \emph {et~al.}(2021)\citenamefont {Hu},
  \citenamefont {Luo}, \citenamefont {Pan}, \citenamefont {Qin}, \citenamefont
  {Zhang}, \citenamefont {Wei}, \citenamefont {Chen}, \citenamefont {Gao},\
  and\ \citenamefont {Li}}]{hu2021collapse}%
  \BibitemOpen
  \bibfield  {author} {\bibinfo {author} {\bibfnamefont {H.}~\bibnamefont
  {Hu}}, \bibinfo {author} {\bibfnamefont {D.}~\bibnamefont {Luo}}, \bibinfo
  {author} {\bibfnamefont {C.}~\bibnamefont {Pan}}, \bibinfo {author}
  {\bibfnamefont {Y.}~\bibnamefont {Qin}}, \bibinfo {author} {\bibfnamefont
  {Y.}~\bibnamefont {Zhang}}, \bibinfo {author} {\bibfnamefont
  {D.}~\bibnamefont {Wei}}, \bibinfo {author} {\bibfnamefont {H.}~\bibnamefont
  {Chen}}, \bibinfo {author} {\bibfnamefont {H.}~\bibnamefont {Gao}}, \ and\
  \bibinfo {author} {\bibfnamefont {F.}~\bibnamefont {Li}},\ }\href@noop {}
  {\bibfield  {journal} {\bibinfo  {journal} {Optics Letters}\ }\textbf
  {\bibinfo {volume} {46}},\ \bibinfo {pages} {2614} (\bibinfo {year}
  {2021})}\BibitemShut {NoStop}%
\bibitem [{\citenamefont {Sroor}\ \emph {et~al.}(2018)\citenamefont {Sroor},
  \citenamefont {Lisa}, \citenamefont {Naidoo}, \citenamefont {Litvin},\ and\
  \citenamefont {Forbes}}]{sroor2018purity}%
  \BibitemOpen
  \bibfield  {author} {\bibinfo {author} {\bibfnamefont {H.}~\bibnamefont
  {Sroor}}, \bibinfo {author} {\bibfnamefont {N.}~\bibnamefont {Lisa}},
  \bibinfo {author} {\bibfnamefont {D.}~\bibnamefont {Naidoo}}, \bibinfo
  {author} {\bibfnamefont {I.}~\bibnamefont {Litvin}}, \ and\ \bibinfo {author}
  {\bibfnamefont {A.}~\bibnamefont {Forbes}},\ }\href@noop {} {\bibfield
  {journal} {\bibinfo  {journal} {Physical Review Applied}\ }\textbf {\bibinfo
  {volume} {9}},\ \bibinfo {pages} {044010} (\bibinfo {year}
  {2018})}\BibitemShut {NoStop}%
\bibitem [{\citenamefont {Meyer}\ \emph {et~al.}(2020)\citenamefont {Meyer},
  \citenamefont {Mamani},\ and\ \citenamefont {Alfano}}]{meyer2020steady}%
  \BibitemOpen
  \bibfield  {author} {\bibinfo {author} {\bibfnamefont {H.~J.}\ \bibnamefont
  {Meyer}}, \bibinfo {author} {\bibfnamefont {S.}~\bibnamefont {Mamani}}, \
  and\ \bibinfo {author} {\bibfnamefont {R.~R.}\ \bibnamefont {Alfano}},\
  }\href@noop {} {\bibfield  {journal} {\bibinfo  {journal} {Applied Optics}\
  }\textbf {\bibinfo {volume} {59}},\ \bibinfo {pages} {6245} (\bibinfo {year}
  {2020})}\BibitemShut {NoStop}%
\bibitem [{\citenamefont {Far{\'\i}as}\ \emph {et~al.}(2015)\citenamefont
  {Far{\'\i}as}, \citenamefont {D'ambrosio}, \citenamefont {Taballione},
  \citenamefont {Bisesto}, \citenamefont {Slussarenko}, \citenamefont {Aolita},
  \citenamefont {Marrucci}, \citenamefont {Walborn},\ and\ \citenamefont
  {Sciarrino}}]{farias2015resilience}%
  \BibitemOpen
  \bibfield  {author} {\bibinfo {author} {\bibfnamefont {O.~J.}\ \bibnamefont
  {Far{\'\i}as}}, \bibinfo {author} {\bibfnamefont {V.}~\bibnamefont
  {D'ambrosio}}, \bibinfo {author} {\bibfnamefont {C.}~\bibnamefont
  {Taballione}}, \bibinfo {author} {\bibfnamefont {F.}~\bibnamefont {Bisesto}},
  \bibinfo {author} {\bibfnamefont {S.}~\bibnamefont {Slussarenko}}, \bibinfo
  {author} {\bibfnamefont {L.}~\bibnamefont {Aolita}}, \bibinfo {author}
  {\bibfnamefont {L.}~\bibnamefont {Marrucci}}, \bibinfo {author}
  {\bibfnamefont {S.~P.}\ \bibnamefont {Walborn}}, \ and\ \bibinfo {author}
  {\bibfnamefont {F.}~\bibnamefont {Sciarrino}},\ }\href@noop {} {\bibfield
  {journal} {\bibinfo  {journal} {Scientific reports}\ }\textbf {\bibinfo
  {volume} {5}},\ \bibinfo {pages} {1} (\bibinfo {year} {2015})}\BibitemShut
  {NoStop}%
\bibitem [{\citenamefont {Ndagano}\ and\ \citenamefont
  {Forbes}(2019)}]{ndagano2019entanglement}%
  \BibitemOpen
  \bibfield  {author} {\bibinfo {author} {\bibfnamefont {B.}~\bibnamefont
  {Ndagano}}\ and\ \bibinfo {author} {\bibfnamefont {A.}~\bibnamefont
  {Forbes}},\ }\href@noop {} {\bibfield  {journal} {\bibinfo  {journal} {APL
  Photonics}\ }\textbf {\bibinfo {volume} {4}},\ \bibinfo {pages} {016103}
  (\bibinfo {year} {2019})}\BibitemShut {NoStop}%
\end{thebibliography}%


\begin{thebibliography}{11}%
\makeatletter
\providecommand \@ifxundefined [1]{%
 \@ifx{#1\undefined}
}%
\providecommand \@ifnum [1]{%
 \ifnum #1\expandafter \@firstoftwo
 \else \expandafter \@secondoftwo
 \fi
}%
\providecommand \@ifx [1]{%
 \ifx #1\expandafter \@firstoftwo
 \else \expandafter \@secondoftwo
 \fi
}%
\providecommand \natexlab [1]{#1}%
\providecommand \enquote  [1]{``#1''}%
\providecommand \bibnamefont  [1]{#1}%
\providecommand \bibfnamefont [1]{#1}%
\providecommand \citenamefont [1]{#1}%
\providecommand \href@noop [0]{\@secondoftwo}%
\providecommand \href [0]{\begingroup \@sanitize@url \@href}%
\providecommand \@href[1]{\@@startlink{#1}\@@href}%
\providecommand \@@href[1]{\endgroup#1\@@endlink}%
\providecommand \@sanitize@url [0]{\catcode `\\12\catcode `\$12\catcode
  `\&12\catcode `\#12\catcode `\^12\catcode `\_12\catcode `\%12\relax}%
\providecommand \@@startlink[1]{}%
\providecommand \@@endlink[0]{}%
\providecommand \url  [0]{\begingroup\@sanitize@url \@url }%
\providecommand \@url [1]{\endgroup\@href {#1}{\urlprefix }}%
\providecommand \urlprefix  [0]{URL }%
\providecommand \Eprint [0]{\href }%
\providecommand \doibase [0]{http://dx.doi.org/}%
\providecommand \selectlanguage [0]{\@gobble}%
\providecommand \bibinfo  [0]{\@secondoftwo}%
\providecommand \bibfield  [0]{\@secondoftwo}%
\providecommand \translation [1]{[#1]}%
\providecommand \BibitemOpen [0]{}%
\providecommand \bibitemStop [0]{}%
\providecommand \bibitemNoStop [0]{.\EOS\space}%
\providecommand \EOS [0]{\spacefactor3000\relax}%
\providecommand \BibitemShut  [1]{\csname bibitem#1\endcsname}%
\let\auto@bib@innerbib\@empty
\bibitem [{\citenamefont {Choi}(1975)}]{choi1975completely}%
  \BibitemOpen
  \bibfield  {author} {\bibinfo {author} {\bibfnamefont {M.-D.}\ \bibnamefont
  {Choi}},\ }\href@noop {} {\bibfield  {journal} {\bibinfo  {journal} {Linear
  algebra and its applications}\ }\textbf {\bibinfo {volume} {10}},\ \bibinfo
  {pages} {285} (\bibinfo {year} {1975})}\BibitemShut {NoStop}%
\bibitem [{\citenamefont {Ndagano}\ \emph {et~al.}(2016)\citenamefont
  {Ndagano}, \citenamefont {Sroor}, \citenamefont {McLaren}, \citenamefont
  {Rosales-Guzm{\'a}n},\ and\ \citenamefont {Forbes}}]{ndagano2016beam}%
  \BibitemOpen
  \bibfield  {author} {\bibinfo {author} {\bibfnamefont {B.}~\bibnamefont
  {Ndagano}}, \bibinfo {author} {\bibfnamefont {H.}~\bibnamefont {Sroor}},
  \bibinfo {author} {\bibfnamefont {M.}~\bibnamefont {McLaren}}, \bibinfo
  {author} {\bibfnamefont {C.}~\bibnamefont {Rosales-Guzm{\'a}n}}, \ and\
  \bibinfo {author} {\bibfnamefont {A.}~\bibnamefont {Forbes}},\ }\href@noop {}
  {\bibfield  {journal} {\bibinfo  {journal} {Optics letters}\ }\textbf
  {\bibinfo {volume} {41}},\ \bibinfo {pages} {3407} (\bibinfo {year}
  {2016})}\BibitemShut {NoStop}%
\bibitem [{\citenamefont {Beijersbergen}\ \emph {et~al.}(1993)\citenamefont
  {Beijersbergen}, \citenamefont {Allen}, \citenamefont {van~der Veen},\ and\
  \citenamefont {Woerdman}}]{Beijersbergen1993}%
  \BibitemOpen
  \bibfield  {author} {\bibinfo {author} {\bibfnamefont {M.}~\bibnamefont
  {Beijersbergen}}, \bibinfo {author} {\bibfnamefont {L.}~\bibnamefont
  {Allen}}, \bibinfo {author} {\bibfnamefont {H.}~\bibnamefont {van~der Veen}},
  \ and\ \bibinfo {author} {\bibfnamefont {J.}~\bibnamefont {Woerdman}},\
  }\href {\doibase 10.1016/0030-4018(93)90535-d} {\bibfield  {journal}
  {\bibinfo  {journal} {Optics Communications}\ }\textbf {\bibinfo {volume}
  {96}},\ \bibinfo {pages} {123} (\bibinfo {year} {1993})}\BibitemShut
  {NoStop}%
\bibitem [{\citenamefont {da~Silva}\ \emph {et~al.}(2021)\citenamefont
  {da~Silva}, \citenamefont {Marques}, \citenamefont {Rodrigues}, \citenamefont
  {Ribeiro},\ and\ \citenamefont {Khoury}}]{Silva2021}%
  \BibitemOpen
  \bibfield  {author} {\bibinfo {author} {\bibfnamefont {B.~P.}\ \bibnamefont
  {da~Silva}}, \bibinfo {author} {\bibfnamefont {B.}~\bibnamefont {Marques}},
  \bibinfo {author} {\bibfnamefont {R.}~\bibnamefont {Rodrigues}}, \bibinfo
  {author} {\bibfnamefont {P.~S.}\ \bibnamefont {Ribeiro}}, \ and\ \bibinfo
  {author} {\bibfnamefont {A.}~\bibnamefont {Khoury}},\ }\href@noop {}
  {\bibfield  {journal} {\bibinfo  {journal} {Physical Review A}\ }\textbf
  {\bibinfo {volume} {103}},\ \bibinfo {pages} {063704} (\bibinfo {year}
  {2021})}\BibitemShut {NoStop}%
\bibitem [{\citenamefont {Vaity}\ \emph {et~al.}(2013)\citenamefont {Vaity},
  \citenamefont {Banerji},\ and\ \citenamefont {Singh}}]{Vaity2013}%
  \BibitemOpen
  \bibfield  {author} {\bibinfo {author} {\bibfnamefont {P.}~\bibnamefont
  {Vaity}}, \bibinfo {author} {\bibfnamefont {J.}~\bibnamefont {Banerji}}, \
  and\ \bibinfo {author} {\bibfnamefont {R.}~\bibnamefont {Singh}},\ }\href
  {\doibase 10.1016/j.physleta.2013.02.030} {\bibfield  {journal} {\bibinfo
  {journal} {Physics Letters A}\ }\textbf {\bibinfo {volume} {377}},\ \bibinfo
  {pages} {1154} (\bibinfo {year} {2013})}\BibitemShut {NoStop}%
\bibitem [{\citenamefont {Collins}(1970)}]{Collins1970lens}%
  \BibitemOpen
  \bibfield  {author} {\bibinfo {author} {\bibfnamefont {S.~A.}\ \bibnamefont
  {Collins}},\ }\href@noop {} {\bibfield  {journal} {\bibinfo  {journal}
  {JOSA}\ }\textbf {\bibinfo {volume} {60}},\ \bibinfo {pages} {1168} (\bibinfo
  {year} {1970})}\BibitemShut {NoStop}%
\bibitem [{\citenamefont {Ndagano}\ \emph {et~al.}(2017)\citenamefont
  {Ndagano}, \citenamefont {Perez-Garcia}, \citenamefont {Roux}, \citenamefont
  {McLaren}, \citenamefont {Rosales-Guzman}, \citenamefont {Zhang},
  \citenamefont {Mouane}, \citenamefont {Hernandez-Aranda}, \citenamefont
  {Konrad},\ and\ \citenamefont {Forbes}}]{Ndagano2017}%
  \BibitemOpen
  \bibfield  {author} {\bibinfo {author} {\bibfnamefont {B.}~\bibnamefont
  {Ndagano}}, \bibinfo {author} {\bibfnamefont {B.}~\bibnamefont
  {Perez-Garcia}}, \bibinfo {author} {\bibfnamefont {F.~S.}\ \bibnamefont
  {Roux}}, \bibinfo {author} {\bibfnamefont {M.}~\bibnamefont {McLaren}},
  \bibinfo {author} {\bibfnamefont {C.}~\bibnamefont {Rosales-Guzman}},
  \bibinfo {author} {\bibfnamefont {Y.}~\bibnamefont {Zhang}}, \bibinfo
  {author} {\bibfnamefont {O.}~\bibnamefont {Mouane}}, \bibinfo {author}
  {\bibfnamefont {R.~I.}\ \bibnamefont {Hernandez-Aranda}}, \bibinfo {author}
  {\bibfnamefont {T.}~\bibnamefont {Konrad}}, \ and\ \bibinfo {author}
  {\bibfnamefont {A.}~\bibnamefont {Forbes}},\ }\href {\doibase
  10.1038/nphys4003} {\bibfield  {journal} {\bibinfo  {journal} {Nature
  Physics}\ }\textbf {\bibinfo {volume} {13}},\ \bibinfo {pages} {397}
  (\bibinfo {year} {2017})}\BibitemShut {NoStop}%
\bibitem [{\citenamefont {Selyem}\ \emph {et~al.}(2019)\citenamefont {Selyem},
  \citenamefont {Rosales-Guzm{\'a}n}, \citenamefont {Croke}, \citenamefont
  {Forbes},\ and\ \citenamefont {Franke-Arnold}}]{selyem2019basis}%
  \BibitemOpen
  \bibfield  {author} {\bibinfo {author} {\bibfnamefont {A.}~\bibnamefont
  {Selyem}}, \bibinfo {author} {\bibfnamefont {C.}~\bibnamefont
  {Rosales-Guzm{\'a}n}}, \bibinfo {author} {\bibfnamefont {S.}~\bibnamefont
  {Croke}}, \bibinfo {author} {\bibfnamefont {A.}~\bibnamefont {Forbes}}, \
  and\ \bibinfo {author} {\bibfnamefont {S.}~\bibnamefont {Franke-Arnold}},\
  }\href@noop {} {\bibfield  {journal} {\bibinfo  {journal} {Physical Review
  A}\ }\textbf {\bibinfo {volume} {100}},\ \bibinfo {pages} {063842} (\bibinfo
  {year} {2019})}\BibitemShut {NoStop}%
\bibitem [{\citenamefont {Singh}\ \emph {et~al.}(2020)\citenamefont {Singh},
  \citenamefont {Tabebordbar}, \citenamefont {Forbes},\ and\ \citenamefont
  {Dudley}}]{singh2020digital}%
  \BibitemOpen
  \bibfield  {author} {\bibinfo {author} {\bibfnamefont {K.}~\bibnamefont
  {Singh}}, \bibinfo {author} {\bibfnamefont {N.}~\bibnamefont {Tabebordbar}},
  \bibinfo {author} {\bibfnamefont {A.}~\bibnamefont {Forbes}}, \ and\ \bibinfo
  {author} {\bibfnamefont {A.}~\bibnamefont {Dudley}},\ }\href@noop {}
  {\bibfield  {journal} {\bibinfo  {journal} {JOSA A}\ }\textbf {\bibinfo
  {volume} {37}},\ \bibinfo {pages} {C33} (\bibinfo {year} {2020})}\BibitemShut
  {NoStop}%
\bibitem [{\citenamefont {Fried}(1966)}]{fried1966optical}%
  \BibitemOpen
  \bibfield  {author} {\bibinfo {author} {\bibfnamefont {D.~L.}\ \bibnamefont
  {Fried}},\ }\href@noop {} {\bibfield  {journal} {\bibinfo  {journal} {JOSA}\
  }\textbf {\bibinfo {volume} {56}},\ \bibinfo {pages} {1372} (\bibinfo {year}
  {1966})}\BibitemShut {NoStop}%
\bibitem [{\citenamefont {Lane}\ \emph {et~al.}(1992)\citenamefont {Lane},
  \citenamefont {Glindemann},\ and\ \citenamefont
  {Dainty}}]{lane1992simulation}%
  \BibitemOpen
  \bibfield  {author} {\bibinfo {author} {\bibfnamefont {R.}~\bibnamefont
  {Lane}}, \bibinfo {author} {\bibfnamefont {A.}~\bibnamefont {Glindemann}}, \
  and\ \bibinfo {author} {\bibfnamefont {J.}~\bibnamefont {Dainty}},\
  }\href@noop {} {\bibfield  {journal} {\bibinfo  {journal} {Waves in random
  media}\ }\textbf {\bibinfo {volume} {2}},\ \bibinfo {pages} {209} (\bibinfo
  {year} {1992})}\BibitemShut {NoStop}%
\end{thebibliography}%

\end{document}


\title{Revealing the invariance of vectorial structured light in perturbing media\\ Supplementary Material}

\author{Isaac Nape}
\email[author contributions:]{ These authors contributed equally to the work}
\affiliation{School of Physics, University of the Witwatersrand, Private Bag 3, Wits 2050, South Africa}

\author{Keshaan Singh}
\email[author contributions:]{ These authors contributed equally to the work}
\affiliation{School of Physics, University of the Witwatersrand, Private Bag 3, Wits 2050, South Africa}

\author{Asher Klug}
\affiliation{School of Physics, University of the Witwatersrand, Private Bag 3, Wits 2050, South Africa}

\author{Wagner Buono}
\affiliation{School of Physics, University of the Witwatersrand, Private Bag 3, Wits 2050, South Africa}

\author{Carmelo Rosales-Guzman}
\affiliation{Centro de Investigaciones en \'Optica, A. C., Loma del Bosque 115, Col. Lomas del Campestre, 37150, Le\'on, Gto., M\'exico}
\affiliation{Wang Da-Heng Collaborative Innovation Center for Quantum Manipulation and Control, Harbin University of Science and Technology, Harbin 150080, China}

\author{Sonja Franke-Arnold}
\affiliation{School of Physics and Astronomy, University of Glasgow, Glasgow G12 8QQ, Scotland}

\author{Angela Dudley}
\affiliation{School of Physics, University of the Witwatersrand, Private Bag 3, Wits 2050, South Africa}

\author{Andrew Forbes}
\email[email:]{ andrew.forbes@wits.ac.za}
\affiliation{School of Physics, University of the Witwatersrand, Private Bag 3, Wits 2050, South Africa}
\email[Corresponding author: ]{andrew.forbes@wits.ac.za}
\date{\today}

\maketitle

\section{Transmitting vector beams through unitary single sided channels}
\label{sec:1}

Suppose we have a vector beam that we would like to transmit through a single sided channel, $\mathbb{1}_A \otimes T_{B}$, where $\mathbb{1}_B$ is the identity matrix while $T_B$ is a unitary operator, i.e., $T_{B}^\dagger = T_B^{-1}$.  The subscripts $A,\, B$ refer to the polarization and spatial degrees of freedom respectively. Such channels are known to be completely postive mappings and are trace preserving \cite{choi1975completely}. We can define our vector mode in a compact manner using Dirac notation
\begin{equation}
    \ket{\Psi} = \ket{e_1}_A\ket{u_1}_B + \ket{e_2}_A\ket{u_2}_B,
    \label{eq:inVMmode}
\end{equation}
where the spatial components $\{ \ket{u_1}$ and $\ket{u_2}\}$ are orthogonal. In the basis, $\{ \ket{j}, j = 1, 2, ..d \}$, we can express the spatial components of $\ket{\Psi}$ as 
\begin{align}
 \ket{u_1} = \sum_{j} \alpha_j \ket{j},
\end{align}
and 
\begin{align}
 \ket{u_2} = \sum_{j} \beta_j \ket{j}.
\end{align}
Here 
$\alpha_j$  and $\beta_{j}$ are the expansion coefficients of $u_{12}$, respectively. We can also express $T_B$ in our chosen basis $\{\ket{j}\}$:

\begin{align}
T_B = \sum_{jn} \tau_{jn} \ket{n}\bra{j},
\end{align}
where  the channel maps any state $\ket{j}_B$ onto to the state 
$\sum_{n}\tau_{jn}\ket{n}$ with coefficients $\tau_{jn}$, having the property that $\sum_{n}(\tau_{in})^* \tau_{nj} = \delta_{ij}$, due to the unitary nature of $T_B$.

Upon transferring $\ket{\Psi}$ through the channel, the spatial components become 
\begin{align}
 \ket{v_1} &= \sum_{j} \alpha_j \sum_{n} \tau_{jn} \ket{n} \nonumber,\\ 
             &= \sum_{n}\sum_{j} (\alpha_j \tau_{jn}) \ket{n} \nonumber,\\ 
            &=\sum_{n} \alpha'_n \ket{n},
            \label{eq: pertubedMode1}
\end{align}
and 
\begin{align}
 \ket{v_2} &= \sum_{j} \beta_{j} \sum_{m} \ \tau_{jm} \ket{m}, \nonumber \\
                &= \sum_{m} \sum_{j}( \beta_{j} \tau_{jm} )\ket{m}, \nonumber \\
               &=\sum_{m} \beta'_m \ket{m}.
               \label{eq: out_vectorBeam}
\end{align}
Consequently, after the channel, $\ket{\Psi}$ is mapped onto to the state 
\begin{equation}
    \ket{\Psi}_\text{out} = \ket{e_1}_A\ket{v_1}_B + \ket{e_2}_A\ket{v_2}_B.
    \label{eq:outVMmode}
\end{equation}
The expansion coefficients of the spatial components can be extracted via modal decomposition on each independent polarisation state. 

Importantly, the spatial components remain orthogonal:  $\braket{v_1}{ v_2 } = \bra{v_1} T_B^{\dagger}  T_B\ket{u_2} = \braket{u_1}{u_2} = 0$. 

 \section{Nonseparability of the transformed vector mode}

 The nonseparability or vector quality factor \cite{ndagano2016beam} of a vector beam can range from 0 to 1, for scalar fields (separable) and for completely nonseparable vector beams, respectively. For the vector beam in Eq. (\ref{eq:outVMmode}), the nonseparability will remain 1 provided the final projections are performed in the new basis ${\ket{u^{'}_{1,2}}}$. However when projected back into the initial basis, the vector modes in Eq.~(\ref{eq:inVMmode}) can be mapped onto 
\begin{align}
\ket{\Phi_{\ell}} &=  a  \ket{e_1}_A\ket{u_1}_B + b \ket{e_1}_A \ket{u_2}_B \nonumber \\
&+c \ket{e_2}_A \ket{u_1}_B  + d \ket{e_2}_A\ket{u_2}_B,
\end{align}
with a corresponding degree of nonseparability
\begin{equation}
    \text{V} = |ad - cb|,
    \label{eq:vqf}
\end{equation}
assuming the coefficients satisfy, $(|a|^2  +|b|^2 + |c|^2 +  |d|^2) = 1$.  When the modes are completely scattered, then $a=b=c=d = 1/2$ and hence $V= 0$. This means that the nonseparability will decay if the state is measured in the initial subspace. Otherwise, if there is no modal scattering then  $a=b$ while $cb = 0$, therefore the VQF is 1.

We can show this with the tilted lens channel, where performing the measurements for $\ell=\pm 1$ and $\ell=\pm 4$ 
(now in basis ${\ket{v_{1,2}}}$) using the old basis (${\ket{u_{1,2}}}$), is a form of post-selection onto the state $\ket{\Psi_{out}} = c_1\ket{R}\ket{u_1} +c_2\ket{R}\ket{u_2}+c_3\ket{L}\ket{u_1} + c_4\ket{L}\ket{u_2}$ with a resulting $V$ that can be estimated from $2|c_1c_4 - c_2b_3|$. The scattering coefficients can be computed as $ c_1 = \braket{u_1}{ v_1}$, $c_2 = \braket{u_2}{ v_1}$, $c_3 = \braket{u_1}{ v_2}$  and $c_4 = \braket{u_2}{ v_2}$. For the titled lens, the coefficients are shown in Table~\ref{table:coeff} where the overlap was performed. Expressions of $\ket{u_{1,2}}$ and $\ket{u_{1,2}}$ for the tilted lens can be found in the Supplementary material in terms of the HG basis. As expected, the $\ell =\pm 1$ $V$ does no change since the mapping is within the same subspace, consistent with the experimental plots in Fig.~1(a). In contrast, $\ell = \pm4$ has vectorness values of 0 and 1  for the incorrect and correct measurement subspace, respectively, reaffirming our experimental results.

\begin{table}

 \begin{tabular}{|c| c | c | c | c | c|} 
 \hline
Subspace & $c_1$ & $c_2$ & $c_3$ & $c_4$ & $V$ \\ [0.5ex] 
 \hline\hline
$\ell=\pm1$ ($u_{1,2}$) & 1/2 & i/2 & 1/2 & -i/2 & 1\\ 
 \hline
$\ell=\pm1$ ($v_{1,2}$) & 1/2 & 0 & 0 & 1/1& 1 \\
 \hline
$\ell=\pm4$ ($u_{1,2}$)  & 1/2 & 1/2 & 1/2 & 1/2 & 0 \\
 \hline
$\ell=\pm4$ ($v_{1,2}$) & 1/2 & 0 & 0 & 1/2 & 1 \\
 \hline
 \end{tabular}
 \caption{\textbf{Impact of selecting the measurement subspace}. The theoretical scattering complex coefficients for tilted lens and resulting vectorness ($V$) from measurements in the incorrect ($u_{1,2}$) and correct ($v_{1,2}$) subspace for $\ell=\pm1(4)$. In each case, the VQF is computed from  $2|c_1c_4 - c_2c_3|$.}
 \label{table:coeff}
\end{table}

\section{Undoing the effects of the channel}
\label{sec:2}

We have now seen that a vector beam $\ket{\Psi}$, interacting with the unitary channel, $ \mathbb{1}_A \otimes T_B$, maps onto a new vector beam $\ket{\Psi}_\text{out}$. Here we show that we can recover the vector beam $\ket{\Psi}$ by probing the channel a with a probe state $\ket{\Psi}_\text{probe}$  and later using it's modal content to prepare a new vector beam $\ket{\Phi}$ which will then be converted to $\ket{\Psi}$, our target mode, upon traversing the channel. 

To achieve this we first send in a probe state 
\begin{equation}
    \ket{\Psi}_\text{probe} =  \ket{e_1}_A \sum_{j} \alpha^{*}_{j}\ket{j}_B +  \ket{e_2}_A  \sum_{j} \beta_j^{*} \ket{j}_B,
\end{equation}
that is encoded with the complex conjugated coefficients of $\ket{\Psi}$, through the channel. 
After undergoing the channel the probe state is mapped to 

\begin{align}
      \ket{\Psi}_\text{probe} &\xrightarrow[]{\mathbb{1}_A \otimes T_B} \ket{e_1}_A \sum_{j n} \alpha^{*}_{j} \tau_{j n} \ket{n}_B + \ket{e_2}_A \sum_{j m} \beta_j^{*}  \tau_{j m} \ket{m}_B.
\end{align}
The coefficients of the modal expansion in each field can be measured via modal decomposition. Using the extracted coefficients, we can then send the new mode 
\begin{align}
    \ket{\Phi} &= \ket{e_1}_A \ket{u^{'}_1}_B + \ket{e_2}_A \ket{u^{'}_2}_B,\\
            &= \ket{e_1}_A\sum_{j n} \alpha_{j} \tau^{*}_{nj} \ket{n}_B + \ket{e_2}_A \sum_{j m} \beta_j \tau^{*}_{mj}  \ket{m}_B,
\end{align}
through the channel. Here $\ket{\Phi}$ has the conjugated coefficients of probe state after interacting with channel which are in-fact the coefficients of our desire target state $\ket{\Psi}$ coupled with the scattering coefficients from the channel. To see that $\ket{\Phi}$ maps to the target state $\ket{\Psi}$ after traversing the channel, we can observe the evolution of, $\ket{v_{12}}$  :
 
\begin{align}
T_B  \ket{u^{'}_1} &=\sum_{jn} \alpha_j \tau^{*}_{jn} T_B \ket{n}, \nonumber \\
                &=\sum_{jnk} \alpha_j \tau^{*}_{jn} \tau_{nk}  \ket{k},  \nonumber \\
                &=\sum_{jk} \alpha_j \sum_{n} \tau^{*}_{jn} \tau_{nk}  \ket{k},  \nonumber  \\
                &=\sum_{jk} \alpha_j \delta_{jk}  \ket{k}, \\
                &=\sum_{j} \alpha_j \ket{j}\\
                &=\ket{u_1}.
\end{align}

Similarly, $\ket{u^{'}_2}$, maps onto
\begin{align}
    T_B  \ket{u^{'}_2} &= \sum_{j} \beta_j \ket{j}, \nonumber\\
                      &= \ket{u_2}.
\end{align}
This gives us the spatial components of $\ket{\Psi}$, as desired.
\section{Examples with a titled lens}
\label{sec:exa_tl}
Vector beams can be used to imprint information about a spatially dependent perturbation. If the channel is unitary and single sided, then the spatial information can be imprinted on the polarisation field. The following steps show how one can obtain the transformation corresponding to a unitary channel via the polarisation pattern of vector beam and maybe also undo the evolution.

Suppose we have a vector beam given by 

\begin{equation}
    \ket{\psi}_\text{in} = \ket{R}_A\ket{\ell}_B + \ket{L}_A\ket{-\ell}_B,
\end{equation}

\noindent that is prepared in the right ($R$) and left ($L$) circular polarisation basis and the OAM spatial basis characterised by eigenstates with indices $\pm\ell$. This vector beam can be found on the Higher order Poincar{\'e} Sphere (HoPS) spanned by the basis modes \{ $\ket{R}_A\ket{\ell}_B$ , $\ket{L}_A \ket{-\ell}_B$ \}

When transmitted through a unitary and single sided channel, the OAM modes can scatter the spatial components into arbitrary $\ell$'s, but what remains is that the mode still remains a vector beam due to the invariance of nonseparability to unitary transformations. Since the vector beam can map the channel, we will show that this information can be used to make corrections on the beam with a simple unitary transformation. Lets see an example with a transformation that only induces transitions on a qubit space.

Consider the case where $\ell = \pm1$ for our vector beam.  The unitary mapping follows
\begin{align}
    \ket{\ell} &\xrightarrow{T_B} \frac{\ket{\ell} + i\ket{-\ell}}{\sqrt{2}} \\
    \ket{-\ell} &\xrightarrow{T_B} \frac{\ket{\ell} - i\ket{-\ell}}{\sqrt{2}}
\end{align}
\noindent mapping the spatial components onto another set of orthogonal modes. This can be performed using a titled lens. Once the vector beam interacts with the lens and is subsequently propagated to the far-field, it evolves into the state
\begin{equation}
    \ket{\psi}_\text{out} = \ket{R}_A \left (  \frac{\ket{\ell}_B + i\ket{-\ell}_B}{\sqrt{2}} \right ) + \ket{L}_A \left ( \frac{\ket{\ell}_B - i\ket{-\ell}_B}{\sqrt{2}} \right ),
\end{equation}
\noindent which is still nonseparable since the spatial components are orthogonal. All that the unitary transformation did was to perform a change of basis in the spatial components. With further manipulation, we arrive at
\begin{align}
    \ket{\psi}_\text{out} &=  \left (  \frac{\ket{R}_A + \ket{L}_A}{\sqrt{2}} \right) \ket{\ell}_B + i
    \left (  \frac{\ket{R}_A - \ket{L}_A}{\sqrt{2}} \right )  \ket{-\ell}_B, \\
    &=  \ket{H}_A\ket{\ell}_B + \ket{V}_A \ket{-\ell}_B.
\end{align}
\noindent which resembles the original mode except that the OAM components are now marked by the horizontal ($\ket{H}_A$) and vertical ($\ket{V}_A$) states. We see that the channel performed a simple change of basis in the spatial components which affects  the polarisation field therefore mapping us onto a new HoPs sphere that is spanned by \{ $\ket{H}_A\ket{\ell}_B$ , $\ket{V}_A \ket{-\ell}_B$ \}. Moreover, for this specific case, the relative phases between the spatial coordinates were transferred to the polarisation components. 

To obtain the original state, $\ket{\psi}_\text{in}$, we simply need to apply a unitary transformation, $\hat{U} = T_B^{\dagger}$, so that 
\begin{align}
   \ket{\psi}_\text{in} =  \hat{U} \ket{\psi}_\text{out}.
\end{align}
Therefore cancelling out the effect of the channel. 

For our example, a quarter-wave plate at an angle of $\theta = 45^\circ$ maps the polarisation field back to its original state. This is possible because the relative phase between the spatial modes was uniform. In general the polarisation mapping can be spatially dependent.

Alternatively, since the unitary maps the initial spatial modes basis to a new one, we can use the new modes in the detection system. For example, if we start with the mode 

\begin{equation}
    \ket{\psi}_\text{in} = \ket{R}_A\ket{\ell}_B + \ket{L}_A\ket{-\ell}_B,
\end{equation}, and send it through the channel $T_B$, we arrive at the new mode 

\begin{align}
   \ket{\psi}_\text{out} = \ket{R}_A\ket{\chi_1}_B + \ket{L}_A\ket{\chi_2}_B.
\end{align}

where $T_B\ket{\pm\ell}_B = \ket{\chi_{1,2}}_B$ are the transformed basis modes. In the next sections we describe how the LG modes can generally be mapped onto new modes with the titled lens.

\section{ Titled lens mapping }
	Following \cite{Beijersbergen1993}, the decomposition of LG modes into HG modes can be written as 
	\begin{equation}\label{eq:LgToHg}
		{\rm LG}_{n,m}(x,y,z)=\sum_{N}^{k=0}i^kb(n,m,k){\rm HG}_{N-k,k}(x,y,z)
	\end{equation}
	where $l=n-m$, $p=\min(n,m)$ and $b(n,m,k)$ is given by
	\begin{equation}\begin{split}
	b(n,m,k)= \left(\frac{(N-k)!k!}{2^N n!m!} \right)^{1/2} \frac{1}{k!}\frac{\mathrm{d}^k}{\mathrm{d}t^k}\left[ (1-t)^n(1+t)^m\right]_{t=0} \ \ .	\end{split} 
\end{equation}

	The same $b$ coefficients are also present when an HG mode 
	is decomposed into HG modes at $45^o$ as in
	\begin{equation}
		{\rm HG}_{n,m}(\frac{x+y}{\sqrt{2}},\frac{x-y}{\sqrt{2}},z)=\sum_{N}^{k=0}b(n,m,k){\rm HG}_{N-k,k}(x,y,z) .
	\end{equation}
	Intuitively, converting an LG into an HG mode is a question of rephasing modes. This is possible by exploiting the Gouy phase through a cylindrical lens. By using a lens with different focal lengths for two axis, i.e. astigmatic, it is possible to change the Rayleigh length of two coordinates independently, thus adding a relative phase to modes of a given geometry. It means that it is not possible to convert any mode in any direction, the nodal lines have to be parallel to the axes of the astigmatism. The HG modes are a suitable basis because they are completely separable in the direction of the astigmatism.

At the conversion plane, the cylindrical lens gives a $\pi/2$ phase between each HG mode that composes the input LG mode. In this sense, the action of the tilted lens in a given LG mode can be seen as:

\begin{equation}\label{eq:cLGp}
			{\rm LG}^{l}_p(x,y,z) \rightarrow \sum^{N}_{k=0}b(l+p,p,k){\rm HG}_{N-k,k}(x,y,z)
\end{equation}
for $l>0$ and 
\begin{equation}\label{eq:cLGn}
	{\rm LG}^{l}_p(x,y,z) \rightarrow \sum^{N}_{k=0}b(p,p-l,k){\rm HG}_{N-k,k}(x,y,z)
\end{equation}
otherwise.
	\begin{table}[t]
		\centering
		\begin{tabular}{ | c | c  | }
			\hline
			${\rm LG}^l_p $ & HG modes decomposition  \\
			\hline
			0,0 & $\text{HG}_{0,0}$\\
			\hline
			0,1 & $\frac{1}{\sqrt{2}}\text{HG}_{1,0}+\frac{i }{\sqrt{2}}\text{HG}_{0,1}$\\
			\hline
			0,-1 & $\frac{1}{\sqrt{2}}\text{HG}_{1,0}-\frac{i }{\sqrt{2}}\text{HG}_{0,1}$\\
			\hline
			0,2 & $-\frac{1}{2}\text{HG}_{0,2}+\frac{i
				}{\sqrt{2}}\text{HG}_{1,1}+\frac{1}{2}\text{HG}_{2,0}$\\
			\hline
			0,-2 & $-\frac{1}{2}\text{HG}_{0,2}-\frac{i
				}{\sqrt{2}}\text{HG}_{1,1}+\frac{1}{2}\text{HG}_{2,0}$\\
			\hline
			0,4 & $\frac{1}{4}\text{HG}_{0,4}-\frac{1}{2} i \text{HG}_{1,3}-\frac{1}{2}
			\sqrt{\frac{3}{2}} \text{HG}_{2,2}+\frac{1}{2} i
			\text{HG}_{3,1}+\frac{1}{4}\text{HG}_{4,0}$\\
			\hline
			0,-4 & $\frac{1}{4}\text{HG}_{0,4}+\frac{1}{2} i \text{HG}_{1,3}-\frac{1}{2}
			\sqrt{\frac{3}{2}} \text{HG}_{2,2}-\frac{1}{2} i
			\text{HG}_{3,1}+\frac{1}{4}\text{HG}_{4,0}$\\
			\hline

		\end{tabular}
		\caption{\textbf{Decomposition of LG modes into HG modes. }}
		\label{tab:lg_to_hg}
	\end{table}
	\begin{table}[t]
	\centering
	\begin{tabular}{ | c | c  | }
		\hline
		${\rm LG}^l_p $ & HG modes after conversion  \\
		\hline
		0,0 & $\text{HG}_{0,0}$\\
		\hline
		0,1 & $\frac{1}{\sqrt{2}}\text{HG}_{1,0}+\frac{1 }{\sqrt{2}}\text{HG}_{0,1}$\\
		\hline
		0,-1 & $\frac{1}{\sqrt{2}}\text{HG}_{1,0}-\frac{1 }{\sqrt{2}}\text{HG}_{0,1}$\\
		\hline
		0,2 & $\frac{1}{2}\text{HG}_{0,2}+\frac{1
		}{\sqrt{2}}\text{HG}_{1,1}+\frac{1}{2}\text{HG}_{2,0}$\\
		\hline
		0,-2 & $\frac{1}{2}\text{HG}_{0,2}-\frac{1
		}{\sqrt{2}}\text{HG}_{1,1}+\frac{1}{2}\text{HG}_{2,0}$\\
		\hline
		0,4 & $\frac{1}{4}\text{HG}_{0,4}+\frac{1}{2}  \text{HG}_{1,3}+\frac{1}{2}
		\sqrt{\frac{3}{2}} \text{HG}_{2,2}+\frac{1}{2} 
		\text{HG}_{3,1}+\frac{1}{4}\text{HG}_{4,0}$\\
		\hline
		0,-4 & $\frac{1}{4}\text{HG}_{0,4}-\frac{1}{2}  \text{HG}_{1,3}+\frac{1}{2}
		\sqrt{\frac{3}{2}} \text{HG}_{2,2}-\frac{1}{2} 
		\text{HG}_{3,1}+\frac{1}{4}\text{HG}_{4,0}$\\
		\hline

	\end{tabular}
	\caption{\textbf{Modal decomposition of LG after a tilted lens transformation. }}
	\label{tab:lg_to_tilted}
\end{table}

Alternatively, as described in \cite{Silva2021}, this astigmatic mode converter can be seen as a unitary transformation in each mode order subspace. The definition in terms of HG modes, which are the eigenstates of this transformation, is given by:
\begin{equation}
	L_{D}=\sum_{m=0}^{D-1}e^{i(m-n)\frac{\pi}{4}}\ket{HG_{m,n}}\bra{HG_{m,n}}
\end{equation}
where $D=N+1$ is the dimension of the subspace of the input mode and $n=D-m-1$.

\subsection{Wave optics description}
Following ref \cite{Vaity2013}, the ray transfer matrix for the tilted lens $\textbf{M}_{TL}$ is given by:
	\begin{equation}
		\textbf{M}_{TL}=\textbf{M}_z\textbf{M}_{lens}\textbf{M}_{z_0}= \left(\begin{matrix}
			\textbf{A} & \textbf{B} \\
			-\textbf{C}/f & \textbf{D}
		\end{matrix}\right)
	\end{equation}
where $\textbf{A}$, $\textbf{B}$, $\textbf{C}$, $\textbf{D}$ are $ 2 \times 2 $ diagonal matrices whose elements are $c_1=\sec\theta$, $c_2=\cos\theta$, $a_j=1-zc_j/f$, $d_j=1-z_0c/f$ and $ b_j=z_0+zd $ for $ j=1,2 $. The matrices $M_{z_0}$ and $M_z$ describe free space propagation before and after the lens by distances $z_0$ and $z$ respectively. $M_{lens}$ describes a lens rotated around the $y$ axis (vertical) by an angle $\theta$. How a ray matrix is interpreted as a transfer function can be found in \cite{Collins1970lens}.

From the generalized Huygens-Fresnel integral, an electric field that propagates through this optical element is given by 
\begin{equation}\begin{split}
    E_2(x_2,y_2)=&\frac{i/\lambda}{|B|^{1/2}}\iint \mathrm{d}x\mathrm{d}yE_1(x_1,y_1) \\
    &\times\exp\left[\frac{-(i\pi)}{\lambda}\phi(x_1,x_2,y_1,y_2)\right]\end{split}
\end{equation}
where $|B|$ is the determinant of $\textbf{B}$ and 
\begin{equation}\begin{split}
	\phi(x_1,x_2,y_1,y_2)=&x^2_1a_1/b_1+y^2a_2/b_2+x^2_2d_1/b_1+y^2_2d_2/b_2 \\
		&-2(x_1x_2/b_1+y_1y_2/b_2)	\end{split} 
\end{equation}
\noindent is the applied phase.

\section{Basis dependent and basis independent non-separability measurements}

A classical vector light field represents a hybrid-entanglement system, where the polarization and spatial mode degrees of freedom (DoFs) are non-separable. Since our one sided channel only affects the spatial mode DoF it is expected that a spatial mode basis dependent non-separability measurement should be affected by the channel while a basis independent measurement should remain invariant. Sections \ref{sec:1} and \ref{sec:2} show that the mode dependent non-separability is recoverable if the spatial mode basis is appropriately chosen. This is an important development as it allows for the recovery of mode dependent quantum correlations across such channels \cite{Ndagano2017}. In this section we will outline the basis independent non-separability measurement based on Stokes intensity projections as well as the basis dependent non-separability measurement based on spatial mode projections of orthogonally polarized components - both of which are based on the quantum mechanical concept of concurrence.

\subsection{Basis independent non-separability}

Consider a vector field of the form

\begin{equation}
    |\Psi\rangle = \ket{H}_A\ket{\psi_H}_B+\ket{V}_A\ket{\psi_V}_B
\end{equation}

where $|\psi_{H,V}\rangle_B$ are arbitrary unnormalized spatial modes. We can then expand the spatial modes into some orthonormal two dimensional basis set $|\psi_{\pm}\rangle_B$, giving the general form

\begin{align}
    |\Psi\rangle &= a|H\rangle_A|\psi_+\rangle_B + b|H\rangle_A|\psi_-\rangle|_B \nonumber \\  &+ c|V\rangle_A|\psi_+\rangle_B +  d|V\rangle_A|\psi_-\rangle|_B.
\end{align}

This state has a concurrence given by $C(|\Psi\rangle) = 2|ad - bc|$. By selecting a negligible global phase, we can express the orthonormal states as \cite{selyem2019basis}

\begin{equation}
    |\psi_{\pm}\rangle =\frac{1}{\sqrt{2(1\pm \langle\tilde{\psi}_H|\tilde{\psi}_V\rangle)}}(|\tilde{\psi}_H\rangle\pm|\tilde{\psi}_V\rangle) 
\end{equation}

where $|\tilde{\psi}_{H,V}\rangle=|\psi_{H,V}\rangle/\langle\psi_{H,V}|\psi_{H,V}\rangle$. This gives a concurrence $C(|\Psi\rangle) = 2\sqrt{\langle\psi_H|\psi_H\rangle\langle\psi_V|\psi_V\rangle - |\langle\psi_H|\psi_V\rangle|^2}$. Since we now have the concurrence expressed in terms of the initial arbitrary unnormalized spatial components we can use the spatially varying Stokes parameters $S'_i(\mathbf{r})$ ($\mathbf{r}$ being the rtansverse spatial coordinates), to retrieve the degree of non-separability using the global Stokes parameters, $S_i = \int\text{d}^2\mathbf{r}\, S'_i$, by exploiting the following relationships:

\begin{align}
    S_0 &= \langle\psi_H|\psi_H\rangle + \langle\psi_V|\psi_V\rangle \\
    S_1 &= \langle\psi_H|\psi_H\rangle - \langle\psi_V|\psi_V\rangle \\
    S_2 &= \langle\psi_H|\psi_V\rangle + \langle\psi_V|\psi_H\rangle\\
    S_3 &= i(\langle\psi_H|\psi_V\rangle - \langle\psi_V|\psi_H\rangle)
\end{align}

This convenient result allows us to determine the non-separability in a basis independent manner according to $C(|\Psi\rangle) = \sqrt{1-(S_1^2+S_2^2+S_3^2)/S_0^2}$. In this work the Stokes parameters were determined (for convenience) by the measurement of the reduced set of four Stokes intensities ($I_H,I_D,I_R$ and $I_L$), where $I_i=\langle\psi_i|\psi_i\rangle$) - according to 

\begin{align}
    S_0 &= I_R + I_L \\
    S_1 &= 2I_H - S_0 \\
    S_2 &= 2I_D - S_0\\
    S_3 &= I_R - I_L
\end{align}

The four intensity projections were acquired through the use of a linear polarizer (for $I_H$ and $I_D$) together with a quater-wave plate (for $I_R$ and $I_L$) \cite{singh2020digital}.

\subsection{Basis dependent non-separability}

Consider a vector field of the form

\begin{equation}
    |\Psi\rangle = \ket{H}_A\ket{\psi_H}_B+\ket{V}_A\ket{\psi_V}_B.
\end{equation}

\noindent The concurrence can be determined from $C(\Psi) = \sqrt{1 - s^2}$ where $s$ is the length of the vector when maps the state to a Bloch sphere defined by a chosen orthogonal set of orthonormal basis spatial modes $|\psi_{1,2}\rangle$ (resulting in the measurements dependence on this choice of basis) - given by

\begin{equation}
    s=\left(\sum\limits_i\langle\sigma_i\rangle^2\right)^{1/2},
\end{equation}

where $\langle\sigma_i\rangle$ are the expectation values of the Pauli matrices \cite{ndagano2016beam}. We can determine these expectation values using projections $\langle P|$ into superpositions of the spatial basis components described by

\begin{equation}
    \langle P_j| = \alpha_j\langle\psi_1| + \beta_j\langle\psi_2|\,.
\end{equation}

With $(\alpha,\beta)=\{(1,0),(0,1),\frac{1}{\sqrt{2}}(1,1),\frac{1}{\sqrt{2}}(1,-1),\\\frac{1}{\sqrt{2}}(1,i),\frac{1}{\sqrt{2}}(1,-i)\}$ for both $|H\rangle$ and $|V\rangle$. These 12 on axis intensity projections are used to calculate the length of the Bloch vector according to

\begin{align}
    \langle\sigma_1\rangle &= (I_{13}+I_{23}) - (I_{14}+I_{24}) \\
    \langle\sigma_2\rangle &= (I_{15}+I_{25}) - (I_{16}+I_{26})  \\
    \langle\sigma_3\rangle &= (I_{11}+I_{21}) - (I_{11}+I_{22}) 
\end{align}

where the $i$ index of $I_{ij}$ corresponds to the $\langle H,V|$ polarization projections and the $j$ index represents the spatial mode projections defined above. In this work the projections into the horizontal and vertical polarization components was achieved using a linear polarizer. The subsequent spatial mode projections were performed using a correlation filter encoded into a digital micro-mirror device (DMD), and a Fourier lens to produce on-axis intensities $I_{ij}$.
\begin{figure}
	\includegraphics[width=\linewidth]{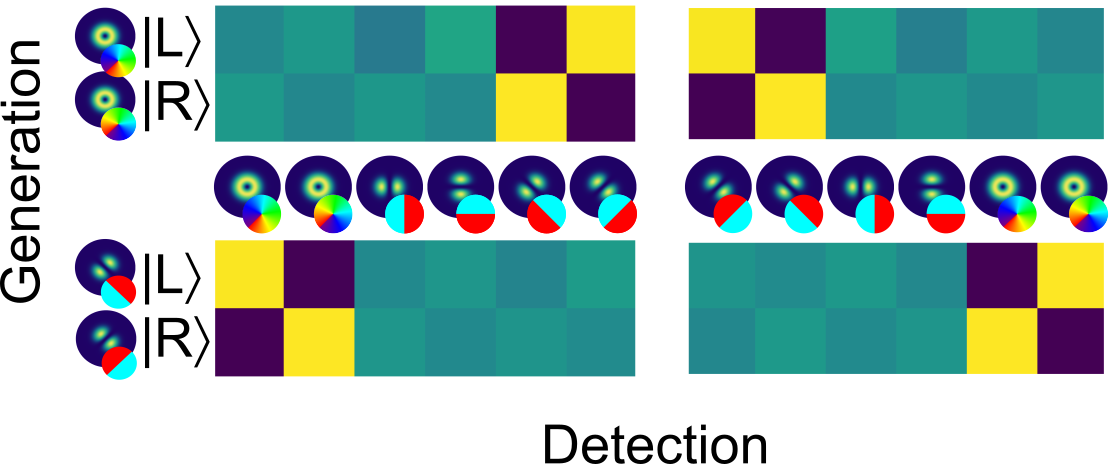}
	\caption{Basis dependent measurement for tilted lens with input mode of $\ell=1$.\label{fig:CT1}}
\end{figure}

\begin{figure}
	\includegraphics[width=\linewidth]{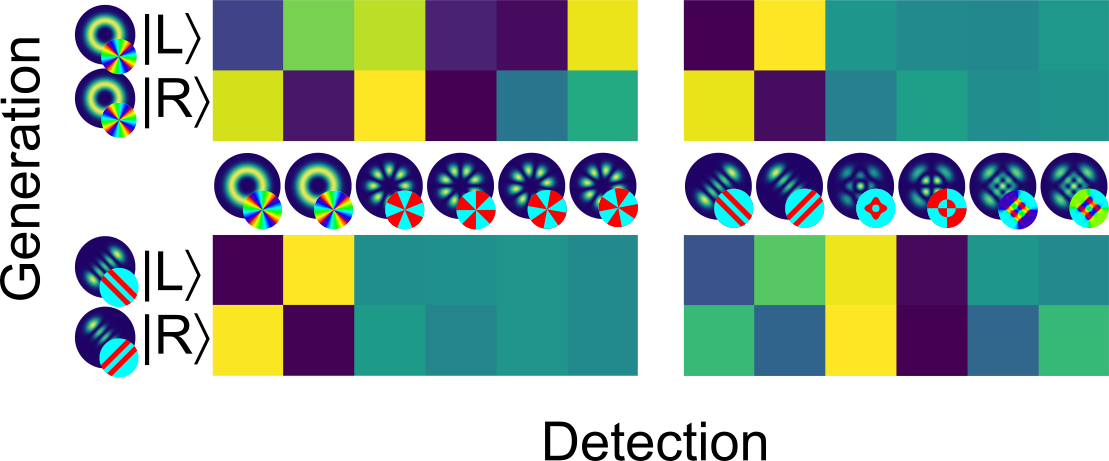}
	\caption{Basis dependent measurement for tilted lens with input mode of $\ell=4$.\label{fig:CT4}}
\end{figure}
In figures \ref{fig:CT1} and \ref{fig:CT4} we can see basis dependent measurements for the vector mode depicted in section \ref{sec:exa_tl} when passed through a tilted lens. While in figure \ref{fig:CT1} the OAM state is $\ell=1$ and we can see a simple shift in basis, in figure \ref{fig:CT4} the OAM state is $\ell=4$ there is considerable crosstalk. These measurements were used to calculate the VQF values in figure 5(a).

\section{Simulating turbulence}
\begin{figure}[h]
    \centering
    \includegraphics[width=\linewidth]{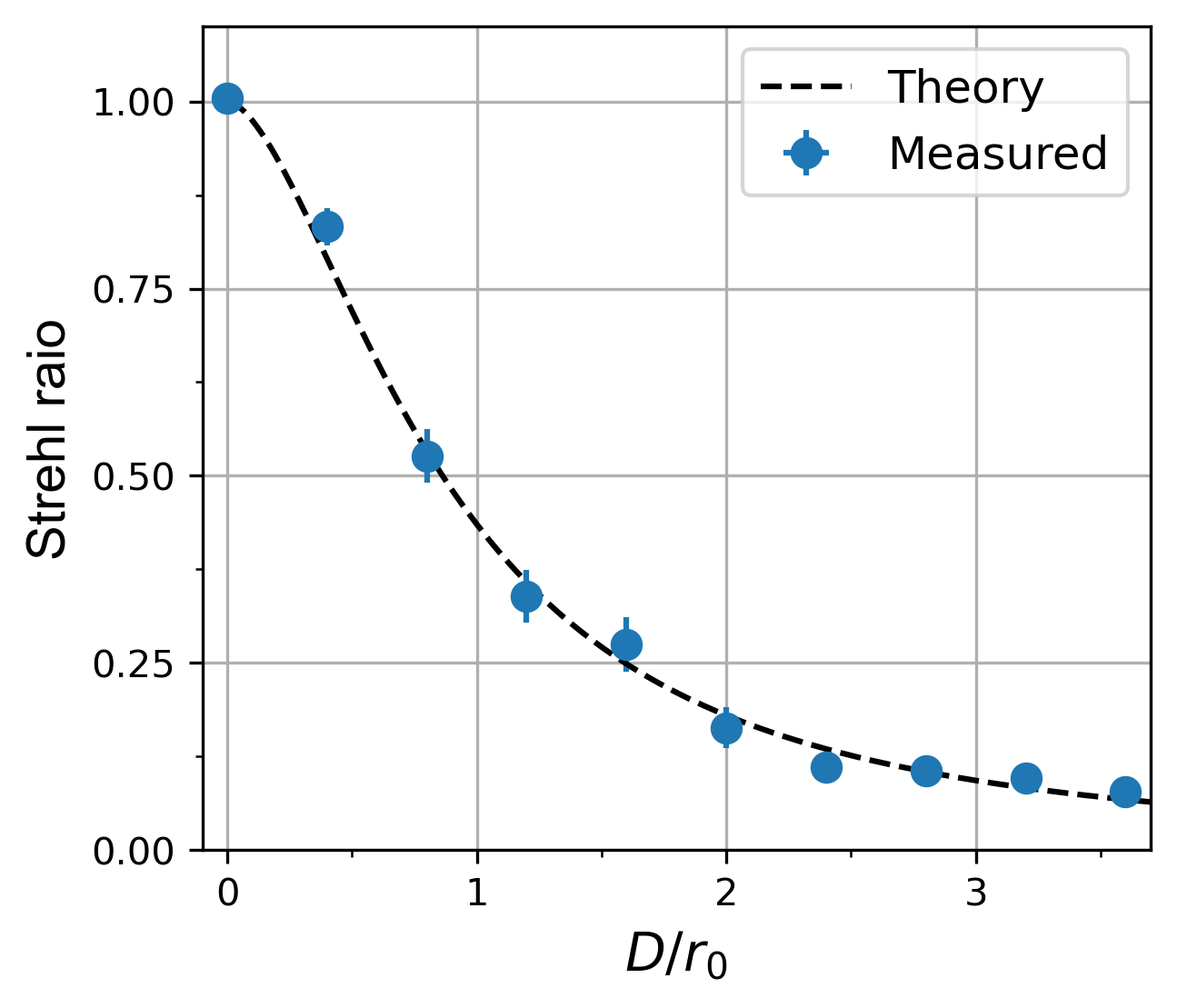}
    \caption{The Strehl ratio (SR) for increasing turbulence strength $D/r_0$. Theoretical (programmed) values are plotted as a solid line and experimentally measured points are represented as blue markers.}
    \label{fig:SR}
\end{figure}

We model turbulence as a random phase screen $\Theta$ that captures the refractive index variations in the atmosphere. The statistics of such a screen are well known. Two useful characterisations of $\Theta$ are the phase structure function, defined as
\begin{equation}\label{eq:psf}
    D(\rr) = \left< |\Theta(\rr'+\rr)-\Theta(\rr')|^2\right>,
\end{equation}
and the Weiner spectrum $\Phi$ which is related to $D(\rr)$ by 
\begin{equation}\label{eq:weiner spectrum}
    D(\rr) = 2\int\dd[2]{\mathbf{k}}\,\Phi(\mathbf{k})[1-\cos(2\pi\mathbf{k}\cdot\rr)],
\end{equation}
where $\langle\cdot\rangle$ denotes the ensemble average. In the Kolmogorov model, these expressions are
\begin{equation}\label{eq:Kolmogorov psf}
    D(\rr) = 6.88\left(\frac{\norm{\rr}}{r_0}\right)^{5/3} ,
\end{equation}
and
\begin{equation}\label{eq:Kolmogorov weiner spectrum}
    \Phi(\mathbf{k}) = \frac{0.023}{r_0^{5/3}}\norm{\mathbf{k}}^{-11/3},
\end{equation}
where $r_0$ is the Fried parameter\cite{fried1966optical}. These quantities, which are taken over many iterations of $\Theta$, are not the same as $\Theta$ itself, which is a random variable, or its equally random power spectrum
\begin{equation}\label{eq:power spectrum}
    P(\mathbf{k}) = \mathcal{F}\{\Theta\} = \int\dd[2]{\rr}\, \Theta\exp\left(-i2\pi\mathbf{k}\cdot\rr\right),
\end{equation}
where $\mathcal{F}$ denotes the Fourier transform. In fact, $\Phi(\mathbf{k})$ is the covariance function of $P(\mathbf{k})$. To simulate a turbulent screen $\Theta$, a matrix of random values must be generated whose statistics match Eq.~\ref{eq:Kolmogorov psf} and Eq.~\ref{eq:Kolmogorov weiner spectrum}.  If we could find $P(\mathbf{k})$, we could extract $\Theta$ using the inverse Fourier transform. We sample $P(\mathbf{k})$ from a random distribution with zero mean and variance equal to $\Phi(\mathbf{k})$. Such a method\cite{lane1992simulation} samples the Kolmogorov Weiner spectrum over a square grid of $N\times N$ points with indices $(i,j)$ as
\begin{equation}\label{eq:lane sampled }
    \Phi(i,j) = 0.023\left(\frac{2D}{r_0}\right)^{5/3}(i^2+j^2)^{-11/3},
\end{equation}
where $D$ is the length of the aperture over which we are considering $\Theta$. We would then multiply $\sqrt{\Phi}$ by a matrix $\mathbf{M}$ of random, complex values with zero mean and variance 1, giving us an instance of $P(\mathbf{k})$, and inverse Fourier transform the resultant matrix to produce the phase screen $\Theta$. However, this does not properly describe the $\norm{\mathbf{k}}^{-11/3}$ behaviour since the smallest sampled frequency is $1/D$.  Frequencies with periods greater than $D$ (corresponding to indices smaller than 1) are thus not included. To counteract this, subharmonic terms must be included by evaluating $\Phi$ at frequencies whose periods are greater than $D$ and multiplying by a weighting factor which takes into account the fractional sampling contribution. Taking the real component only gives
\begin{equation}\label{eq:FFT screen}
    \Theta = \mathfrak{R}\{\mathcal{F}^{-1}\{ \mathbf{M}\sqrt{\Phi} \}\}
\end{equation}
where $\Phi$ includes the subharmonic contributions. 

The phase screens must be calibrated to ensure that their turbulence strength is consistent with the programmed value of $D/r_0$. The Strehl ratio (SR) provides a convenient link between measured intensities and turbulence strength $D/r_0$. It is defined as the ratio of the average on axis intensity with, $\langle I(\mathbf{0})\rangle$, and without, $I_0(\mathbf{0})$, turbulence. For a plane wave in Kolmogorov theory it is
\begin{equation}\label{eq:SR}
    \text{SR} = \frac{\left<I(\mathbf{0})\right>}{I_0(\mathbf{0})} \approx \frac{1}{\left(1+(D/r_0)^{5/3}\right)^{6/5}}.
\end{equation}
Experimentally, one would calculate the SR for a range of turbulence strengths $D/r_0$ while adjusting the parameter $D$ of the phase screens so that the measured ($\langle I(\mathbf{0})\rangle/I_0(\mathbf{0})$) and programmed values agree. Fig.~\ref{fig:SR} shows such a calibration curve, where the solid line represents the theoretical (programmed) strength and the markers denote experimentally measured values. 

\section{Basis independent VQF propagation of uncertainty}

To record our intensity images, we used a FLIR Grasshopper3 CCD camera operating at 16-bit depth. Therefore the measurement uncertainty in a single pixel of a given normalized intensity image is $\Delta_{meas}\approx7.63\times10^{-6}$. The propagation of this uncertainty into a pixel in a calculated Stokes parameter is then $\Delta_{S'_i} = \sqrt{2\Delta_{meas}^2}$, while the uncertainty in the global parameters gains dependence in the number of pixels ($N$) integrated over $\Delta_{S_i} = \sqrt{2\Delta_{meas}^2 N^2}$. Finally the uncertainty in the VQF is a function of the measured VQF, $N$ and $\Delta I$ according to

\begin{equation}
    \Delta_{VQF} = \frac{\sqrt{8\Delta_{meas}^2 N^2}}{VQF}\,.
\end{equation}

In this work $N\in [100, 480]$ to accommodate changes in beam size (e.g. due to propagation from the tilted lens). Therefor our $\Delta_{VQF} \in [2.16\times10^{-3}, 1.04\times10^{-2}]$.


\begin{table}[t]
\centering
\begin{tabular}{ | c | c  | c | }
    \hline
	Mode & VQF value & VQF error  \\
    \hline
    \multicolumn{3}{|c|}{Figure 1(a)}   \\
	\hline
	$\ell=1$ & $0.996952$ & $\pm 0.002164$ \\
	$\ell=4$ & $0.999133$ & $\pm 0.003239$ \\
	\hline
    \multicolumn{3}{|c|}{Figure 1(c)}   \\
	\hline
	$z_1$,$\ell=1$ & $0.991797$ & $\pm 0.003263$ \\
	$z_1$,$\ell=4$ & $0.961124$ & $\pm 0.004715$ \\
	\hline
	$z_2$,$\ell=1$ & $0.974518$ & $\pm 0.004871$ \\
	$z_2$,$\ell=4$ & $0.960012$ & $\pm 0.006294$ \\
	\hline
	$z_3$,$\ell=1$ & $0.931651$ & $\pm 0.008339$ \\
	$z_3$,$\ell=4$ & $0.952852$ & $\pm 0.010871$ \\
	\hline
    \multicolumn{3}{|c|}{Figure 2(f)}   \\
	\hline
	In (Row 1) & $0.999133$ & $\pm 0.003239$\\
    Out (Row 1) & $ 0.960012$ & $ \pm 0.006294$\\
    In (Row 2) & $ 0.989600$ & $ \pm 0.003271$\\
    Out (Row 2) & $ 0.991154$ & $ \pm 0.003266$\\
	\hline
    \multicolumn{3}{|c|}{Figure 2(g)}   \\
	\hline
    Input & $ 0.999707 $ & $ \pm 0.008634 $\\
    Tilted Lens & $ 0.998234 $ & $ \pm 0.008647 $\\
    QWP & $ 0.988016 $ & $ \pm 0.008737 $ \\
	\hline
    \multicolumn{3}{|c|}{Figure 4(e)}   \\
	\hline
	In & $0.999418$ & $\pm 0.002807$ \\
	Out & $0.991506$ & $\pm 0.002829$ \\
	\hline
    \multicolumn{3}{|c|}{Figure 4(f)}   \\
	\hline
	In & $0.995465$ & $\pm 0.002818$\\
	Out & $0.995430$ & $\pm 0.002818$\\
	\hline
	\end{tabular}
	\caption{\textbf{VQF Values and errors. }}
	\label{tab:VQFvalues}
\end{table}

\section{Stokes Parameters}
In order to represent spatially non uniform polarisation, the following Stokes parameters were calculated using the intensity data from polarisation projections, as described in the Methods section.
\begin{figure*}
	\includegraphics[width=0.245\linewidth]{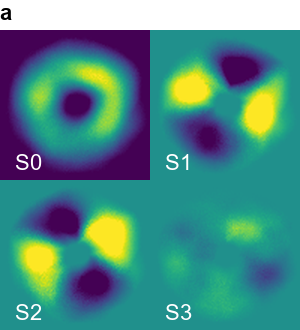}	
	\includegraphics[width=0.245\linewidth]{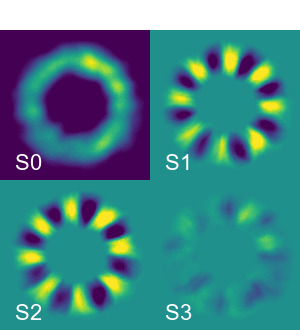} 
	\includegraphics[scale=1.045]{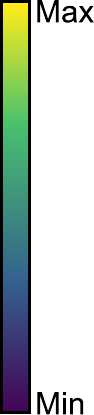}\\
	
	\includegraphics[width=0.245\linewidth]{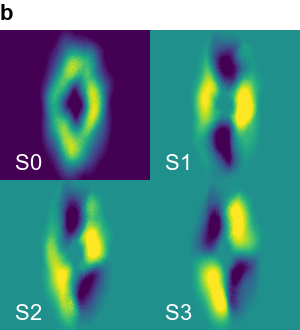}
	\includegraphics[width=0.245\linewidth]{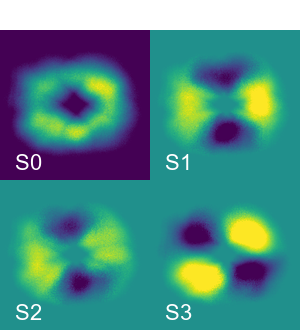}
	\includegraphics[width=0.245\linewidth]{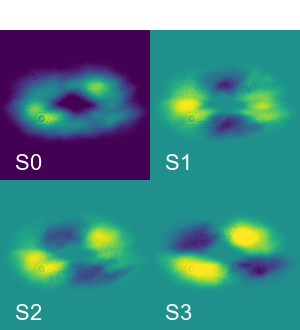}\\

	\includegraphics[width=0.245\linewidth]{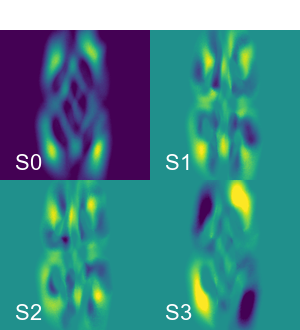}
	\includegraphics[width=0.245\linewidth]{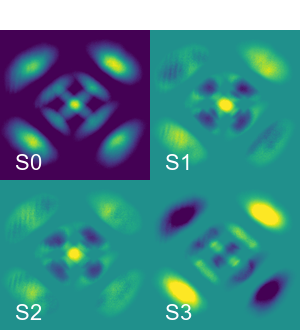}
	\includegraphics[width=0.245\linewidth]{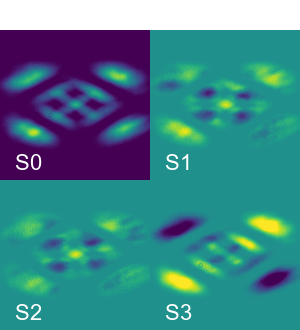}
	\caption{\textbf{Stokes Parameters for transverse modes going through a tilted lens in figure 2.} Column (a) Shows the parameters for the modes on the inset of 2(a). The set of figures (b) show the parameters for the modes depicted in figure 2(c) in sequence.The minimum and maximum values are $ \{0,1\} $ for $ S_0 $ and $ \{-1,1\} $ for $ S_{1,2,3} $.\label{fig:stokes1} }
\end{figure*}
In figure \ref{fig:stokes1} are the Stokes parameters for figure 2. The insets show the input modes and the respective parameters are in \ref{fig:stokes1}(a). When propagating through a tilted lens, the transformation of the polarisation structure can be seen in figure 2(c) and their respective parameters can be seen in \ref{fig:stokes1}(b).
\begin{figure*}
	\includegraphics[width=0.245\linewidth]{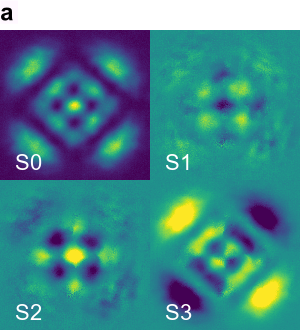}
	\includegraphics[width=0.245\linewidth]{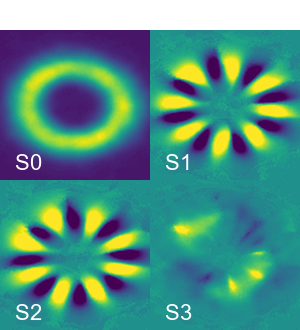}
	\includegraphics[scale=1.045]{SI_figs/colorbar4.pdf}\\
	\includegraphics[width=0.245\linewidth]{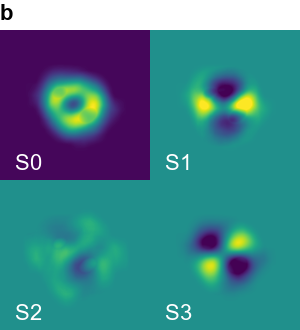}
	\includegraphics[width=0.245\linewidth]{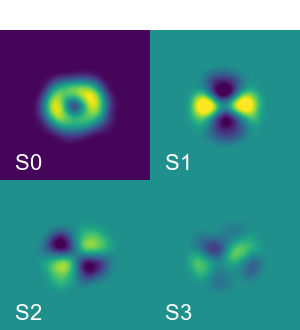}
	\includegraphics[scale=1.045]{SI_figs/colorbar4.pdf}
	\caption{\textbf{Stokes Parameters for the correction of modes depicted in figure 4.} Row (a) Shows the parameters for figure 4(c). On the second row (b) are the parameters for figure 4(d). The minimum and maximum values are $ \{0,1\} $ for $ S_0 $ and $ \{-1,1\} $ for $ S_{1,2,3} $. \label{fig:stokes2}}
\end{figure*}

Figure \ref{fig:stokes2} shows the stokes parameters for modes in figure 4. In figure 4(c) we show pre-channel correction, with the respective parameters in \ref{fig:stokes2}(a). In figure 4(d) we show post-channel correction and the respective parameters are in \ref{fig:stokes2}(b).
\begin{figure*}
	\includegraphics[width=0.245\linewidth]{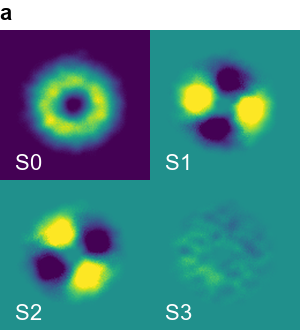}
	\includegraphics[width=0.245\linewidth]{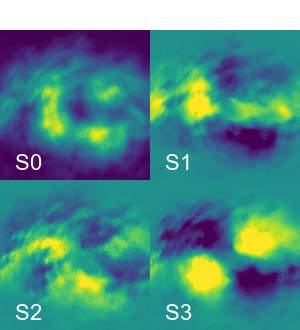}\includegraphics[scale=1.045]{SI_figs/colorbar4.pdf}\\
	\includegraphics[width=0.245\linewidth]{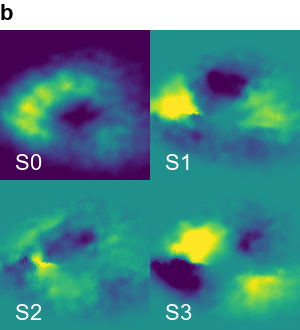}
	\includegraphics[width=0.245\linewidth]{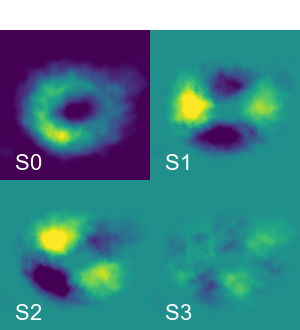}
	\includegraphics[scale=1.045]{SI_figs/colorbar4.pdf}
	\caption{\textbf{Stokes Parameters for modes afected by turbulence depicted in figure 6.} (a) Shows the parameters for figure 6(e) In (left) and Out (right). (b) Shows the parameters for figure 6(f) In (left) and Out (right). The minimum and maximum values are $ \{0,1\} $ for $ S_0 $ and $ \{-1,1\} $ for $ S_{1,2,3} $.\label{fig:stokes3} }
\end{figure*}

In figure \ref{fig:stokes3} are the parameters used to show the polarisation structure in figure 6(e) and (f). In (a) are the parameters for figure 6(e) where left corresponds to "In" mode and right to "Out" mode. In (b) are the parameters for figure6(f) where left is "In" and right is "Out".

\clearpage
\newpage
\bibliography{mybib_2}